\documentclass[aps,pra,superscriptaddress,reprint,nolongbibliography]{revtex4-2}

\usepackage{graphicx}
\usepackage{makecell}
\usepackage{lineno}
\usepackage{amsmath,amssymb}
\usepackage[table]{xcolor}
\usepackage{physics}
\usepackage{placeins}
\usepackage{adjustbox}
\usepackage{caption}
\usepackage{multirow}

\newcommand{\red}[1]{\textcolor{black}{#1}}
\usepackage[colorlinks=true,linkcolor=blue,citecolor=blue,urlcolor=blue]{hyperref}

\newcommand{\threeJ}[6]{\left(\begin{array}{ccc} #1 & #3 & #5 \\ #2 & #4 & #6 \end{array}\right )}

\begin{document}

\title{Production and spectroscopy of cold radioactive molecules}

\newcommand{\caltech}{\affiliation{Division of Physics, Mathematics, and Astronomy, California Institute of Technology;  Pasadena, CA, 91125, USA}}
\newcommand{\jhu}{\affiliation{Department of Chemistry, Johns Hopkins University; Baltimore, MD, 21218, USA}}

\author{Chandler J. Conn}
\thanks{These two authors contributed equally.}
\caltech

\author{Phelan Yu}
\thanks{These two authors contributed equally.}
\caltech

\author{Madison I. Howard}
\caltech

\author{Yuxi Yang}
\caltech

\author{Chaoqun Zhang}
\thanks{Present address: Department of Chemistry, Yale University; New Haven, CT, 06511, USA}
\jhu

\author{Arian Jadbabaie}
\thanks{Present address: Department of Physics, Massachusetts Institute of Technology; Cambridge, MA, 02139, USA}
\caltech

\author{Aikaterini Gorou}
\thanks{Present address: Department of Chemistry, University of California; Berkeley, CA, 94720, USA}
\caltech

\author{Alyssa N. Gaiser}
\affiliation{Department of Chemistry, Michigan State University; East Lansing, MI, 48824, USA}
\affiliation{Facility for Rare Isotope Beams, Michigan State University; East Lansing, MI, 48824, USA}

\author{Timothy C. Steimle}
\caltech

\author{Lan Cheng}
\jhu

\author{Nicholas R. Hutzler}
\email[]{hutzler@caltech.edu}
\caltech

\date{\today}

\begin{abstract}
Molecules with heavy, radioactive nuclei promise extreme sensitivity to fundamental nuclear and particle physics. However, these nuclei are available in limited quantities, which challenges their use in precision measurements.  Here we demonstrate the gas-phase synthesis, cryogenic cooling, and high-resolution laser spectroscopy of radium monohydroxide, monodeuteroxide, and monofluoride molecules ($^{226}$RaOH, $^{226}$RaOD, and $^{226}$RaF) in a tabletop apparatus by combining novel radioactive target production protocols, optically driven chemistry in a cryogenic buffer gas, and low-background spectroscopic detection methods. The molecules are cooled in the lab frame, creating conditions that are the same starting points as many current molecular precision measurement and quantum information experiments.  This approach is readily applied to a wide range of species and establishes key capabilities for molecular quantum sensing of exotic nuclei.

\end{abstract}

\maketitle

\begin{figure*}
    \centering
    \includegraphics[width=\linewidth]{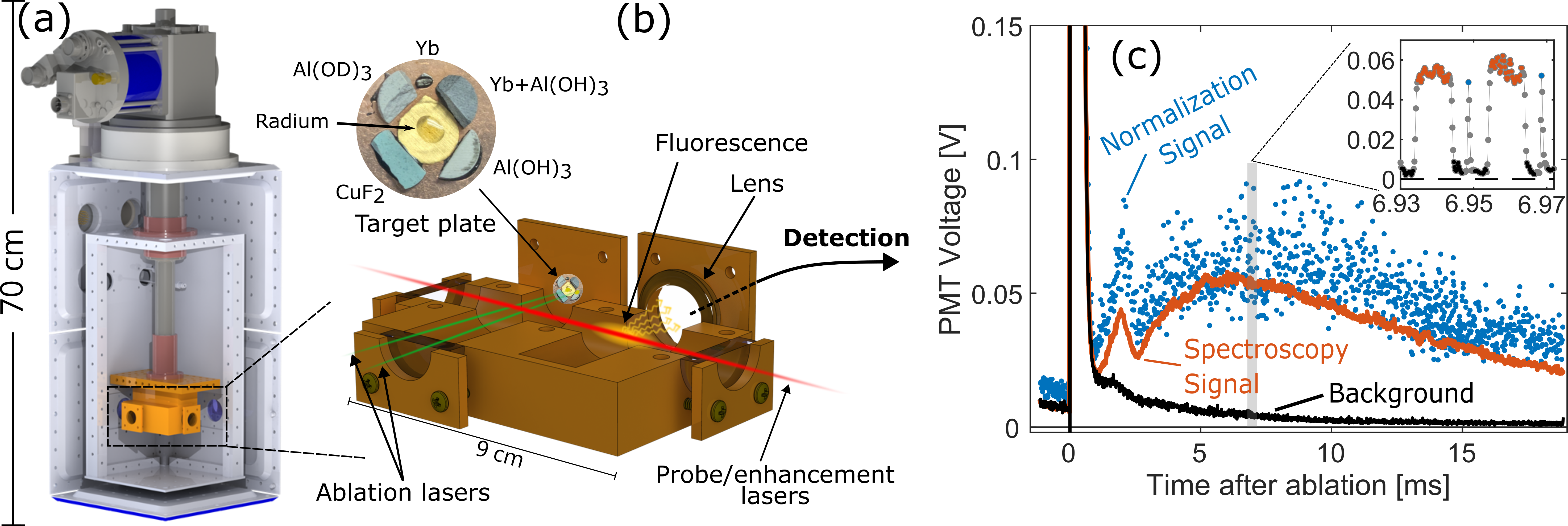}
    \caption{The experimental apparatus. (a) A cutaway render of the cryostat without gas or electrical feedthroughs. The entire system is less than a meter tall, and the experiment fits on an optical table. (b) Detail of the copper cell where molecules are produced and studied spectroscopically. Pulsed ablation lasers vaporize reagents off of a target plate, which fill the cell interior and thermalize with cold helium buffer gas. Probe lasers are sent through an orthogonal axis. A light collection lens, parallel to the ablation axis but offset by 4~cm, collimates laser-induced fluorescence to be detected by a photomultiplier tube external to the cryostat. The cell is vacuum-sealed to mitigate the spread of radioisotope contamination to the larger cryostat. The example target plate shows the drop cast radium target on gold foil surrounded by various reagents and Yb metal for testing.  (c) Resonant laser-induced fluorescence of RaOH molecules recorded after ablation while addressing the $\tilde{C}^2\Sigma^+ - \tilde{X}^2\Sigma^+$ band system. Time-dependent traces depict concurrently recorded fluorescence from fixed-frequency ``normalization" laser excitation (blue, top), and scanning ``spectroscopy" laser excitation (orange, middle), as well as non-resonant scatter from ablation and plasma backgrounds (black, bottom).  The inset is zoomed-in and depicts two cycles of amplitude-modulated fluorescence data, with the short, pulsed normalization probe (blue) interspersed with the longer, CW spectroscopy probe (orange). See supplemental fig.~\ref{fig:cycle} for further discussion.}
    \label{fig:fig1}
\end{figure*}

Radioactive atoms and molecules are sought for use in studying a wide range of physics including nuclear structure, fundamental symmetries, stellar processes, and more~\cite{RadMolWhitepaper2024}.  However, limited availability and safety hazards result in difficulties which inhibit every step of their study. For this reason, radioactive molecules in particular remain a relatively unexplored area with many unrealized applications. Recently, there has been great interest in molecules containing heavy, octupole-deformed (``pear-shaped'') nuclei, which amplify sensitivity to nuclear symmetry violations due to new physics by more than three orders of magnitude compared to spherical nuclei~\cite{dobaczewski2005nuclear,GarciaRuiz2020RaF,RadMolWhitepaper2024}. Combined with the roughly thousand-fold enhancement of molecules over atoms~\cite{Safronova2018Review}, these species are extremely sensitive probes of fundamental nuclear properties and physics beyond the Standard Model (BSM). 

Among the possible octupole-deformed candidates, radium-containing molecules are compelling due to the large and relatively well-characterized octupole shape deformation of radium~\cite{Gaffney2013,Dobaczewski2018Correlating}, their large molecular sensitivity to fundamental symmetry violations via a nuclear Schiff moment~\cite{Auerbach1996OctDev,Flambaum2002NSMAtoms,dobaczewski2005nuclear,Kudashov2014}, and their fairly unique ability to create optically controllable, laser-coolable molecules~\cite{Isaev2013RaF,Kudashov2014,Isaev2017RaOH,zhang2023intensity}.  By combining modern quantum tools with the significant sensitivity enhancements to BSM physics afforded by molecules containing heavy, deformed nuclei, one can probe far above TeV energy scales, complementing and extending the reach of state-of-the-art colliders and precision measurements~\cite{RadMolWhitepaper2024,Alarcon2022Snowmass}.

For precision atomic and molecular experiments, achieving a high degree of motional and internal quantum state control is an essential prerequisite as it enables long interrogation times as well as coherent, quantum state-resolved preparation, manipulation, and readout. Recent approaches have made progress in this direction with radium-containing molecules, including RaF spectroscopy using accelerated radioisotope beams~\cite{GarciaRuiz2020RaF, udrescu2021isotope, udrescu2024precision, athanasakis2025electron} and the trapping of radium-containing polyatomic molecular ions~\cite{fan2021optical}, both with an eye toward precision measurement of fundamental symmetries~\cite{Isaev2013RaF,yu2021probing}.

Here we report the production, cooling, and high-resolution laser spectroscopy of radium-226 monohydroxide, monodeuteroxide, and monofluoride molecules ($^{226}$RaOH, $^{226}$RaOD, and $^{226}$RaF) at $\sim 4$~K temperatures in a tabletop apparatus. 
Using pulsed laser ablation of fabricated radium targets, cryogenic buffer gas cooling~\cite{hutzler2012buffer}, resonant optical driving of state-selective chemical reactions~\cite{jadbabaie2020enhanced}, and high-sensitivity low-background detection methods, we create and study cold and low-velocity ($<$ 30 m/s) samples. 
These species have structures amenable to laser cooling and optical trapping for long coherence times and high-fidelity quantum state control and readout~\cite{fitch2021laser,Augenbraun2023PolyLCReview,Isaev2013RaF,Kudashov2014,Isaev2017RaOH,zhang2023intensity}.  The polyatomic species furthermore feature near-degenerate states of opposite parity which enable advanced protocols for precision measurement~\cite{Hutzler2020PolyReview,kozyryev2017precision,takahashi2023engineering,anderegg2023quantum} and quantum information~\cite{Yu2019PolyQIS}.
This work illustrates a pathway for applying precision molecular science tools -- particularly in the areas of BSM searches~\cite{DeMille2024QSReview,Chupp2019Review,Safronova2018Review} and quantum information~\cite{Cornish2024Review,Langen2024Review} -- to short-lived, radioactive systems.
More generally, the availability of radioactive molecules for tabletop experiments in a university (or larger) setting enables wide-ranging applications in fundamental physics~\cite{RadMolWhitepaper2024}.

In addition to the usual difficulties of working with molecules for precision measurements~\cite{Safronova2018Review}, the primary challenge limiting access to radioactive molecules is low material quantity, often less than a few micrograms. A compounding challenge is the large theoretical uncertainties for these very heavy molecules; the electronic energy levels can be predicted~\cite{zhang2023relativistic} with $\sim$~10~THz uncertainty, which is many orders of magnitude larger than the $\sim100$~MHz linewidth of individual quantum states at cryogenic temperatures.  In combination, these challenges mean that molecule production and measurement strategies must be efficient. By using an array of broadly tunable continuous-wave (CW) and pulsed dye lasers with a hierarchy of linewidths, we demonstrate material-efficient spectral searches ranging over $>10$~THz for {\it a priori} unknown optical transitions from previously unobserved radioactive molecules. Subsequent laser excitation with progressively decreasing laser linewidth down to single longitudinal mode ($\Delta\nu\leq$ 500 kHz) enables us to directly resolve individual, low-lying $(N\leq 9)$ rotational transitions on molecular vibronic bands with $\sim 100$ MHz Doppler-limited linewidth and few-MHz resolution. Our developments on the radiochemical target preparation, cryogenic source design, chemical production, spectroscopic methods, and detection approaches are readily extendable to a broad range of short-lived radioisotopes and complex molecular structures.
\vspace{-0.1 cm} 
\section*{Methods}
\begin{figure}
    \centering
    \includegraphics[width=\linewidth]{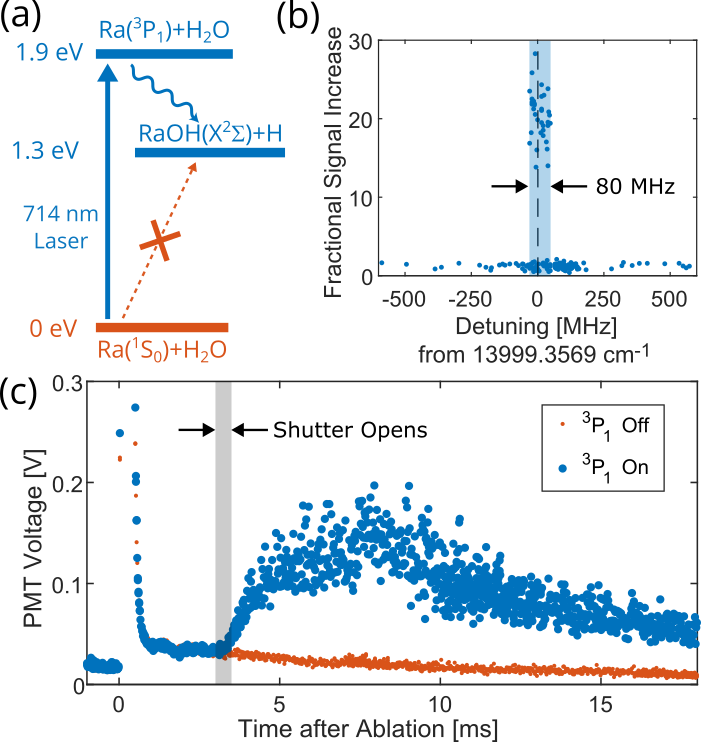}
    \caption{Molecular production via resonant, optically driven chemistry in the cryogenic gas. (a) Energy levels of Ra reacting with water. The ground $^1S_0$ state cannot react to form RaOH, but the excited $^3P_1$ state can, so we excite the atoms with a strong 714~nm laser.  (b) The dependence of RaOH production enhancement on the $^3P_1$ laser detuning from resonance. (c) Single-shot fluorescence traces demonstrating resonant production of RaOD.  When the Ra $^3P_1$ laser is off-resonant (``off''), we see some production of RaOD, which is enhanced in the presence of an on-resonant (``on'') laser.  We open a shutter around 3~ms after the ablation pulse to distinguish between molecules created in the ablation plume versus by the resonant chemistry. Following an initial ablation shot on a new drop cast target spot, we estimate that up to $\sim 10^{10}$ total molecules are produced (see~\cite{supmat} for details on the number estimation).}
    \label{fig:fig2}
\end{figure}
Cold gas-phase radioactive atoms and molecules are produced inside a copper cryogenic buffer gas cell approximately 48 cm$^3$ in volume, which is held at a base temperature between 4~K and 7~K by a commercial closed-cycle cryocooler. The geometry and optical paths of the cryogenic cell are depicted in Fig. \ref{fig:fig1}.  First, atomic and molecular precursors are vaporized by laser ablation of solid targets.  Next, molecules are formed by optically driven chemical reactions between radium atoms and reagents.  Finally, laser-induced fluorescence from tunable lasers is collected onto a photodetector to measure molecular spectra.  These steps are described in detail below.

Within the cell, atomic and molecular precursors are ablated by focused, nanosecond 532 nm Nd:YAG lasers delivering 5-20 mJ of energy per pulse, depending on the target. Molecule production occurs via ablation of separate targets for the radioisotope and the ligand species, allowing for species-selective production by switching between targets \textit{in-situ}. Radioisotope targets containing $10-50$~$\mu$Ci ($0.37-1.85$ MBq, $44-221$ nmol) of radium chloride (RaCl$_2$) or nitrate (Ra(NO$_3$)$_2$) salts are fabricated in-house via manual pipetting (``drop casting'') of weakly acidic aqueous salt solution onto a heated gold surface. A small amount of xylitol is dissolved into the solution to improve ablation target consistency and adhesion following evaporation. Reagent ``co-targets'' are solid pellets formed by hydraulic pressing of fluoride-, hydroxide-, or deuteroxide-containing powders. Production of different molecules is achieved by steering the ablation laser to specific co-targets, of which there can be many (ten or more) in a single cell (Fig.~\ref{fig:fig1}b), with significant chemical diversity, limited only by the available space on the target plate.

After ablation, the gas-phase atomic and molecular products rapidly cool to cryogenic temperatures via collisions with pre-loaded helium buffer gas.  This means that many possible chemical reactions which would form the desired molecules become energetically forbidden, stifling molecule production (Fig.~\ref{fig:fig2}a).  To overcome this,
radium atoms are excited from the $^1S_0$ ground state into the $^3P_1$ metastable state ($\lambda = 714$~nm, $\tau \approx$ 422 ns~\cite{scielzo2006measurement}) using 1.5~W of CW laser light operating close to population saturation. The $^3P_1 - ^1S_0$ transition is spin-forbidden, but made allowed by extremely strong spin-orbit mixing. In the $^3P_1$ state, the Ra atoms have enough internal energy to overcome additional reaction barriers, such as the one shown in Fig.~\ref{fig:fig2}a, resulting in a factor of $\sim 10-30$ increase in molecule number (Figs.~\ref{fig:fig2}b,c).  Note that the resonant behavior of this reaction gives a strong systematic check that the molecules are radium-containing even in the presence of significant chemical contamination or spectral congestion~\cite{Pilgram2021YbOHOdd}. This method works to enhance production of RaOH, RaOD, and RaF, as well as BaOH and BaF, which are used as test and calibration species in this apparatus. 

Atomic Ra densities inside the cell are measured via resonant optical absorption on the strong $^1P_1 - ^1S_0$ line at 483~nm  ($\tau \sim 6$ ns~\cite{dzuba2006calculation}). Approximately $10^{10}-10^{12}$ cold Ra atoms are generated per ablation pulse. Each radium target yields $\sim 0.5 - 2\times 10^4$ shots before being depleted, depending on the amount deposited and the deposition procedure (see Fig.~\ref{fig:Ra_target}). 

Molecular products are detected via laser-induced fluorescence from both pulsed and CW lasers. An integrated lens in the cryogenic cell assembly collects light from a focal plane in the middle of the cell bore for real-time readout via a low-noise photomultiplier tube (PMT) module.  This detection presents several challenges: the constrained geometry results in significant light scatter from the various lasers off of surfaces; non-resonant scattering occurs off of gas-phase clusters and macromolecules created from ablation; and metastable excited states of various atoms and molecules created from the ablation emit broadband fluorescence. We treat the cell with low-reflectance surface coatings and detect fluorescence at a different wavelength from any excitation laser so that scattered laser light can be blocked with optical filters.  Resonant and non-resonant scattering are distinguished by scanning the laser frequencies to directly observe resonant behavior.  Broadband background fluorescence is differentiated from laser-induced fluorescence signals by using amplitude-modulated lasers, typically in the 50-100~kHz range, combined with lock-in detection.

The lock-in detection approach also gives a useful tool to normalize against shot-to-shot fluctuations in atomic and molecular yield in a material-efficient way.  By temporally interleaving two lasers (Fig.~\ref{fig:fig1}c) within each molecule pulse, one with a varying frequency for measuring a spectrum and the other having a fixed frequency at a known molecular resonance for signal normalization, we can monitor molecule production to reduce noise without needing to average many ablation shots. This method also verifies that any lack of spectral signal was not simply a result of poor molecule production due to ablation of a depleted target spot.

\begin{figure*}
    \centering
    \includegraphics[width=1\linewidth]{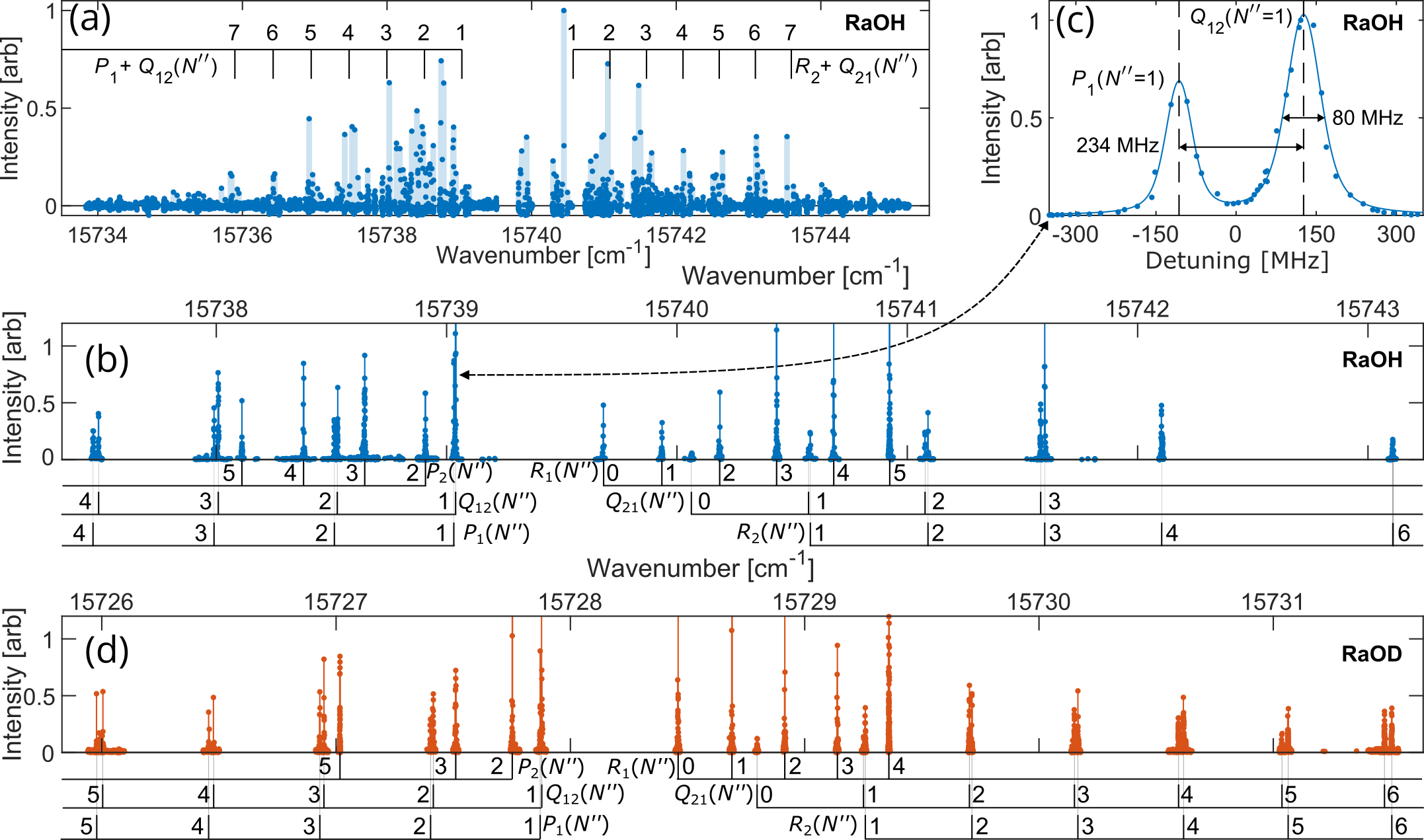}
    \caption{Laser-induced fluorescence spectra of the RaOH and RaOD $\tilde{C}^2\Sigma^+-\tilde{X}^2\Sigma^+$ origin band systems.  Each point represents a single ablation pulse. (a) Pulsed-dye laser fluorescence spectrum of RaOH (without normalization) at $\sim$~2.5~GHz resolution set by the laser linewidth.  (b) CW dye laser fluorescence spectrum of RaOH with $\sim$~100~MHz linewidth set by the Doppler temperature and radiative broadening. (c) Zoom-in on a spin-rotation doublet from (b) showing the lineshape with Voigt fit.  (d) CW dye laser fluorescence spectrum of RaOD. Table \ref{tab:sm-raoxlines} in the Supplemental Text contains the full line list and comparison to prediction for both RaOH and RaOD.}
    \label{fig:fig3}
\end{figure*}
\begin{table*}
\caption{\label{tab:raoh_params}
Fitted parameters for RaOH and RaOD (Figs. \ref{fig:fig3}b,d). 
Theory values for band origins ($T_0$) are taken from Ref. ~\onlinecite{zhang2023relativistic}. Theory values for rotational constants ($B_0$) and spin-rotation constants ($\gamma$)
were obtained from relativistic exact two-component equation-of-motion coupled-cluster (X2C-EOM-CC) \cite{Zhang22,Asthana19} calculations
(see Supplemental Text regarding the methods and comparison to experiment). Values in parentheses are standard errors (1-sigma level) from fitting, while those in brackets are due to wavemeter uncertainties.
}
\begin{ruledtabular}
\begin{tabular}{l|llll|l|lllll}
RaOH & \multicolumn{2}{c}{$\tilde{X}^2\Sigma^+_{1/2}(000)$} & \multicolumn{2}{c|}{$\tilde{C}^2\Sigma^+_{1/2}(000)$} & RaOD & \multicolumn{2}{c}{$\tilde{X}^2\Sigma^+_{1/2}(000)$} & \multicolumn{2}{c}{$\tilde{C}^2\Sigma^+_{1/2}(000)$}\\
\cline{2-3} \cline{4-5} \cline{7-8} \cline{9-10}
\textit{(high-res)} & Measured & Theory & Measured & Theory & \textit{(high-res)} & Measured & Theory & Measured & Theory\\
\hline
$T_0/$cm$^{-1}$ & 0 & 0 & 15739.4220(2)[20] & 15749 & $T_0/$cm$^{-1}$ & 0 & 0 & 15728.2246(2)[20] & 15730\\
$B_0/$MHz & 5814.3(7) & 5818 & 5780.3(6) & 5783 & $B_0/$MHz  & 5256.3(6) & 5258 & 5208.9(6) & 5226\\
$\gamma/$MHz & 151(3) & 165 &--7641(4) & --13574 & $\gamma/$MHz & 143(2) & 149 & --6739(3) & --12342
\end{tabular}
\end{ruledtabular}
\end{table*}

\section*{Results and Discussion}

Laser spectroscopy of RaOH and RaOD focused on the $\tilde{C}^2\Sigma^+-\tilde{X}^2\Sigma^+$ electronic transition, which approximately corresponds to the single $ p\sigma \leftarrow s\sigma$ excitation of a valence, metal-localized electron \cite{ivanov2019towards, Augenbraun2023PolyLCReview}.  A probe laser excites $\tilde{C}^2\Sigma^+(000) \leftarrow \tilde{X}^2\Sigma^+(000)$, and fluorescence from the decay $\tilde{C}^2\Sigma^+(000)\rightarrow \tilde{X}^2\Sigma^+(\nu_1\nu_2\nu_3)$ is detected.  Here $(\nu_1\nu_2\nu_3)$ labels the number of quanta in the (Ra--O stretch, Ra--O--H bend, O--H stretch) vibrational modes~\cite{kozyryev2017precision}, and the relative decay rates to each state are governed by the magnitudes of Franck-Condon factors~\cite{Augenbraun2023PolyLCReview}.  Because these molecules are predicted to be laser-coolable, the electronic and vibrational degrees of freedom are largely decoupled and the dominant decay is back down to (000)~\cite{zhang2023relativistic}; however, since that decay wavelength is the same as the intense probe laser, it is optically filtered so that only decays to excited vibrational states $\nu_1=1$ and $\nu_2=2$ are detected.

In order to efficiently find and explore the spectra given the large theoretical uncertainty and limited material,
spectra were collected in phases with increasing resolution. 
The initial broadband CW survey spectroscopy for the RaOH $\tilde{C}$ state was guided by prior relativistic coupled-cluster electronic structure calculations~\cite{zhang2023relativistic}, which pointed to a search window on the order of 100~cm$^{-1}$ near 15750~cm$^{-1}$. An efficient search strategy was devised to scan over the full theory uncertainty range with a single 10 $\mu$Ci radioisotope target.
We used a spectrally broad ($\Delta\nu\sim 30$~GHz $\sim$ 1~cm$^{-1}$) tunable CW laser to scan for laser-induced fluorescence, which enabled coverage of this large area in only a few hours and a few hundred ablation shots.  By using a CW laser we utilized the full few-ms duration of the molecular pulse.
From this initial survey data, a region of excess fluorescence at 15740 cm$^{-1}$ was located (see Fig. \ref{fig:RaOH_lowres} in Supplemental Text), which we tentatively assign as the $\tilde{C}^2\Sigma^+ \to \tilde{X}^2\Sigma^+$ origin system.

Subsequent scans with increasing resolution were performed over increasingly narrow regions.  Medium resolution ($\Delta\nu\sim 2.5$ GHz) pulsed dye laser excitation revealed a tightly bunched band head with isolated, rotationally resolved progressions in the wings, as depicted in Fig. \ref{fig:fig3}a.  The pulsed dye laser was operated with 50~kHz repetition rate, providing $\sim$1000 fluorescence pulses over the length of the molecular pulse. The clustered band head appearance, atypical for a highly diagonal $\Sigma^+ - \Sigma^+$ band, is attributable to the large, negative excited state spin rotation of the $\tilde{C}^2\Sigma^+$ state, which pushes rotational $R_1$ and $P_2$ branch progressions towards the band origins. This is likely a consequence of large spin-orbit coupling ($A_{SO}\sim 1500$ cm$^{-1}$) to the $|\Omega| = 1/2$ component of the $\tilde{A}^2\Pi$ electronic manifold that is predicted to lie $\sim 2000$ cm$^{-1}$ below the $\tilde{C}^2\Sigma^+$ manifold~\cite{zhang2023relativistic} (see Supplemental Text).
Finally, individual lines were resolved at high resolution with a single-frequency ($\Delta\nu\sim 0.5$~MHz) tunable CW dye laser, giving linewidths of $\sim$100~MHz FWHM due to Doppler and natural broadening, and enabling the determination of line centers to $\sim$~MHz, as depicted in Figs. \ref{fig:fig3}c and \ref{fig:RaOD} in the Supplemental Text.

A similar spectroscopic search was performed for the deuterated isotopologue RaOD, locating a band $\sim 9$ cm$^{-1}$ to the red of the main origin feature for RaOH, in reasonable agreement with electronic structure predictions for vibrational isotope shifts. The features at both bands show clear dependence on ablation of the (un)deuterated co-target as well as resonant pumping into the Ra $^3P_1$ state, indicative of optically driven chemical production (see Fig. \ref{fig:fig2}c and Fig. \ref{fig:RaOD}d in the Supplemental Text). Substituting H$\rightarrow$D (or even H$\rightarrow$T) should lower the bending vibration energy, thereby increasing the spontaneous lifetime of the symmetry-lowered ``science state''~\cite{kozyryev2017precision,Vilas2023BBR}.

\begin{figure}
    \centering
    \includegraphics[width=\linewidth]{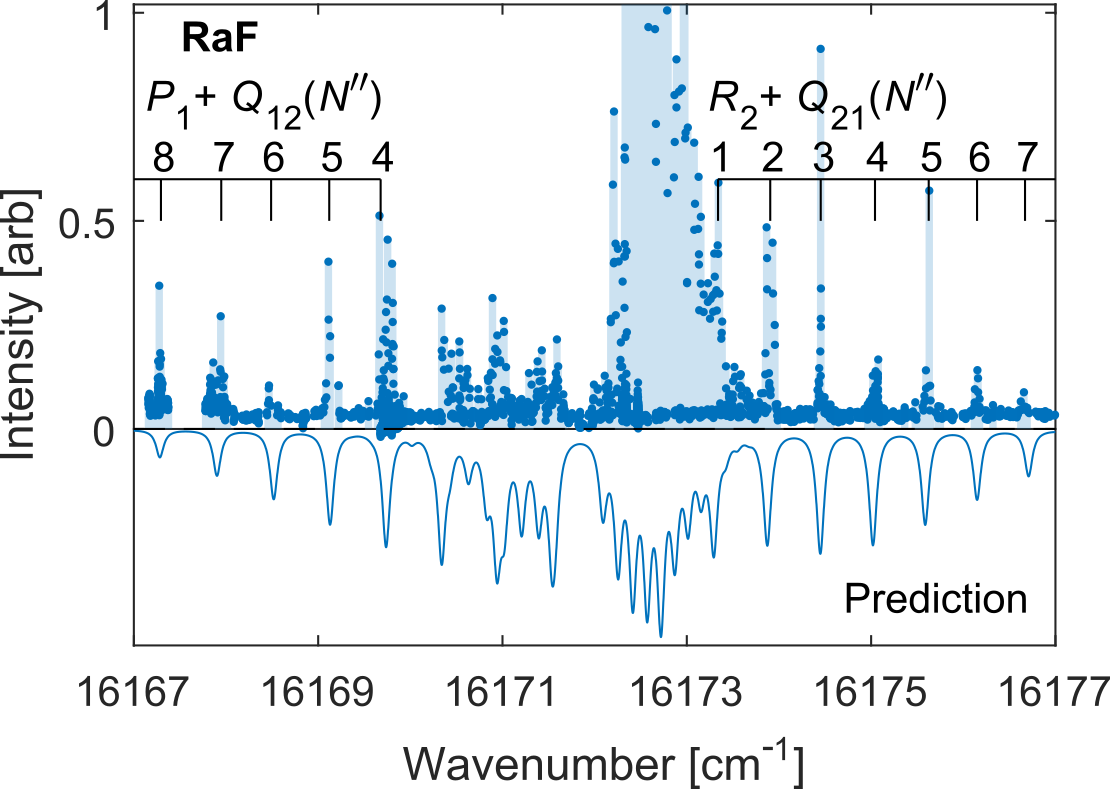}
    \caption{Fluorescence spectrum without normalization of the RaF $C^2\Sigma^+_{1/2}(v' = 0) - X^2\Sigma^+_{1/2} (v''= 1)$ vibronic band system, recorded via pulsed dye excitation at $\Delta\nu\sim 2.5$ GHz laser linewidth. The top panel contains background-subtracted fluorescence spectra, identifying resolved lines from the $P_1$ and $R_2$ branch progressions. The bottom panel shows simulated band features based on fit parameters (see Table \ref{tab:raf_params}), laser linewidth, and rotational temperature $T=7$~K.}
    \label{fig:fig4}

\captionof{table}{\label{tab:raf_params}
Fitted parameters for the RaF $C^2\Sigma^+$ state compared with X2C-EOM-CC calculated values (see Supplemental Text). $T_0$ is referenced to the $C^2\Sigma^+(v'=0) \to X^2\Sigma^+(v''=1)$ vibronic transition (Fig. \ref{fig:fig4}), whose theory values are from Ref. \onlinecite{zhang2023relativistic}.}

\begin{ruledtabular}
\begin{tabular}{l|lll}
\textit{(mid-res)} & $T_0/$cm$^{-1}$ & $B_0/$MHz & $\gamma/$MHz\\
\hline
Measured & 16171.920(8) & 5663(6) & --12388(90)\\
Theory & 16199  & 5647 &  --13443\\
\end{tabular}
\end{ruledtabular}

\end{figure}

Twenty-seven and thirty-two low-$N$ lines on the RaOH and deuterated RaOD $\tilde{C}^2\Sigma^+ - \tilde{X}^2\Sigma^+$ systems, respectively, were recorded via narrowband CW laser excitation.  Here $N$ is the angular momentum of the molecule not including electron or nuclear spin.
A list of line centers determined via Voigt profile fit can be found in Table \ref{tab:sm-raoxlines}. Splittings due to the spin-rotation interaction are resolved for all lines down to the lowest $N = 1$ rotational states. Hyperfine coupling from the proton and deuteron spin is expected on the order of $\sim 1$ MHz, which is below the present linewidth and therefore unresolved. In contrast to the severe perturbations observed in the first excited $^2\Sigma^+$ state of BaOH~\cite{kinsey1986rotational, gustavsson1991perturbations}, the low-$N$ progressions into the $\tilde{C}^2\Sigma^+$ manifold in RaOH/D appear largely unperturbed.  However, as discussed in the Supplemental Text, the intensities of some features and the excited state spin-rotation constants are indicative of spin-orbit induced mixing of the $\tilde{C}^2\Sigma^+$ state with adjacent states.

The complete RaOH/D high-resolution line lists were fit to a 5-parameter molecular fine-structure Hamiltonian model with rotation ($B_0$) and spin rotation ($\gamma$), whose operator form can be written as:
 \begin{align}
     \hat{H}_g &= B_0''\hat{N_g}^2 + \gamma'' \hat{N_g}\cdot \hat{S_g}\\
     \hat{H}_e &= T_0 + B_0'\hat{N_e}^2 + \gamma' \hat{N_e}\cdot \hat{S_e}
 \end{align}
where the double and single primes as well as $g$, $e$ subscript reference the ground and excited states, respectively, and $T_0$ is the excited state origin. The angular momentum matrix elements of the Hamiltonians are written in Section \ref{sec:effective_ham} of the Supplemental Text.

Rotational constants and deuterated isotope shifts for both the $\tilde{X}^2\Sigma^+$ and $\tilde{C}^2\Sigma^+$ states are in excellent agreement with bond lengths and moments of inertia calculated in~\cite{zhang2023relativistic}. A full set of extracted molecular parameters with \textit{ab initio} theory comparisons are listed in Table \ref{tab:raoh_params}. Good agreement is reached between observation and calculation of the ground spin-rotation constants, while a discrepancy exists in the magnitude of the excited state value. This suggests the influence of additional perturbations on the excited manifold from adjacent electronic states, which have been observed in similar systems~\cite{kinsey1986rotational, gustavsson1991perturbations} but whose effects are not captured in the present electronic structure calculations.

A medium-resolution spectrum of the analogous $C^2\Sigma^+ - X^2\Sigma^+$ transition in RaF, which was first observed in \cite{GarciaRuiz2020RaF, athanasakis2025electron}, was recorded via pulsed dye excitation of the $v'' = 1 \to v' = 0$ band system (see Fig. \ref{fig:fig4}). Detection was performed on the strong $v' = 0 \to v'' = 0$ decay via optical filtering. Similar to the RaOH data, isolated low-$N$ features in the $Q_{21} + R_2$ and $P_1 + Q_{12}$ progressions were resolved at the pulsed dye linewidth. As the ground state $X^2\Sigma^+$ parameters for RaF have already been determined~\cite{udrescu2024precision}, these features can be used to extract molecular parameters for the $C^2\Sigma^+$ state, which are listed in Table \ref{tab:raf_params}.

\section*{Outlook}

We have demonstrated cryogenic production and high-resolution laser spectroscopy of cold radioactive molecules, as well as achieved the detection and characterization of the first neutral radium-containing polyatomic molecules, putting them in position to benefit from a wide range of modern quantum science tools. The techniques developed in this work are broadly applicable, establishing a new pathway for precision study of a wide range of radioactive species~\cite{RadMolWhitepaper2024} in a tabletop setting. 

With the generation of cold samples, advanced spectroscopic techniques currently used on stable species to improve bandwidth and resolution of spectra are now applicable for the study of radioactive atoms and molecules. 
Broadband light sources in combination with high-optical-dispersion detectors can enable fast, multi-THz broadband spectral acquisition while preserving rotational and fine structure resolution~\cite{patel2025rapid}. 

Formation of molecular beams~\cite{hutzler2012buffer} will enable laser cooling, which is expected to be feasible in these species~\cite{Isaev2013RaF,Isaev2017RaOH,zhang2023intensity,udrescu2024precision}, and therefore open up a wide range of precision measurement applications.  These molecules have the electronic structure for which many advanced optical trapping~\cite{anderegg2019optical, vilas2024optical} and precision measurement schemes~\cite{hudson2011improved,kozyryev2017precision,yu2021probing,takahashi2023engineering, anderegg2023quantum} have been both proposed and demonstrated.
Our approach can be used for many radioactive systems, including 
molecular isotopologues \cite{udrescu2021isotope, wilkins2025observation} containing the spinful, octupole-deformed $^{225}$Ra ($I = 1/2$) and $^{223}$Ra ($I = 3/2$) isotopes, which we anticipate being able to study using this approach given the demonstrated signal-to-noise ratio, and which would enable highly sensitive searches for $CP$-violating hadronic physics at the frontiers of current experimental limits. 

\begin{acknowledgments}

The authors thank Haick Issaian and Andr\'e Jefferson for extensive radiation safety support and advice throughout the planning and operation of the project, as well as Sonjong Hwang and the Caltech Solid State NMR Facility for verification of synthesized reagents. C. J. C. and P. Y. thank Ana Duarte for assistance in obtaining laser hardware, Yi Zeng for contributions to the control system and prototyping lasers, Mumtaz Gababa and Zitian Ye for assistance with developing the cryogenic apparatus, as well as Chi Zhang for advice on low-noise RF design. The authors also thank John M. Doyle, Ronald F. Garcia Ruiz, and the RaX Collaboration for helpful discussions and comments on the manuscript. We acknowledge Oak Ridge National Laboratory and Eckert \& Ziegler for supplying the isotopes used in this work.  Some of the isotopes used in this research were supplied by the U.S. Department of Energy Isotope Program, managed by the Office of Science for Isotope R\&D and Production. Computational modeling of chemical reactions were conducted in the Resnick High Performance Computing Center, a facility supported by Resnick Sustainability Institute at Caltech. 

\textbf{Funding:} Experimental work at Caltech was supported by the National Science Foundation (PHY-2309361), a De Logi Science and Technology Grant, the Heising-Simons Foundation (2022-3361), and an Alfred P. Sloan Research Fellowship. Computational and theoretical work at Johns Hopkins University (C. Z. and L. C)
was supported by the National Science Foundation, under Grant No. PHY-2309253.
P. Y. acknowledges support from a Caltech Eddleman Graduate Fellowship. M. I. H. acknowledges support from an NSF Graduate Research Fellowship and the Caltech Dominic Orr Graduate Fellowship. A.~G. acknowledges support from the Thomas Lauritsen Caltech SURF Fellowship. 

\textbf{Author contributions:} C.~J.~C., P.~Y., and N.~R.~H. conceived of the project. C.~J.~C. and P.~Y. designed and built the experiment and with M.~I.~H. and Y.~Y. collected and analyzed the data. C.~Z. and L.~C. developed and implemented electronic structure calculations. T.~C.~S. performed spectral modeling and fits and assisted with lasers. A.~N.~G. designed and implemented wet radiochemical procedures.  A.~J. assisted with measurement protocol development. A.~G. performed computational modeling of chemical reaction pathways and optically driven production. N.~R.~H. supervised the project. 

\textbf{Competing interests:} A patent application covering inventions described in this report (serial \# 19/551,368) has been filed in the U.S. with the following listed inventors: N.~R.~H., C.~J.~C., P.~Y., M.~I.~H., Y.~Y. 

\textbf{Data and materials availability:} Underlying data and analysis code produced in this work is available on the Caltech Research Data Repository~\cite{hutzler_conn_yu_2026}.

\end{acknowledgments}

\bibliography{references}

\begin{thebibliography}{82}%
\makeatletter
\providecommand \@ifxundefined [1]{%
 \@ifx{#1\undefined}
}%
\providecommand \@ifnum [1]{%
 \ifnum #1\expandafter \@firstoftwo
 \else \expandafter \@secondoftwo
 \fi
}%
\providecommand \@ifx [1]{%
 \ifx #1\expandafter \@firstoftwo
 \else \expandafter \@secondoftwo
 \fi
}%
\providecommand \natexlab [1]{#1}%
\providecommand \enquote  [1]{``#1''}%
\providecommand \bibnamefont  [1]{#1}%
\providecommand \bibfnamefont [1]{#1}%
\providecommand \citenamefont [1]{#1}%
\providecommand \href@noop [0]{\@secondoftwo}%
\providecommand \href [0]{\begingroup \@sanitize@url \@href}%
\providecommand \@href[1]{\@@startlink{#1}\@@href}%
\providecommand \@@href[1]{\endgroup#1\@@endlink}%
\providecommand \@sanitize@url [0]{\catcode `\\12\catcode `\$12\catcode `\&12\catcode `\#12\catcode `\^12\catcode `\_12\catcode `\%12\relax}%
\providecommand \@@startlink[1]{}%
\providecommand \@@endlink[0]{}%
\providecommand \url  [0]{\begingroup\@sanitize@url \@url }%
\providecommand \@url [1]{\endgroup\@href {#1}{\urlprefix }}%
\providecommand \urlprefix  [0]{URL }%
\providecommand \Eprint [0]{\href }%
\providecommand \doibase [0]{https://doi.org/}%
\providecommand \selectlanguage [0]{\@gobble}%
\providecommand \bibinfo  [0]{\@secondoftwo}%
\providecommand \bibfield  [0]{\@secondoftwo}%
\providecommand \translation [1]{[#1]}%
\providecommand \BibitemOpen [0]{}%
\providecommand \bibitemStop [0]{}%
\providecommand \bibitemNoStop [0]{.\EOS\space}%
\providecommand \EOS [0]{\spacefactor3000\relax}%
\providecommand \BibitemShut  [1]{\csname bibitem#1\endcsname}%
\let\auto@bib@innerbib\@empty
\bibitem [{\citenamefont {Arrowsmith-Kron}\ \emph {et~al.}(2024)\citenamefont {Arrowsmith-Kron}, \citenamefont {Athanasakis-Kaklamanakis}, \citenamefont {Au}, \citenamefont {Ballof}, \citenamefont {Berger}, \citenamefont {Borschevsky}, \citenamefont {Breier}, \citenamefont {Buchinger}, \citenamefont {Budker}, \citenamefont {Caldwell}, \citenamefont {Charles}, \citenamefont {Dattani}, \citenamefont {de~Groote}, \citenamefont {DeMille}, \citenamefont {Dickel}, \citenamefont {Dobaczewski}, \citenamefont {D{\"u}llmann}, \citenamefont {Eliav}, \citenamefont {Engel}, \citenamefont {Fan}, \citenamefont {Flambaum}, \citenamefont {Flanagan}, \citenamefont {Gaiser}, \citenamefont {Ruiz}, \citenamefont {Gaul}, \citenamefont {Giesen}, \citenamefont {Ginges}, \citenamefont {Gottberg}, \citenamefont {Gwinner}, \citenamefont {Heinke}, \citenamefont {Hoekstra}, \citenamefont {Holt}, \citenamefont {Hutzler}, \citenamefont {Jayich}, \citenamefont {Karthein}, \citenamefont {Leach}, \citenamefont {Madison}, \citenamefont
  {Malbrunot-Ettenauer}, \citenamefont {Miyagi}, \citenamefont {Moore}, \citenamefont {Moroch}, \citenamefont {Navratil}, \citenamefont {Nazarewicz}, \citenamefont {Neyens}, \citenamefont {Norrgard}, \citenamefont {Nusgart}, \citenamefont {Pa{\v{s}}teka}, \citenamefont {Petrov}, \citenamefont {Pla{\ss}}, \citenamefont {Ready}, \citenamefont {Reiter}, \citenamefont {Reponen}, \citenamefont {Rothe}, \citenamefont {Safronova}, \citenamefont {Scheidenerger}, \citenamefont {Shindler}, \citenamefont {Singh}, \citenamefont {Skripnikov}, \citenamefont {Titov}, \citenamefont {Udrescu}, \citenamefont {Wilkins},\ and\ \citenamefont {Yang}}]{RadMolWhitepaper2024}%
  \BibitemOpen
  \bibfield  {author} {\bibinfo {author} {\bibfnamefont {G.}~\bibnamefont {Arrowsmith-Kron}}, \bibinfo {author} {\bibfnamefont {M.}~\bibnamefont {Athanasakis-Kaklamanakis}}, \bibinfo {author} {\bibfnamefont {M.}~\bibnamefont {Au}}, \bibinfo {author} {\bibfnamefont {J.}~\bibnamefont {Ballof}}, \bibinfo {author} {\bibfnamefont {R.}~\bibnamefont {Berger}}, \bibinfo {author} {\bibfnamefont {A.}~\bibnamefont {Borschevsky}}, \bibinfo {author} {\bibfnamefont {A.~A.}\ \bibnamefont {Breier}}, \bibinfo {author} {\bibfnamefont {F.}~\bibnamefont {Buchinger}}, \bibinfo {author} {\bibfnamefont {D.}~\bibnamefont {Budker}}, \bibinfo {author} {\bibfnamefont {L.}~\bibnamefont {Caldwell}}, \bibinfo {author} {\bibfnamefont {C.}~\bibnamefont {Charles}}, \bibinfo {author} {\bibfnamefont {N.}~\bibnamefont {Dattani}}, \bibinfo {author} {\bibfnamefont {R.~P.}\ \bibnamefont {de~Groote}}, \bibinfo {author} {\bibfnamefont {D.}~\bibnamefont {DeMille}}, \bibinfo {author} {\bibfnamefont {T.}~\bibnamefont {Dickel}}, \bibinfo {author}
  {\bibfnamefont {J.}~\bibnamefont {Dobaczewski}}, \bibinfo {author} {\bibfnamefont {C.~E.}\ \bibnamefont {D{\"u}llmann}}, \bibinfo {author} {\bibfnamefont {E.}~\bibnamefont {Eliav}}, \bibinfo {author} {\bibfnamefont {J.}~\bibnamefont {Engel}}, \bibinfo {author} {\bibfnamefont {M.}~\bibnamefont {Fan}}, \bibinfo {author} {\bibfnamefont {V.}~\bibnamefont {Flambaum}}, \bibinfo {author} {\bibfnamefont {K.~T.}\ \bibnamefont {Flanagan}}, \bibinfo {author} {\bibfnamefont {A.~N.}\ \bibnamefont {Gaiser}}, \bibinfo {author} {\bibfnamefont {R.~F.~G.}\ \bibnamefont {Ruiz}}, \bibinfo {author} {\bibfnamefont {K.}~\bibnamefont {Gaul}}, \bibinfo {author} {\bibfnamefont {T.~F.}\ \bibnamefont {Giesen}}, \bibinfo {author} {\bibfnamefont {J.~S.~M.}\ \bibnamefont {Ginges}}, \bibinfo {author} {\bibfnamefont {A.}~\bibnamefont {Gottberg}}, \bibinfo {author} {\bibfnamefont {G.}~\bibnamefont {Gwinner}}, \bibinfo {author} {\bibfnamefont {R.}~\bibnamefont {Heinke}}, \bibinfo {author} {\bibfnamefont {S.}~\bibnamefont {Hoekstra}},
  \bibinfo {author} {\bibfnamefont {J.~D.}\ \bibnamefont {Holt}}, \bibinfo {author} {\bibfnamefont {N.~R.}\ \bibnamefont {Hutzler}}, \bibinfo {author} {\bibfnamefont {A.}~\bibnamefont {Jayich}}, \bibinfo {author} {\bibfnamefont {J.}~\bibnamefont {Karthein}}, \bibinfo {author} {\bibfnamefont {K.~G.}\ \bibnamefont {Leach}}, \bibinfo {author} {\bibfnamefont {K.~W.}\ \bibnamefont {Madison}}, \bibinfo {author} {\bibfnamefont {S.}~\bibnamefont {Malbrunot-Ettenauer}}, \bibinfo {author} {\bibfnamefont {T.}~\bibnamefont {Miyagi}}, \bibinfo {author} {\bibfnamefont {I.~D.}\ \bibnamefont {Moore}}, \bibinfo {author} {\bibfnamefont {S.}~\bibnamefont {Moroch}}, \bibinfo {author} {\bibfnamefont {P.}~\bibnamefont {Navratil}}, \bibinfo {author} {\bibfnamefont {W.}~\bibnamefont {Nazarewicz}}, \bibinfo {author} {\bibfnamefont {G.}~\bibnamefont {Neyens}}, \bibinfo {author} {\bibfnamefont {E.~B.}\ \bibnamefont {Norrgard}}, \bibinfo {author} {\bibfnamefont {N.}~\bibnamefont {Nusgart}}, \bibinfo {author} {\bibfnamefont {L.~F.}\
  \bibnamefont {Pa{\v{s}}teka}}, \bibinfo {author} {\bibfnamefont {A.~N.}\ \bibnamefont {Petrov}}, \bibinfo {author} {\bibfnamefont {W.~R.}\ \bibnamefont {Pla{\ss}}}, \bibinfo {author} {\bibfnamefont {R.~A.}\ \bibnamefont {Ready}}, \bibinfo {author} {\bibfnamefont {M.~P.}\ \bibnamefont {Reiter}}, \bibinfo {author} {\bibfnamefont {M.}~\bibnamefont {Reponen}}, \bibinfo {author} {\bibfnamefont {S.}~\bibnamefont {Rothe}}, \bibinfo {author} {\bibfnamefont {M.~S.}\ \bibnamefont {Safronova}}, \bibinfo {author} {\bibfnamefont {C.}~\bibnamefont {Scheidenerger}}, \bibinfo {author} {\bibfnamefont {A.}~\bibnamefont {Shindler}}, \bibinfo {author} {\bibfnamefont {J.~T.}\ \bibnamefont {Singh}}, \bibinfo {author} {\bibfnamefont {L.~V.}\ \bibnamefont {Skripnikov}}, \bibinfo {author} {\bibfnamefont {A.~V.}\ \bibnamefont {Titov}}, \bibinfo {author} {\bibfnamefont {S.-M.}\ \bibnamefont {Udrescu}}, \bibinfo {author} {\bibfnamefont {S.~G.}\ \bibnamefont {Wilkins}},\ and\ \bibinfo {author} {\bibfnamefont {X.}~\bibnamefont {Yang}},\
  }\href {https://doi.org/10.1088/1361-6633/ad1e39} {\bibfield  {journal} {\bibinfo  {journal} {Reports on Progress in Physics}\ }\textbf {\bibinfo {volume} {87}},\ \bibinfo {pages} {084301} (\bibinfo {year} {2024})}\BibitemShut {NoStop}%
\bibitem [{\citenamefont {Dobaczewski}\ and\ \citenamefont {Engel}(2005)}]{dobaczewski2005nuclear}%
  \BibitemOpen
  \bibfield  {author} {\bibinfo {author} {\bibfnamefont {J.}~\bibnamefont {Dobaczewski}}\ and\ \bibinfo {author} {\bibfnamefont {J.}~\bibnamefont {Engel}},\ }\href {https://doi.org/10.1103/PhysRevLett.94.232502} {\bibfield  {journal} {\bibinfo  {journal} {Physical Review Letters}\ }\textbf {\bibinfo {volume} {94}},\ \bibinfo {pages} {232502} (\bibinfo {year} {2005})}\BibitemShut {NoStop}%
\bibitem [{\citenamefont {Garcia~Ruiz}\ \emph {et~al.}(2020)\citenamefont {Garcia~Ruiz}, \citenamefont {Berger}, \citenamefont {Billowes}, \citenamefont {Binnersley}, \citenamefont {Bissell}, \citenamefont {Breier}, \citenamefont {Brinson}, \citenamefont {Chrysalidis}, \citenamefont {Cocolios}, \citenamefont {Cooper}, \citenamefont {Flanagan}, \citenamefont {Giesen}, \citenamefont {de~Groote}, \citenamefont {Franchoo}, \citenamefont {Gustafsson}, \citenamefont {Isaev}, \citenamefont {Koszor{\'u}s}, \citenamefont {Neyens}, \citenamefont {Perrett}, \citenamefont {Ricketts}, \citenamefont {Rothe}, \citenamefont {Schweikhard}, \citenamefont {Vernon}, \citenamefont {Wendt}, \citenamefont {Wienholtz}, \citenamefont {Wilkins},\ and\ \citenamefont {Yang}}]{GarciaRuiz2020RaF}%
  \BibitemOpen
  \bibfield  {author} {\bibinfo {author} {\bibfnamefont {R.~F.}\ \bibnamefont {Garcia~Ruiz}}, \bibinfo {author} {\bibfnamefont {R.}~\bibnamefont {Berger}}, \bibinfo {author} {\bibfnamefont {J.}~\bibnamefont {Billowes}}, \bibinfo {author} {\bibfnamefont {C.~L.}\ \bibnamefont {Binnersley}}, \bibinfo {author} {\bibfnamefont {M.~L.}\ \bibnamefont {Bissell}}, \bibinfo {author} {\bibfnamefont {A.~A.}\ \bibnamefont {Breier}}, \bibinfo {author} {\bibfnamefont {A.~J.}\ \bibnamefont {Brinson}}, \bibinfo {author} {\bibfnamefont {K.}~\bibnamefont {Chrysalidis}}, \bibinfo {author} {\bibfnamefont {T.~E.}\ \bibnamefont {Cocolios}}, \bibinfo {author} {\bibfnamefont {B.~S.}\ \bibnamefont {Cooper}}, \bibinfo {author} {\bibfnamefont {K.~T.}\ \bibnamefont {Flanagan}}, \bibinfo {author} {\bibfnamefont {T.~F.}\ \bibnamefont {Giesen}}, \bibinfo {author} {\bibfnamefont {R.~P.}\ \bibnamefont {de~Groote}}, \bibinfo {author} {\bibfnamefont {S.}~\bibnamefont {Franchoo}}, \bibinfo {author} {\bibfnamefont {F.~P.}\ \bibnamefont
  {Gustafsson}}, \bibinfo {author} {\bibfnamefont {T.~A.}\ \bibnamefont {Isaev}}, \bibinfo {author} {\bibfnamefont {{\'A}.}~\bibnamefont {Koszor{\'u}s}}, \bibinfo {author} {\bibfnamefont {G.}~\bibnamefont {Neyens}}, \bibinfo {author} {\bibfnamefont {H.~A.}\ \bibnamefont {Perrett}}, \bibinfo {author} {\bibfnamefont {C.~M.}\ \bibnamefont {Ricketts}}, \bibinfo {author} {\bibfnamefont {S.}~\bibnamefont {Rothe}}, \bibinfo {author} {\bibfnamefont {L.}~\bibnamefont {Schweikhard}}, \bibinfo {author} {\bibfnamefont {A.~R.}\ \bibnamefont {Vernon}}, \bibinfo {author} {\bibfnamefont {K.~D.~A.}\ \bibnamefont {Wendt}}, \bibinfo {author} {\bibfnamefont {F.}~\bibnamefont {Wienholtz}}, \bibinfo {author} {\bibfnamefont {S.~G.}\ \bibnamefont {Wilkins}},\ and\ \bibinfo {author} {\bibfnamefont {X.~F.}\ \bibnamefont {Yang}},\ }\href {https://doi.org/10.1038/s41586-020-2299-4} {\bibfield  {journal} {\bibinfo  {journal} {Nature}\ }\textbf {\bibinfo {volume} {581}},\ \bibinfo {pages} {396} (\bibinfo {year} {2020})}\BibitemShut
  {NoStop}%
\bibitem [{\citenamefont {Safronova}\ \emph {et~al.}(2018)\citenamefont {Safronova}, \citenamefont {Budker}, \citenamefont {DeMille}, \citenamefont {Kimball}, \citenamefont {Derevianko},\ and\ \citenamefont {Clark}}]{Safronova2018Review}%
  \BibitemOpen
  \bibfield  {author} {\bibinfo {author} {\bibfnamefont {M.~S.}\ \bibnamefont {Safronova}}, \bibinfo {author} {\bibfnamefont {D.}~\bibnamefont {Budker}}, \bibinfo {author} {\bibfnamefont {D.}~\bibnamefont {DeMille}}, \bibinfo {author} {\bibfnamefont {D.~F.~J.}\ \bibnamefont {Kimball}}, \bibinfo {author} {\bibfnamefont {A.}~\bibnamefont {Derevianko}},\ and\ \bibinfo {author} {\bibfnamefont {C.~W.}\ \bibnamefont {Clark}},\ }\href {https://doi.org/10.1103/RevModPhys.90.025008} {\bibfield  {journal} {\bibinfo  {journal} {Reviews of Modern Physics}\ }\textbf {\bibinfo {volume} {90}},\ \bibinfo {pages} {025008} (\bibinfo {year} {2018})}\BibitemShut {NoStop}%
\bibitem [{\citenamefont {Gaffney}\ \emph {et~al.}(2013)\citenamefont {Gaffney}, \citenamefont {Butler}, \citenamefont {Scheck}, \citenamefont {Hayes}, \citenamefont {Wenander}, \citenamefont {Albers}, \citenamefont {Bastin}, \citenamefont {Bauer}, \citenamefont {Blazhev}, \citenamefont {Bönig}, \citenamefont {Bree}, \citenamefont {Cederkäll}, \citenamefont {Chupp}, \citenamefont {Cline}, \citenamefont {Cocolios}, \citenamefont {Davinson}, \citenamefont {Witte}, \citenamefont {Diriken}, \citenamefont {Grahn}, \citenamefont {Herzan}, \citenamefont {Huyse}, \citenamefont {Jenkins}, \citenamefont {Joss}, \citenamefont {Kesteloot}, \citenamefont {Konki}, \citenamefont {Kowalczyk}, \citenamefont {Kröll}, \citenamefont {Kwan}, \citenamefont {Lutter}, \citenamefont {Moschner}, \citenamefont {Napiorkowski}, \citenamefont {Pakarinen}, \citenamefont {Pfeiffer}, \citenamefont {Radeck}, \citenamefont {Reiter}, \citenamefont {Reynders}, \citenamefont {Rigby}, \citenamefont {Robledo}, \citenamefont {Rudigier},
  \citenamefont {Sambi}, \citenamefont {Seidlitz}, \citenamefont {Siebeck}, \citenamefont {Stora}, \citenamefont {Thoele}, \citenamefont {Duppen}, \citenamefont {Vermeulen}, \citenamefont {von Schmid}, \citenamefont {Voulot}, \citenamefont {Warr}, \citenamefont {Wimmer}, \citenamefont {Wrzosek-Lipska}, \citenamefont {Wu},\ and\ \citenamefont {Zielinska}}]{Gaffney2013}%
  \BibitemOpen
  \bibfield  {author} {\bibinfo {author} {\bibfnamefont {L.~P.}\ \bibnamefont {Gaffney}}, \bibinfo {author} {\bibfnamefont {P.~A.}\ \bibnamefont {Butler}}, \bibinfo {author} {\bibfnamefont {M.}~\bibnamefont {Scheck}}, \bibinfo {author} {\bibfnamefont {A.~B.}\ \bibnamefont {Hayes}}, \bibinfo {author} {\bibfnamefont {F.}~\bibnamefont {Wenander}}, \bibinfo {author} {\bibfnamefont {M.}~\bibnamefont {Albers}}, \bibinfo {author} {\bibfnamefont {B.}~\bibnamefont {Bastin}}, \bibinfo {author} {\bibfnamefont {C.}~\bibnamefont {Bauer}}, \bibinfo {author} {\bibfnamefont {A.}~\bibnamefont {Blazhev}}, \bibinfo {author} {\bibfnamefont {S.}~\bibnamefont {Bönig}}, \bibinfo {author} {\bibfnamefont {N.}~\bibnamefont {Bree}}, \bibinfo {author} {\bibfnamefont {J.}~\bibnamefont {Cederkäll}}, \bibinfo {author} {\bibfnamefont {T.}~\bibnamefont {Chupp}}, \bibinfo {author} {\bibfnamefont {D.}~\bibnamefont {Cline}}, \bibinfo {author} {\bibfnamefont {T.~E.}\ \bibnamefont {Cocolios}}, \bibinfo {author} {\bibfnamefont {T.}~\bibnamefont
  {Davinson}}, \bibinfo {author} {\bibfnamefont {H.~D.}\ \bibnamefont {Witte}}, \bibinfo {author} {\bibfnamefont {J.}~\bibnamefont {Diriken}}, \bibinfo {author} {\bibfnamefont {T.}~\bibnamefont {Grahn}}, \bibinfo {author} {\bibfnamefont {A.}~\bibnamefont {Herzan}}, \bibinfo {author} {\bibfnamefont {M.}~\bibnamefont {Huyse}}, \bibinfo {author} {\bibfnamefont {D.~G.}\ \bibnamefont {Jenkins}}, \bibinfo {author} {\bibfnamefont {D.~T.}\ \bibnamefont {Joss}}, \bibinfo {author} {\bibfnamefont {N.}~\bibnamefont {Kesteloot}}, \bibinfo {author} {\bibfnamefont {J.}~\bibnamefont {Konki}}, \bibinfo {author} {\bibfnamefont {M.}~\bibnamefont {Kowalczyk}}, \bibinfo {author} {\bibfnamefont {T.}~\bibnamefont {Kröll}}, \bibinfo {author} {\bibfnamefont {E.}~\bibnamefont {Kwan}}, \bibinfo {author} {\bibfnamefont {R.}~\bibnamefont {Lutter}}, \bibinfo {author} {\bibfnamefont {K.}~\bibnamefont {Moschner}}, \bibinfo {author} {\bibfnamefont {P.}~\bibnamefont {Napiorkowski}}, \bibinfo {author} {\bibfnamefont {J.}~\bibnamefont
  {Pakarinen}}, \bibinfo {author} {\bibfnamefont {M.}~\bibnamefont {Pfeiffer}}, \bibinfo {author} {\bibfnamefont {D.}~\bibnamefont {Radeck}}, \bibinfo {author} {\bibfnamefont {P.}~\bibnamefont {Reiter}}, \bibinfo {author} {\bibfnamefont {K.}~\bibnamefont {Reynders}}, \bibinfo {author} {\bibfnamefont {S.~V.}\ \bibnamefont {Rigby}}, \bibinfo {author} {\bibfnamefont {L.~M.}\ \bibnamefont {Robledo}}, \bibinfo {author} {\bibfnamefont {M.}~\bibnamefont {Rudigier}}, \bibinfo {author} {\bibfnamefont {S.}~\bibnamefont {Sambi}}, \bibinfo {author} {\bibfnamefont {M.}~\bibnamefont {Seidlitz}}, \bibinfo {author} {\bibfnamefont {B.}~\bibnamefont {Siebeck}}, \bibinfo {author} {\bibfnamefont {T.}~\bibnamefont {Stora}}, \bibinfo {author} {\bibfnamefont {P.}~\bibnamefont {Thoele}}, \bibinfo {author} {\bibfnamefont {P.~V.}\ \bibnamefont {Duppen}}, \bibinfo {author} {\bibfnamefont {M.~J.}\ \bibnamefont {Vermeulen}}, \bibinfo {author} {\bibfnamefont {M.}~\bibnamefont {von Schmid}}, \bibinfo {author} {\bibfnamefont
  {D.}~\bibnamefont {Voulot}}, \bibinfo {author} {\bibfnamefont {N.}~\bibnamefont {Warr}}, \bibinfo {author} {\bibfnamefont {K.}~\bibnamefont {Wimmer}}, \bibinfo {author} {\bibfnamefont {K.}~\bibnamefont {Wrzosek-Lipska}}, \bibinfo {author} {\bibfnamefont {C.~Y.}\ \bibnamefont {Wu}},\ and\ \bibinfo {author} {\bibfnamefont {M.}~\bibnamefont {Zielinska}},\ }\href {https://doi.org/10.1038/nature12073} {\bibfield  {journal} {\bibinfo  {journal} {Nature}\ }\textbf {\bibinfo {volume} {497}},\ \bibinfo {pages} {199} (\bibinfo {year} {2013})}\BibitemShut {NoStop}%
\bibitem [{\citenamefont {Dobaczewski}\ \emph {et~al.}(2018)\citenamefont {Dobaczewski}, \citenamefont {Engel}, \citenamefont {Kortelainen},\ and\ \citenamefont {Becker}}]{Dobaczewski2018Correlating}%
  \BibitemOpen
  \bibfield  {author} {\bibinfo {author} {\bibfnamefont {J.}~\bibnamefont {Dobaczewski}}, \bibinfo {author} {\bibfnamefont {J.}~\bibnamefont {Engel}}, \bibinfo {author} {\bibfnamefont {M.}~\bibnamefont {Kortelainen}},\ and\ \bibinfo {author} {\bibfnamefont {P.}~\bibnamefont {Becker}},\ }\href {https://doi.org/10.1103/PhysRevLett.121.232501} {\bibfield  {journal} {\bibinfo  {journal} {Physical Review Letters}\ }\textbf {\bibinfo {volume} {121}},\ \bibinfo {pages} {232501} (\bibinfo {year} {2018})}\BibitemShut {NoStop}%
\bibitem [{\citenamefont {Auerbach}\ \emph {et~al.}(1996)\citenamefont {Auerbach}, \citenamefont {Flambaum},\ and\ \citenamefont {Spevak}}]{Auerbach1996OctDev}%
  \BibitemOpen
  \bibfield  {author} {\bibinfo {author} {\bibfnamefont {N.}~\bibnamefont {Auerbach}}, \bibinfo {author} {\bibfnamefont {V.~V.}\ \bibnamefont {Flambaum}},\ and\ \bibinfo {author} {\bibfnamefont {V.}~\bibnamefont {Spevak}},\ }\href {https://doi.org/10.1103/PhysRevLett.76.4316} {\bibfield  {journal} {\bibinfo  {journal} {Physical Review Letters}\ }\textbf {\bibinfo {volume} {76}},\ \bibinfo {pages} {4316} (\bibinfo {year} {1996})}\BibitemShut {NoStop}%
\bibitem [{\citenamefont {Flambaum}\ and\ \citenamefont {Ginges}(2002)}]{Flambaum2002NSMAtoms}%
  \BibitemOpen
  \bibfield  {author} {\bibinfo {author} {\bibfnamefont {V.~V.}\ \bibnamefont {Flambaum}}\ and\ \bibinfo {author} {\bibfnamefont {J.~S.~M.}\ \bibnamefont {Ginges}},\ }\href {https://doi.org/10.1103/PhysRevA.65.032113} {\bibfield  {journal} {\bibinfo  {journal} {Physical Review A}\ }\textbf {\bibinfo {volume} {65}},\ \bibinfo {pages} {032113} (\bibinfo {year} {2002})}\BibitemShut {NoStop}%
\bibitem [{\citenamefont {Kudashov}\ \emph {et~al.}(2014)\citenamefont {Kudashov}, \citenamefont {Petrov}, \citenamefont {Skripnikov}, \citenamefont {Mosyagin}, \citenamefont {Isaev}, \citenamefont {Berger},\ and\ \citenamefont {Titov}}]{Kudashov2014}%
  \BibitemOpen
  \bibfield  {author} {\bibinfo {author} {\bibfnamefont {A.~D.}\ \bibnamefont {Kudashov}}, \bibinfo {author} {\bibfnamefont {A.~N.}\ \bibnamefont {Petrov}}, \bibinfo {author} {\bibfnamefont {L.~V.}\ \bibnamefont {Skripnikov}}, \bibinfo {author} {\bibfnamefont {N.~S.}\ \bibnamefont {Mosyagin}}, \bibinfo {author} {\bibfnamefont {T.~A.}\ \bibnamefont {Isaev}}, \bibinfo {author} {\bibfnamefont {R.}~\bibnamefont {Berger}},\ and\ \bibinfo {author} {\bibfnamefont {A.~V.}\ \bibnamefont {Titov}},\ }\href {https://doi.org/10.1103/PhysRevA.90.052513} {\bibfield  {journal} {\bibinfo  {journal} {Physical Review A}\ }\textbf {\bibinfo {volume} {90}},\ \bibinfo {pages} {052513} (\bibinfo {year} {2014})}\BibitemShut {NoStop}%
\bibitem [{\citenamefont {Isaev}\ and\ \citenamefont {Berger}(2013)}]{Isaev2013RaF}%
  \BibitemOpen
  \bibfield  {author} {\bibinfo {author} {\bibfnamefont {T.}~\bibnamefont {Isaev}}\ and\ \bibinfo {author} {\bibfnamefont {R.}~\bibnamefont {Berger}},\ }\href {https://arxiv.org/abs/1302.5682} {\bibfield  {journal} {\bibinfo  {journal} {arXiv:1302.5682}\ } (\bibinfo {year} {2013})}\BibitemShut {NoStop}%
\bibitem [{\citenamefont {Isaev}\ \emph {et~al.}(2017)\citenamefont {Isaev}, \citenamefont {Zaitsevskii},\ and\ \citenamefont {Eliav}}]{Isaev2017RaOH}%
  \BibitemOpen
  \bibfield  {author} {\bibinfo {author} {\bibfnamefont {T.~A.}\ \bibnamefont {Isaev}}, \bibinfo {author} {\bibfnamefont {A.~V.}\ \bibnamefont {Zaitsevskii}},\ and\ \bibinfo {author} {\bibfnamefont {E.}~\bibnamefont {Eliav}},\ }\href {https://doi.org/10.1088/1361-6455/aa8f34} {\bibfield  {journal} {\bibinfo  {journal} {Journal of Physics B}\ }\textbf {\bibinfo {volume} {50}},\ \bibinfo {pages} {225101} (\bibinfo {year} {2017})}\BibitemShut {NoStop}%
\bibitem [{\citenamefont {Zhang}\ \emph {et~al.}(2023{\natexlab{a}})\citenamefont {Zhang}, \citenamefont {Hutzler},\ and\ \citenamefont {Cheng}}]{zhang2023intensity}%
  \BibitemOpen
  \bibfield  {author} {\bibinfo {author} {\bibfnamefont {C.}~\bibnamefont {Zhang}}, \bibinfo {author} {\bibfnamefont {N.~R.}\ \bibnamefont {Hutzler}},\ and\ \bibinfo {author} {\bibfnamefont {L.}~\bibnamefont {Cheng}},\ }\href {https://doi.org/10.1021/acs.jctc.3c00408} {\bibfield  {journal} {\bibinfo  {journal} {Journal of Chemical Theory and Computation}\ }\textbf {\bibinfo {volume} {19}},\ \bibinfo {pages} {4136} (\bibinfo {year} {2023}{\natexlab{a}})}\BibitemShut {NoStop}%
\bibitem [{\citenamefont {Alarcon}\ \emph {et~al.}(2022)\citenamefont {Alarcon}, \citenamefont {Alexander}, \citenamefont {Anastassopoulos}, \citenamefont {Aoki}, \citenamefont {Baartman}, \citenamefont {Baeßler}, \citenamefont {Bartoszek}, \citenamefont {Beck}, \citenamefont {Bedeschi}, \citenamefont {Berger}, \citenamefont {Berz}, \citenamefont {Bethlem}, \citenamefont {Bhattacharya}, \citenamefont {Blaskiewicz}, \citenamefont {Blum}, \citenamefont {Bowcock}, \citenamefont {Borschevsky}, \citenamefont {Brown}, \citenamefont {Budker}, \citenamefont {Burdin}, \citenamefont {Casey}, \citenamefont {Casse}, \citenamefont {Cantatore}, \citenamefont {Cheng}, \citenamefont {Chupp}, \citenamefont {Cianciolo}, \citenamefont {Cirigliano}, \citenamefont {Clayton}, \citenamefont {Crawford}, \citenamefont {Das}, \citenamefont {Davoudiasl}, \citenamefont {de~Vries}, \citenamefont {DeMille}, \citenamefont {Denisov}, \citenamefont {Diwan}, \citenamefont {Doyle}, \citenamefont {Engel}, \citenamefont {Fanourakis},
  \citenamefont {Fatemi}, \citenamefont {Filippone}, \citenamefont {Flambaum}, \citenamefont {Fleig}, \citenamefont {Fomin}, \citenamefont {Fischer}, \citenamefont {Gabrielse}, \citenamefont {Ruiz}, \citenamefont {Gardikiotis}, \citenamefont {Gatti}, \citenamefont {Geraci}, \citenamefont {Gooding}, \citenamefont {Golub}, \citenamefont {Graham}, \citenamefont {Gray}, \citenamefont {Griffith}, \citenamefont {Haciomeroglu}, \citenamefont {Gwinner}, \citenamefont {Hoekstra}, \citenamefont {Hoffstaetter}, \citenamefont {Huang}, \citenamefont {Hutzler}, \citenamefont {Incagli}, \citenamefont {Ito}, \citenamefont {Izubuchi}, \citenamefont {Jayich}, \citenamefont {Jeong}, \citenamefont {Kaplan}, \citenamefont {Karuza}, \citenamefont {Kawall}, \citenamefont {Kim}, \citenamefont {Koop}, \citenamefont {Korsch}, \citenamefont {Korobkina}, \citenamefont {Lebedev}, \citenamefont {Lee}, \citenamefont {Lee}, \citenamefont {Lehnert}, \citenamefont {Leung}, \citenamefont {Liu}, \citenamefont {Long}, \citenamefont {Lusiani},
  \citenamefont {Marciano}, \citenamefont {Maroudas}, \citenamefont {Matlashov}, \citenamefont {Matsumoto}, \citenamefont {Mawhorter}, \citenamefont {Meot}, \citenamefont {Mereghetti}, \citenamefont {Miller}, \citenamefont {Morse}, \citenamefont {Mott}, \citenamefont {Omarov}, \citenamefont {Orozco}, \citenamefont {O'Shaughnessy}, \citenamefont {Ozben}, \citenamefont {Park}, \citenamefont {Pattie}, \citenamefont {Petrov}, \citenamefont {Piacentino}, \citenamefont {Plaster}, \citenamefont {Podobedov}, \citenamefont {Poelker}, \citenamefont {Pocanic}, \citenamefont {Prasannaa}, \citenamefont {Price}, \citenamefont {Ramsey-Musolf}, \citenamefont {Raparia}, \citenamefont {Rajendran}, \citenamefont {Reece}, \citenamefont {Reid}, \citenamefont {Rescia}, \citenamefont {Ritz}, \citenamefont {Roberts}, \citenamefont {Safronova}, \citenamefont {Sakemi}, \citenamefont {Schmidt-Wellenburg}, \citenamefont {Shindler}, \citenamefont {Semertzidis}, \citenamefont {Silenko}, \citenamefont {Singh}, \citenamefont {Skripnikov},
  \citenamefont {Soni}, \citenamefont {Stephenson}, \citenamefont {Suleiman}, \citenamefont {Sunaga}, \citenamefont {Syphers}, \citenamefont {Syritsyn}, \citenamefont {Tarbutt}, \citenamefont {Thoerngren}, \citenamefont {Timmermans}, \citenamefont {Tishchenko}, \citenamefont {Titov}, \citenamefont {Tsoupas}, \citenamefont {Tzamarias}, \citenamefont {Variola}, \citenamefont {Venanzoni}, \citenamefont {Vilella}, \citenamefont {Vossebeld}, \citenamefont {Winter}, \citenamefont {Won}, \citenamefont {Zelenski}, \citenamefont {Zelevinsky}, \citenamefont {Zhou},\ and\ \citenamefont {Zioutas}}]{Alarcon2022Snowmass}%
  \BibitemOpen
  \bibfield  {author} {\bibinfo {author} {\bibfnamefont {R.}~\bibnamefont {Alarcon}}, \bibinfo {author} {\bibfnamefont {J.}~\bibnamefont {Alexander}}, \bibinfo {author} {\bibfnamefont {V.}~\bibnamefont {Anastassopoulos}}, \bibinfo {author} {\bibfnamefont {T.}~\bibnamefont {Aoki}}, \bibinfo {author} {\bibfnamefont {R.}~\bibnamefont {Baartman}}, \bibinfo {author} {\bibfnamefont {S.}~\bibnamefont {Baeßler}}, \bibinfo {author} {\bibfnamefont {L.}~\bibnamefont {Bartoszek}}, \bibinfo {author} {\bibfnamefont {D.~H.}\ \bibnamefont {Beck}}, \bibinfo {author} {\bibfnamefont {F.}~\bibnamefont {Bedeschi}}, \bibinfo {author} {\bibfnamefont {R.}~\bibnamefont {Berger}}, \bibinfo {author} {\bibfnamefont {M.}~\bibnamefont {Berz}}, \bibinfo {author} {\bibfnamefont {H.~L.}\ \bibnamefont {Bethlem}}, \bibinfo {author} {\bibfnamefont {T.}~\bibnamefont {Bhattacharya}}, \bibinfo {author} {\bibfnamefont {M.}~\bibnamefont {Blaskiewicz}}, \bibinfo {author} {\bibfnamefont {T.}~\bibnamefont {Blum}}, \bibinfo {author} {\bibfnamefont
  {T.}~\bibnamefont {Bowcock}}, \bibinfo {author} {\bibfnamefont {A.}~\bibnamefont {Borschevsky}}, \bibinfo {author} {\bibfnamefont {K.}~\bibnamefont {Brown}}, \bibinfo {author} {\bibfnamefont {D.}~\bibnamefont {Budker}}, \bibinfo {author} {\bibfnamefont {S.}~\bibnamefont {Burdin}}, \bibinfo {author} {\bibfnamefont {B.~C.}\ \bibnamefont {Casey}}, \bibinfo {author} {\bibfnamefont {G.}~\bibnamefont {Casse}}, \bibinfo {author} {\bibfnamefont {G.}~\bibnamefont {Cantatore}}, \bibinfo {author} {\bibfnamefont {L.}~\bibnamefont {Cheng}}, \bibinfo {author} {\bibfnamefont {T.}~\bibnamefont {Chupp}}, \bibinfo {author} {\bibfnamefont {V.}~\bibnamefont {Cianciolo}}, \bibinfo {author} {\bibfnamefont {V.}~\bibnamefont {Cirigliano}}, \bibinfo {author} {\bibfnamefont {S.~M.}\ \bibnamefont {Clayton}}, \bibinfo {author} {\bibfnamefont {C.}~\bibnamefont {Crawford}}, \bibinfo {author} {\bibfnamefont {B.~P.}\ \bibnamefont {Das}}, \bibinfo {author} {\bibfnamefont {H.}~\bibnamefont {Davoudiasl}}, \bibinfo {author} {\bibfnamefont
  {J.}~\bibnamefont {de~Vries}}, \bibinfo {author} {\bibfnamefont {D.}~\bibnamefont {DeMille}}, \bibinfo {author} {\bibfnamefont {D.}~\bibnamefont {Denisov}}, \bibinfo {author} {\bibfnamefont {M.~V.}\ \bibnamefont {Diwan}}, \bibinfo {author} {\bibfnamefont {J.~M.}\ \bibnamefont {Doyle}}, \bibinfo {author} {\bibfnamefont {J.}~\bibnamefont {Engel}}, \bibinfo {author} {\bibfnamefont {G.}~\bibnamefont {Fanourakis}}, \bibinfo {author} {\bibfnamefont {R.}~\bibnamefont {Fatemi}}, \bibinfo {author} {\bibfnamefont {B.~W.}\ \bibnamefont {Filippone}}, \bibinfo {author} {\bibfnamefont {V.~V.}\ \bibnamefont {Flambaum}}, \bibinfo {author} {\bibfnamefont {T.}~\bibnamefont {Fleig}}, \bibinfo {author} {\bibfnamefont {N.}~\bibnamefont {Fomin}}, \bibinfo {author} {\bibfnamefont {W.}~\bibnamefont {Fischer}}, \bibinfo {author} {\bibfnamefont {G.}~\bibnamefont {Gabrielse}}, \bibinfo {author} {\bibfnamefont {R.~F.~G.}\ \bibnamefont {Ruiz}}, \bibinfo {author} {\bibfnamefont {A.}~\bibnamefont {Gardikiotis}}, \bibinfo {author}
  {\bibfnamefont {C.}~\bibnamefont {Gatti}}, \bibinfo {author} {\bibfnamefont {A.}~\bibnamefont {Geraci}}, \bibinfo {author} {\bibfnamefont {J.}~\bibnamefont {Gooding}}, \bibinfo {author} {\bibfnamefont {B.}~\bibnamefont {Golub}}, \bibinfo {author} {\bibfnamefont {P.}~\bibnamefont {Graham}}, \bibinfo {author} {\bibfnamefont {F.}~\bibnamefont {Gray}}, \bibinfo {author} {\bibfnamefont {W.~C.}\ \bibnamefont {Griffith}}, \bibinfo {author} {\bibfnamefont {S.}~\bibnamefont {Haciomeroglu}}, \bibinfo {author} {\bibfnamefont {G.}~\bibnamefont {Gwinner}}, \bibinfo {author} {\bibfnamefont {S.}~\bibnamefont {Hoekstra}}, \bibinfo {author} {\bibfnamefont {G.~H.}\ \bibnamefont {Hoffstaetter}}, \bibinfo {author} {\bibfnamefont {H.}~\bibnamefont {Huang}}, \bibinfo {author} {\bibfnamefont {N.~R.}\ \bibnamefont {Hutzler}}, \bibinfo {author} {\bibfnamefont {M.}~\bibnamefont {Incagli}}, \bibinfo {author} {\bibfnamefont {T.~M.}\ \bibnamefont {Ito}}, \bibinfo {author} {\bibfnamefont {T.}~\bibnamefont {Izubuchi}}, \bibinfo {author}
  {\bibfnamefont {A.~M.}\ \bibnamefont {Jayich}}, \bibinfo {author} {\bibfnamefont {H.}~\bibnamefont {Jeong}}, \bibinfo {author} {\bibfnamefont {D.}~\bibnamefont {Kaplan}}, \bibinfo {author} {\bibfnamefont {M.}~\bibnamefont {Karuza}}, \bibinfo {author} {\bibfnamefont {D.}~\bibnamefont {Kawall}}, \bibinfo {author} {\bibfnamefont {O.}~\bibnamefont {Kim}}, \bibinfo {author} {\bibfnamefont {I.}~\bibnamefont {Koop}}, \bibinfo {author} {\bibfnamefont {W.}~\bibnamefont {Korsch}}, \bibinfo {author} {\bibfnamefont {E.}~\bibnamefont {Korobkina}}, \bibinfo {author} {\bibfnamefont {V.}~\bibnamefont {Lebedev}}, \bibinfo {author} {\bibfnamefont {J.}~\bibnamefont {Lee}}, \bibinfo {author} {\bibfnamefont {S.}~\bibnamefont {Lee}}, \bibinfo {author} {\bibfnamefont {R.}~\bibnamefont {Lehnert}}, \bibinfo {author} {\bibfnamefont {K.~K.~H.}\ \bibnamefont {Leung}}, \bibinfo {author} {\bibfnamefont {C.-Y.}\ \bibnamefont {Liu}}, \bibinfo {author} {\bibfnamefont {J.}~\bibnamefont {Long}}, \bibinfo {author} {\bibfnamefont
  {A.}~\bibnamefont {Lusiani}}, \bibinfo {author} {\bibfnamefont {W.~J.}\ \bibnamefont {Marciano}}, \bibinfo {author} {\bibfnamefont {M.}~\bibnamefont {Maroudas}}, \bibinfo {author} {\bibfnamefont {A.}~\bibnamefont {Matlashov}}, \bibinfo {author} {\bibfnamefont {N.}~\bibnamefont {Matsumoto}}, \bibinfo {author} {\bibfnamefont {R.}~\bibnamefont {Mawhorter}}, \bibinfo {author} {\bibfnamefont {F.}~\bibnamefont {Meot}}, \bibinfo {author} {\bibfnamefont {E.}~\bibnamefont {Mereghetti}}, \bibinfo {author} {\bibfnamefont {J.~P.}\ \bibnamefont {Miller}}, \bibinfo {author} {\bibfnamefont {W.~M.}\ \bibnamefont {Morse}}, \bibinfo {author} {\bibfnamefont {J.}~\bibnamefont {Mott}}, \bibinfo {author} {\bibfnamefont {Z.}~\bibnamefont {Omarov}}, \bibinfo {author} {\bibfnamefont {L.~A.}\ \bibnamefont {Orozco}}, \bibinfo {author} {\bibfnamefont {C.~M.}\ \bibnamefont {O'Shaughnessy}}, \bibinfo {author} {\bibfnamefont {C.}~\bibnamefont {Ozben}}, \bibinfo {author} {\bibfnamefont {S.}~\bibnamefont {Park}}, \bibinfo {author}
  {\bibfnamefont {R.~W.}\ \bibnamefont {Pattie}}, \bibinfo {author} {\bibfnamefont {A.~N.}\ \bibnamefont {Petrov}}, \bibinfo {author} {\bibfnamefont {G.~M.}\ \bibnamefont {Piacentino}}, \bibinfo {author} {\bibfnamefont {B.~R.}\ \bibnamefont {Plaster}}, \bibinfo {author} {\bibfnamefont {B.}~\bibnamefont {Podobedov}}, \bibinfo {author} {\bibfnamefont {M.}~\bibnamefont {Poelker}}, \bibinfo {author} {\bibfnamefont {D.}~\bibnamefont {Pocanic}}, \bibinfo {author} {\bibfnamefont {V.~S.}\ \bibnamefont {Prasannaa}}, \bibinfo {author} {\bibfnamefont {J.}~\bibnamefont {Price}}, \bibinfo {author} {\bibfnamefont {M.~J.}\ \bibnamefont {Ramsey-Musolf}}, \bibinfo {author} {\bibfnamefont {D.}~\bibnamefont {Raparia}}, \bibinfo {author} {\bibfnamefont {S.}~\bibnamefont {Rajendran}}, \bibinfo {author} {\bibfnamefont {M.}~\bibnamefont {Reece}}, \bibinfo {author} {\bibfnamefont {A.}~\bibnamefont {Reid}}, \bibinfo {author} {\bibfnamefont {S.}~\bibnamefont {Rescia}}, \bibinfo {author} {\bibfnamefont {A.}~\bibnamefont {Ritz}},
  \bibinfo {author} {\bibfnamefont {B.~L.}\ \bibnamefont {Roberts}}, \bibinfo {author} {\bibfnamefont {M.~S.}\ \bibnamefont {Safronova}}, \bibinfo {author} {\bibfnamefont {Y.}~\bibnamefont {Sakemi}}, \bibinfo {author} {\bibfnamefont {P.}~\bibnamefont {Schmidt-Wellenburg}}, \bibinfo {author} {\bibfnamefont {A.}~\bibnamefont {Shindler}}, \bibinfo {author} {\bibfnamefont {Y.~K.}\ \bibnamefont {Semertzidis}}, \bibinfo {author} {\bibfnamefont {A.}~\bibnamefont {Silenko}}, \bibinfo {author} {\bibfnamefont {J.~T.}\ \bibnamefont {Singh}}, \bibinfo {author} {\bibfnamefont {L.~V.}\ \bibnamefont {Skripnikov}}, \bibinfo {author} {\bibfnamefont {A.}~\bibnamefont {Soni}}, \bibinfo {author} {\bibfnamefont {E.}~\bibnamefont {Stephenson}}, \bibinfo {author} {\bibfnamefont {R.}~\bibnamefont {Suleiman}}, \bibinfo {author} {\bibfnamefont {A.}~\bibnamefont {Sunaga}}, \bibinfo {author} {\bibfnamefont {M.}~\bibnamefont {Syphers}}, \bibinfo {author} {\bibfnamefont {S.}~\bibnamefont {Syritsyn}}, \bibinfo {author} {\bibfnamefont
  {M.~R.}\ \bibnamefont {Tarbutt}}, \bibinfo {author} {\bibfnamefont {P.}~\bibnamefont {Thoerngren}}, \bibinfo {author} {\bibfnamefont {R.~G.~E.}\ \bibnamefont {Timmermans}}, \bibinfo {author} {\bibfnamefont {V.}~\bibnamefont {Tishchenko}}, \bibinfo {author} {\bibfnamefont {A.~V.}\ \bibnamefont {Titov}}, \bibinfo {author} {\bibfnamefont {N.}~\bibnamefont {Tsoupas}}, \bibinfo {author} {\bibfnamefont {S.}~\bibnamefont {Tzamarias}}, \bibinfo {author} {\bibfnamefont {A.}~\bibnamefont {Variola}}, \bibinfo {author} {\bibfnamefont {G.}~\bibnamefont {Venanzoni}}, \bibinfo {author} {\bibfnamefont {E.}~\bibnamefont {Vilella}}, \bibinfo {author} {\bibfnamefont {J.}~\bibnamefont {Vossebeld}}, \bibinfo {author} {\bibfnamefont {P.}~\bibnamefont {Winter}}, \bibinfo {author} {\bibfnamefont {E.}~\bibnamefont {Won}}, \bibinfo {author} {\bibfnamefont {A.}~\bibnamefont {Zelenski}}, \bibinfo {author} {\bibfnamefont {T.}~\bibnamefont {Zelevinsky}}, \bibinfo {author} {\bibfnamefont {Y.}~\bibnamefont {Zhou}},\ and\ \bibinfo {author}
  {\bibfnamefont {K.}~\bibnamefont {Zioutas}},\ }\href {http://arxiv.org/abs/2203.08103} {\bibfield  {journal} {\bibinfo  {journal} {arXiv:2203.08103}\ } (\bibinfo {year} {2022})}\BibitemShut {NoStop}%
\bibitem [{\citenamefont {Udrescu}\ \emph {et~al.}(2021)\citenamefont {Udrescu}, \citenamefont {Brinson}, \citenamefont {Ruiz}, \citenamefont {Gaul}, \citenamefont {Berger}, \citenamefont {Billowes}, \citenamefont {Binnersley}, \citenamefont {Bissell}, \citenamefont {Breier}, \citenamefont {Chrysalidis}, \citenamefont {Cocolios}, \citenamefont {Cooper}, \citenamefont {Flanagan}, \citenamefont {Giesen}, \citenamefont {De~Groote}, \citenamefont {Franchoo}, \citenamefont {Gustafsson}, \citenamefont {Isaev}, \citenamefont {Koszor{\'u}s}, \citenamefont {Neyens}, \citenamefont {Perrett}, \citenamefont {Ricketts}, \citenamefont {Rothe}, \citenamefont {Vernon}, \citenamefont {Wendt}, \citenamefont {Wienholtz}, \citenamefont {Wilkins},\ and\ \citenamefont {Yang}}]{udrescu2021isotope}%
  \BibitemOpen
  \bibfield  {author} {\bibinfo {author} {\bibfnamefont {S.~M.}\ \bibnamefont {Udrescu}}, \bibinfo {author} {\bibfnamefont {A.~J.}\ \bibnamefont {Brinson}}, \bibinfo {author} {\bibfnamefont {R.~F.~G.}\ \bibnamefont {Ruiz}}, \bibinfo {author} {\bibfnamefont {K.}~\bibnamefont {Gaul}}, \bibinfo {author} {\bibfnamefont {R.}~\bibnamefont {Berger}}, \bibinfo {author} {\bibfnamefont {J.}~\bibnamefont {Billowes}}, \bibinfo {author} {\bibfnamefont {C.~L.}\ \bibnamefont {Binnersley}}, \bibinfo {author} {\bibfnamefont {M.~L.}\ \bibnamefont {Bissell}}, \bibinfo {author} {\bibfnamefont {A.~A.}\ \bibnamefont {Breier}}, \bibinfo {author} {\bibfnamefont {K.}~\bibnamefont {Chrysalidis}}, \bibinfo {author} {\bibfnamefont {T.~E.}\ \bibnamefont {Cocolios}}, \bibinfo {author} {\bibfnamefont {B.~S.}\ \bibnamefont {Cooper}}, \bibinfo {author} {\bibfnamefont {K.~T.}\ \bibnamefont {Flanagan}}, \bibinfo {author} {\bibfnamefont {T.~F.}\ \bibnamefont {Giesen}}, \bibinfo {author} {\bibfnamefont {R.~P.}\ \bibnamefont {De~Groote}},
  \bibinfo {author} {\bibfnamefont {S.}~\bibnamefont {Franchoo}}, \bibinfo {author} {\bibfnamefont {F.~P.}\ \bibnamefont {Gustafsson}}, \bibinfo {author} {\bibfnamefont {T.~A.}\ \bibnamefont {Isaev}}, \bibinfo {author} {\bibfnamefont {{\'A}.}~\bibnamefont {Koszor{\'u}s}}, \bibinfo {author} {\bibfnamefont {G.}~\bibnamefont {Neyens}}, \bibinfo {author} {\bibfnamefont {H.~A.}\ \bibnamefont {Perrett}}, \bibinfo {author} {\bibfnamefont {C.~M.}\ \bibnamefont {Ricketts}}, \bibinfo {author} {\bibfnamefont {S.}~\bibnamefont {Rothe}}, \bibinfo {author} {\bibfnamefont {A.~R.}\ \bibnamefont {Vernon}}, \bibinfo {author} {\bibfnamefont {K.~D.~A.}\ \bibnamefont {Wendt}}, \bibinfo {author} {\bibfnamefont {F.}~\bibnamefont {Wienholtz}}, \bibinfo {author} {\bibfnamefont {S.~G.}\ \bibnamefont {Wilkins}},\ and\ \bibinfo {author} {\bibfnamefont {X.~F.}\ \bibnamefont {Yang}},\ }\href {https://doi.org/10.1103/PhysRevLett.127.033001} {\bibfield  {journal} {\bibinfo  {journal} {Physical Review Letters}\ }\textbf {\bibinfo {volume}
  {127}},\ \bibinfo {pages} {033001} (\bibinfo {year} {2021})}\BibitemShut {NoStop}%
\bibitem [{\citenamefont {Udrescu}\ \emph {et~al.}(2024)\citenamefont {Udrescu}, \citenamefont {Wilkins}, \citenamefont {Breier}, \citenamefont {Athanasakis-Kaklamanakis}, \citenamefont {Garcia~Ruiz}, \citenamefont {Au}, \citenamefont {Belo{\v{s}}evi{\'c}}, \citenamefont {Berger}, \citenamefont {Bissell}, \citenamefont {Binnersley}, \citenamefont {Brinson}, \citenamefont {Chrysalidis}, \citenamefont {Cocolios}, \citenamefont {De~Groote}, \citenamefont {Dorne}, \citenamefont {Flanagan}, \citenamefont {Franchoo}, \citenamefont {Gaul}, \citenamefont {Geldhof}, \citenamefont {Giesen}, \citenamefont {Hanstorp}, \citenamefont {Heinke}, \citenamefont {Koszor{\'u}s}, \citenamefont {Kujanp{\"a}{\"a}}, \citenamefont {Lalanne}, \citenamefont {Neyens}, \citenamefont {Nichols}, \citenamefont {Perrett}, \citenamefont {Reilly}, \citenamefont {Rothe}, \citenamefont {Van Den~Borne}, \citenamefont {Vernon}, \citenamefont {Wang}, \citenamefont {Wessolek}, \citenamefont {Yang},\ and\ \citenamefont
  {Z{\"u}lch}}]{udrescu2024precision}%
  \BibitemOpen
  \bibfield  {author} {\bibinfo {author} {\bibfnamefont {S.~M.}\ \bibnamefont {Udrescu}}, \bibinfo {author} {\bibfnamefont {S.~G.}\ \bibnamefont {Wilkins}}, \bibinfo {author} {\bibfnamefont {A.~A.}\ \bibnamefont {Breier}}, \bibinfo {author} {\bibfnamefont {M.}~\bibnamefont {Athanasakis-Kaklamanakis}}, \bibinfo {author} {\bibfnamefont {R.~F.}\ \bibnamefont {Garcia~Ruiz}}, \bibinfo {author} {\bibfnamefont {M.}~\bibnamefont {Au}}, \bibinfo {author} {\bibfnamefont {I.}~\bibnamefont {Belo{\v{s}}evi{\'c}}}, \bibinfo {author} {\bibfnamefont {R.}~\bibnamefont {Berger}}, \bibinfo {author} {\bibfnamefont {M.~L.}\ \bibnamefont {Bissell}}, \bibinfo {author} {\bibfnamefont {C.~L.}\ \bibnamefont {Binnersley}}, \bibinfo {author} {\bibfnamefont {A.~J.}\ \bibnamefont {Brinson}}, \bibinfo {author} {\bibfnamefont {K.}~\bibnamefont {Chrysalidis}}, \bibinfo {author} {\bibfnamefont {T.~E.}\ \bibnamefont {Cocolios}}, \bibinfo {author} {\bibfnamefont {R.~P.}\ \bibnamefont {De~Groote}}, \bibinfo {author} {\bibfnamefont
  {A.}~\bibnamefont {Dorne}}, \bibinfo {author} {\bibfnamefont {K.~T.}\ \bibnamefont {Flanagan}}, \bibinfo {author} {\bibfnamefont {S.}~\bibnamefont {Franchoo}}, \bibinfo {author} {\bibfnamefont {K.}~\bibnamefont {Gaul}}, \bibinfo {author} {\bibfnamefont {S.}~\bibnamefont {Geldhof}}, \bibinfo {author} {\bibfnamefont {T.~F.}\ \bibnamefont {Giesen}}, \bibinfo {author} {\bibfnamefont {D.}~\bibnamefont {Hanstorp}}, \bibinfo {author} {\bibfnamefont {R.}~\bibnamefont {Heinke}}, \bibinfo {author} {\bibfnamefont {{\'A}.}~\bibnamefont {Koszor{\'u}s}}, \bibinfo {author} {\bibfnamefont {S.}~\bibnamefont {Kujanp{\"a}{\"a}}}, \bibinfo {author} {\bibfnamefont {L.}~\bibnamefont {Lalanne}}, \bibinfo {author} {\bibfnamefont {G.}~\bibnamefont {Neyens}}, \bibinfo {author} {\bibfnamefont {M.}~\bibnamefont {Nichols}}, \bibinfo {author} {\bibfnamefont {H.~A.}\ \bibnamefont {Perrett}}, \bibinfo {author} {\bibfnamefont {J.~R.}\ \bibnamefont {Reilly}}, \bibinfo {author} {\bibfnamefont {S.}~\bibnamefont {Rothe}}, \bibinfo {author}
  {\bibfnamefont {B.}~\bibnamefont {Van Den~Borne}}, \bibinfo {author} {\bibfnamefont {A.~R.}\ \bibnamefont {Vernon}}, \bibinfo {author} {\bibfnamefont {Q.}~\bibnamefont {Wang}}, \bibinfo {author} {\bibfnamefont {J.}~\bibnamefont {Wessolek}}, \bibinfo {author} {\bibfnamefont {X.~F.}\ \bibnamefont {Yang}},\ and\ \bibinfo {author} {\bibfnamefont {C.}~\bibnamefont {Z{\"u}lch}},\ }\href {https://doi.org/10.1038/s41567-023-02296-w} {\bibfield  {journal} {\bibinfo  {journal} {Nature Physics}\ }\textbf {\bibinfo {volume} {20}},\ \bibinfo {pages} {202} (\bibinfo {year} {2024})}\BibitemShut {NoStop}%
\bibitem [{\citenamefont {Athanasakis-Kaklamanakis}\ \emph {et~al.}(2025)\citenamefont {Athanasakis-Kaklamanakis}, \citenamefont {Wilkins}, \citenamefont {Skripnikov}, \citenamefont {Koszor{\'u}s}, \citenamefont {Breier}, \citenamefont {Ahmad}, \citenamefont {Au}, \citenamefont {Bai}, \citenamefont {Belo{\v{s}}evi{\'c}}, \citenamefont {Berbalk}, \citenamefont {Berger}, \citenamefont {Bernerd}, \citenamefont {Bissell}, \citenamefont {Borschevsky}, \citenamefont {Brinson}, \citenamefont {Chrysalidis}, \citenamefont {Cocolios}, \citenamefont {De~Groote}, \citenamefont {Dorne}, \citenamefont {Fajardo-Zambrano}, \citenamefont {Field}, \citenamefont {Flanagan}, \citenamefont {Franchoo}, \citenamefont {Garcia~Ruiz}, \citenamefont {Gaul}, \citenamefont {Geldhof}, \citenamefont {Giesen}, \citenamefont {Hanstorp}, \citenamefont {Heinke}, \citenamefont {Imgram}, \citenamefont {Isaev}, \citenamefont {Kyuberis}, \citenamefont {Kujanp{\"a}{\"a}}, \citenamefont {Lalanne}, \citenamefont {Lass{\`e}gues}, \citenamefont {Lim},
  \citenamefont {Liu}, \citenamefont {Lynch}, \citenamefont {McGlone}, \citenamefont {Mei}, \citenamefont {Neyens}, \citenamefont {Nichols}, \citenamefont {Nies}, \citenamefont {Pa{\v{s}}teka}, \citenamefont {Perrett}, \citenamefont {Raggio}, \citenamefont {Reilly}, \citenamefont {Rothe}, \citenamefont {Smets}, \citenamefont {Udrescu}, \citenamefont {Van Den~Borne}, \citenamefont {Wang}, \citenamefont {Warbinek}, \citenamefont {Wessolek}, \citenamefont {Yang},\ and\ \citenamefont {Z{\"u}lch}}]{athanasakis2025electron}%
  \BibitemOpen
  \bibfield  {author} {\bibinfo {author} {\bibfnamefont {M.}~\bibnamefont {Athanasakis-Kaklamanakis}}, \bibinfo {author} {\bibfnamefont {S.~G.}\ \bibnamefont {Wilkins}}, \bibinfo {author} {\bibfnamefont {L.~V.}\ \bibnamefont {Skripnikov}}, \bibinfo {author} {\bibfnamefont {{\'A}.}~\bibnamefont {Koszor{\'u}s}}, \bibinfo {author} {\bibfnamefont {A.~A.}\ \bibnamefont {Breier}}, \bibinfo {author} {\bibfnamefont {O.}~\bibnamefont {Ahmad}}, \bibinfo {author} {\bibfnamefont {M.}~\bibnamefont {Au}}, \bibinfo {author} {\bibfnamefont {S.~W.}\ \bibnamefont {Bai}}, \bibinfo {author} {\bibfnamefont {I.}~\bibnamefont {Belo{\v{s}}evi{\'c}}}, \bibinfo {author} {\bibfnamefont {J.}~\bibnamefont {Berbalk}}, \bibinfo {author} {\bibfnamefont {R.}~\bibnamefont {Berger}}, \bibinfo {author} {\bibfnamefont {C.}~\bibnamefont {Bernerd}}, \bibinfo {author} {\bibfnamefont {M.~L.}\ \bibnamefont {Bissell}}, \bibinfo {author} {\bibfnamefont {A.}~\bibnamefont {Borschevsky}}, \bibinfo {author} {\bibfnamefont {A.}~\bibnamefont {Brinson}},
  \bibinfo {author} {\bibfnamefont {K.}~\bibnamefont {Chrysalidis}}, \bibinfo {author} {\bibfnamefont {T.~E.}\ \bibnamefont {Cocolios}}, \bibinfo {author} {\bibfnamefont {R.~P.}\ \bibnamefont {De~Groote}}, \bibinfo {author} {\bibfnamefont {A.}~\bibnamefont {Dorne}}, \bibinfo {author} {\bibfnamefont {C.~M.}\ \bibnamefont {Fajardo-Zambrano}}, \bibinfo {author} {\bibfnamefont {R.~W.}\ \bibnamefont {Field}}, \bibinfo {author} {\bibfnamefont {K.~T.}\ \bibnamefont {Flanagan}}, \bibinfo {author} {\bibfnamefont {S.}~\bibnamefont {Franchoo}}, \bibinfo {author} {\bibfnamefont {R.~F.}\ \bibnamefont {Garcia~Ruiz}}, \bibinfo {author} {\bibfnamefont {K.}~\bibnamefont {Gaul}}, \bibinfo {author} {\bibfnamefont {S.}~\bibnamefont {Geldhof}}, \bibinfo {author} {\bibfnamefont {T.~F.}\ \bibnamefont {Giesen}}, \bibinfo {author} {\bibfnamefont {D.}~\bibnamefont {Hanstorp}}, \bibinfo {author} {\bibfnamefont {R.}~\bibnamefont {Heinke}}, \bibinfo {author} {\bibfnamefont {P.}~\bibnamefont {Imgram}}, \bibinfo {author} {\bibfnamefont
  {T.~A.}\ \bibnamefont {Isaev}}, \bibinfo {author} {\bibfnamefont {A.~A.}\ \bibnamefont {Kyuberis}}, \bibinfo {author} {\bibfnamefont {S.}~\bibnamefont {Kujanp{\"a}{\"a}}}, \bibinfo {author} {\bibfnamefont {L.}~\bibnamefont {Lalanne}}, \bibinfo {author} {\bibfnamefont {P.}~\bibnamefont {Lass{\`e}gues}}, \bibinfo {author} {\bibfnamefont {J.}~\bibnamefont {Lim}}, \bibinfo {author} {\bibfnamefont {Y.~C.}\ \bibnamefont {Liu}}, \bibinfo {author} {\bibfnamefont {K.~M.}\ \bibnamefont {Lynch}}, \bibinfo {author} {\bibfnamefont {A.}~\bibnamefont {McGlone}}, \bibinfo {author} {\bibfnamefont {W.~C.}\ \bibnamefont {Mei}}, \bibinfo {author} {\bibfnamefont {G.}~\bibnamefont {Neyens}}, \bibinfo {author} {\bibfnamefont {M.}~\bibnamefont {Nichols}}, \bibinfo {author} {\bibfnamefont {L.}~\bibnamefont {Nies}}, \bibinfo {author} {\bibfnamefont {L.~F.}\ \bibnamefont {Pa{\v{s}}teka}}, \bibinfo {author} {\bibfnamefont {H.~A.}\ \bibnamefont {Perrett}}, \bibinfo {author} {\bibfnamefont {A.}~\bibnamefont {Raggio}}, \bibinfo {author}
  {\bibfnamefont {J.~R.}\ \bibnamefont {Reilly}}, \bibinfo {author} {\bibfnamefont {S.}~\bibnamefont {Rothe}}, \bibinfo {author} {\bibfnamefont {E.}~\bibnamefont {Smets}}, \bibinfo {author} {\bibfnamefont {S.-M.}\ \bibnamefont {Udrescu}}, \bibinfo {author} {\bibfnamefont {B.}~\bibnamefont {Van Den~Borne}}, \bibinfo {author} {\bibfnamefont {Q.}~\bibnamefont {Wang}}, \bibinfo {author} {\bibfnamefont {J.}~\bibnamefont {Warbinek}}, \bibinfo {author} {\bibfnamefont {J.}~\bibnamefont {Wessolek}}, \bibinfo {author} {\bibfnamefont {X.~F.}\ \bibnamefont {Yang}},\ and\ \bibinfo {author} {\bibfnamefont {C.}~\bibnamefont {Z{\"u}lch}},\ }\href {https://doi.org/10.1038/s41467-025-55977-w} {\bibfield  {journal} {\bibinfo  {journal} {Nature Communications}\ }\textbf {\bibinfo {volume} {16}},\ \bibinfo {pages} {2139} (\bibinfo {year} {2025})}\BibitemShut {NoStop}%
\bibitem [{\citenamefont {Fan}\ \emph {et~al.}(2021)\citenamefont {Fan}, \citenamefont {Holliman}, \citenamefont {Shi}, \citenamefont {Zhang}, \citenamefont {Straus}, \citenamefont {Li}, \citenamefont {Buechele},\ and\ \citenamefont {Jayich}}]{fan2021optical}%
  \BibitemOpen
  \bibfield  {author} {\bibinfo {author} {\bibfnamefont {M.}~\bibnamefont {Fan}}, \bibinfo {author} {\bibfnamefont {C.}~\bibnamefont {Holliman}}, \bibinfo {author} {\bibfnamefont {X.}~\bibnamefont {Shi}}, \bibinfo {author} {\bibfnamefont {H.}~\bibnamefont {Zhang}}, \bibinfo {author} {\bibfnamefont {M.}~\bibnamefont {Straus}}, \bibinfo {author} {\bibfnamefont {X.}~\bibnamefont {Li}}, \bibinfo {author} {\bibfnamefont {S.}~\bibnamefont {Buechele}},\ and\ \bibinfo {author} {\bibfnamefont {A.}~\bibnamefont {Jayich}},\ }\href {https://doi.org/10.1103/PhysRevLett.126.023002} {\bibfield  {journal} {\bibinfo  {journal} {Physical Review Letters}\ }\textbf {\bibinfo {volume} {126}},\ \bibinfo {pages} {023002} (\bibinfo {year} {2021})}\BibitemShut {NoStop}%
\bibitem [{\citenamefont {Yu}\ and\ \citenamefont {Hutzler}(2021)}]{yu2021probing}%
  \BibitemOpen
  \bibfield  {author} {\bibinfo {author} {\bibfnamefont {P.}~\bibnamefont {Yu}}\ and\ \bibinfo {author} {\bibfnamefont {N.~R.}\ \bibnamefont {Hutzler}},\ }\href {https://doi.org/10.1103/PhysRevLett.126.023003} {\bibfield  {journal} {\bibinfo  {journal} {Physical Review Letters}\ }\textbf {\bibinfo {volume} {126}},\ \bibinfo {pages} {023003} (\bibinfo {year} {2021})}\BibitemShut {NoStop}%
\bibitem [{\citenamefont {Hutzler}\ \emph {et~al.}(2012)\citenamefont {Hutzler}, \citenamefont {Lu},\ and\ \citenamefont {Doyle}}]{hutzler2012buffer}%
  \BibitemOpen
  \bibfield  {author} {\bibinfo {author} {\bibfnamefont {N.~R.}\ \bibnamefont {Hutzler}}, \bibinfo {author} {\bibfnamefont {H.-I.}\ \bibnamefont {Lu}},\ and\ \bibinfo {author} {\bibfnamefont {J.~M.}\ \bibnamefont {Doyle}},\ }\href {https://doi.org/10.1021/cr200362u} {\bibfield  {journal} {\bibinfo  {journal} {Chemical Reviews}\ }\textbf {\bibinfo {volume} {112}},\ \bibinfo {pages} {4803} (\bibinfo {year} {2012})}\BibitemShut {NoStop}%
\bibitem [{\citenamefont {Jadbabaie}\ \emph {et~al.}(2020)\citenamefont {Jadbabaie}, \citenamefont {Pilgram}, \citenamefont {K{\l}os}, \citenamefont {Kotochigova},\ and\ \citenamefont {Hutzler}}]{jadbabaie2020enhanced}%
  \BibitemOpen
  \bibfield  {author} {\bibinfo {author} {\bibfnamefont {A.}~\bibnamefont {Jadbabaie}}, \bibinfo {author} {\bibfnamefont {N.~H.}\ \bibnamefont {Pilgram}}, \bibinfo {author} {\bibfnamefont {J.}~\bibnamefont {K{\l}os}}, \bibinfo {author} {\bibfnamefont {S.}~\bibnamefont {Kotochigova}},\ and\ \bibinfo {author} {\bibfnamefont {N.~R.}\ \bibnamefont {Hutzler}},\ }\href {https://doi.org/10.1088/1367-2630/ab6eae} {\bibfield  {journal} {\bibinfo  {journal} {New Journal of Physics}\ }\textbf {\bibinfo {volume} {22}},\ \bibinfo {pages} {022002} (\bibinfo {year} {2020})}\BibitemShut {NoStop}%
\bibitem [{\citenamefont {Fitch}\ and\ \citenamefont {Tarbutt}(2021)}]{fitch2021laser}%
  \BibitemOpen
  \bibfield  {author} {\bibinfo {author} {\bibfnamefont {N.}~\bibnamefont {Fitch}}\ and\ \bibinfo {author} {\bibfnamefont {M.}~\bibnamefont {Tarbutt}},\ }\href {https://doi.org/10.1016/bs.aamop.2021.04.003} {\bibfield  {journal} {\bibinfo  {journal} {Advances in Atomic, Molecular, and Optical Physics}\ }\textbf {\bibinfo {volume} {70}},\ \bibinfo {pages} {157} (\bibinfo {year} {2021})}\BibitemShut {NoStop}%
\bibitem [{\citenamefont {Augenbraun}\ \emph {et~al.}(2023)\citenamefont {Augenbraun}, \citenamefont {Anderegg}, \citenamefont {Hallas}, \citenamefont {Lasner}, \citenamefont {Vilas},\ and\ \citenamefont {Doyle}}]{Augenbraun2023PolyLCReview}%
  \BibitemOpen
  \bibfield  {author} {\bibinfo {author} {\bibfnamefont {B.~L.}\ \bibnamefont {Augenbraun}}, \bibinfo {author} {\bibfnamefont {L.}~\bibnamefont {Anderegg}}, \bibinfo {author} {\bibfnamefont {C.}~\bibnamefont {Hallas}}, \bibinfo {author} {\bibfnamefont {Z.~D.}\ \bibnamefont {Lasner}}, \bibinfo {author} {\bibfnamefont {N.~B.}\ \bibnamefont {Vilas}},\ and\ \bibinfo {author} {\bibfnamefont {J.~M.}\ \bibnamefont {Doyle}},\ }\href {https://doi.org/10.1016/bs.aamop.2023.04.005} {\bibfield  {journal} {\bibinfo  {journal} {Advances in Atomic, Molecular and Optical Physics}\ }\textbf {\bibinfo {volume} {72}},\ \bibinfo {pages} {89} (\bibinfo {year} {2023})}\BibitemShut {NoStop}%
\bibitem [{\citenamefont {Hutzler}(2020)}]{Hutzler2020PolyReview}%
  \BibitemOpen
  \bibfield  {author} {\bibinfo {author} {\bibfnamefont {N.~R.}\ \bibnamefont {Hutzler}},\ }\href {https://doi.org/10.1088/2058-9565/abb9c5} {\bibfield  {journal} {\bibinfo  {journal} {Quantum Science and Technology}\ }\textbf {\bibinfo {volume} {5}},\ \bibinfo {pages} {044011} (\bibinfo {year} {2020})}\BibitemShut {NoStop}%
\bibitem [{\citenamefont {Kozyryev}\ and\ \citenamefont {Hutzler}(2017)}]{kozyryev2017precision}%
  \BibitemOpen
  \bibfield  {author} {\bibinfo {author} {\bibfnamefont {I.}~\bibnamefont {Kozyryev}}\ and\ \bibinfo {author} {\bibfnamefont {N.~R.}\ \bibnamefont {Hutzler}},\ }\href {https://doi.org/10.1103/PhysRevLett.119.133002} {\bibfield  {journal} {\bibinfo  {journal} {Physical Review Letters}\ }\textbf {\bibinfo {volume} {119}},\ \bibinfo {pages} {133002} (\bibinfo {year} {2017})}\BibitemShut {NoStop}%
\bibitem [{\citenamefont {Takahashi}\ \emph {et~al.}(2023)\citenamefont {Takahashi}, \citenamefont {Zhang}, \citenamefont {Jadbabaie},\ and\ \citenamefont {Hutzler}}]{takahashi2023engineering}%
  \BibitemOpen
  \bibfield  {author} {\bibinfo {author} {\bibfnamefont {Y.}~\bibnamefont {Takahashi}}, \bibinfo {author} {\bibfnamefont {C.}~\bibnamefont {Zhang}}, \bibinfo {author} {\bibfnamefont {A.}~\bibnamefont {Jadbabaie}},\ and\ \bibinfo {author} {\bibfnamefont {N.~R.}\ \bibnamefont {Hutzler}},\ }\href {https://doi.org/10.1103/PhysRevLett.131.183003} {\bibfield  {journal} {\bibinfo  {journal} {Physical Review Letters}\ }\textbf {\bibinfo {volume} {131}},\ \bibinfo {pages} {183003} (\bibinfo {year} {2023})}\BibitemShut {NoStop}%
\bibitem [{\citenamefont {Anderegg}\ \emph {et~al.}(2023)\citenamefont {Anderegg}, \citenamefont {Vilas}, \citenamefont {Hallas}, \citenamefont {Robichaud}, \citenamefont {Jadbabaie}, \citenamefont {Doyle},\ and\ \citenamefont {Hutzler}}]{anderegg2023quantum}%
  \BibitemOpen
  \bibfield  {author} {\bibinfo {author} {\bibfnamefont {L.}~\bibnamefont {Anderegg}}, \bibinfo {author} {\bibfnamefont {N.~B.}\ \bibnamefont {Vilas}}, \bibinfo {author} {\bibfnamefont {C.}~\bibnamefont {Hallas}}, \bibinfo {author} {\bibfnamefont {P.}~\bibnamefont {Robichaud}}, \bibinfo {author} {\bibfnamefont {A.}~\bibnamefont {Jadbabaie}}, \bibinfo {author} {\bibfnamefont {J.~M.}\ \bibnamefont {Doyle}},\ and\ \bibinfo {author} {\bibfnamefont {N.~R.}\ \bibnamefont {Hutzler}},\ }\href {https://doi.org/10.1126/science.adg8155} {\bibfield  {journal} {\bibinfo  {journal} {Science}\ }\textbf {\bibinfo {volume} {382}},\ \bibinfo {pages} {665} (\bibinfo {year} {2023})}\BibitemShut {NoStop}%
\bibitem [{\citenamefont {Yu}\ \emph {et~al.}(2019)\citenamefont {Yu}, \citenamefont {Cheuk}, \citenamefont {Kozyryev},\ and\ \citenamefont {Doyle}}]{Yu2019PolyQIS}%
  \BibitemOpen
  \bibfield  {author} {\bibinfo {author} {\bibfnamefont {P.}~\bibnamefont {Yu}}, \bibinfo {author} {\bibfnamefont {L.~W.}\ \bibnamefont {Cheuk}}, \bibinfo {author} {\bibfnamefont {I.}~\bibnamefont {Kozyryev}},\ and\ \bibinfo {author} {\bibfnamefont {J.~M.}\ \bibnamefont {Doyle}},\ }\href {https://doi.org/10.1088/1367-2630/ab428d} {\bibfield  {journal} {\bibinfo  {journal} {New Journal of Physics}\ }\textbf {\bibinfo {volume} {21}},\ \bibinfo {pages} {093049} (\bibinfo {year} {2019})}\BibitemShut {NoStop}%
\bibitem [{\citenamefont {DeMille}\ \emph {et~al.}(2024)\citenamefont {DeMille}, \citenamefont {Hutzler}, \citenamefont {Rey},\ and\ \citenamefont {Zelevinsky}}]{DeMille2024QSReview}%
  \BibitemOpen
  \bibfield  {author} {\bibinfo {author} {\bibfnamefont {D.}~\bibnamefont {DeMille}}, \bibinfo {author} {\bibfnamefont {N.~R.}\ \bibnamefont {Hutzler}}, \bibinfo {author} {\bibfnamefont {A.~M.}\ \bibnamefont {Rey}},\ and\ \bibinfo {author} {\bibfnamefont {T.}~\bibnamefont {Zelevinsky}},\ }\href {https://doi.org/10.1038/s41567-024-02499-9} {\bibfield  {journal} {\bibinfo  {journal} {Nature Physics}\ }\textbf {\bibinfo {volume} {20}},\ \bibinfo {pages} {741} (\bibinfo {year} {2024})}\BibitemShut {NoStop}%
\bibitem [{\citenamefont {Chupp}\ \emph {et~al.}(2019)\citenamefont {Chupp}, \citenamefont {Fierlinger}, \citenamefont {Ramsey-Musolf},\ and\ \citenamefont {Singh}}]{Chupp2019Review}%
  \BibitemOpen
  \bibfield  {author} {\bibinfo {author} {\bibfnamefont {T.~E.}\ \bibnamefont {Chupp}}, \bibinfo {author} {\bibfnamefont {P.}~\bibnamefont {Fierlinger}}, \bibinfo {author} {\bibfnamefont {M.~J.}\ \bibnamefont {Ramsey-Musolf}},\ and\ \bibinfo {author} {\bibfnamefont {J.~T.}\ \bibnamefont {Singh}},\ }\href {https://doi.org/10.1103/RevModPhys.91.015001} {\bibfield  {journal} {\bibinfo  {journal} {Reviews of Modern Physics}\ }\textbf {\bibinfo {volume} {91}},\ \bibinfo {pages} {015001} (\bibinfo {year} {2019})}\BibitemShut {NoStop}%
\bibitem [{\citenamefont {Cornish}\ \emph {et~al.}(2024)\citenamefont {Cornish}, \citenamefont {Tarbutt},\ and\ \citenamefont {Hazzard}}]{Cornish2024Review}%
  \BibitemOpen
  \bibfield  {author} {\bibinfo {author} {\bibfnamefont {S.~L.}\ \bibnamefont {Cornish}}, \bibinfo {author} {\bibfnamefont {M.~R.}\ \bibnamefont {Tarbutt}},\ and\ \bibinfo {author} {\bibfnamefont {K.~R.~A.}\ \bibnamefont {Hazzard}},\ }\href {https://doi.org/10.1038/s41567-024-02453-9} {\bibfield  {journal} {\bibinfo  {journal} {Nature Physics}\ }\textbf {\bibinfo {volume} {20}},\ \bibinfo {pages} {730} (\bibinfo {year} {2024})}\BibitemShut {NoStop}%
\bibitem [{\citenamefont {Langen}\ \emph {et~al.}(2024)\citenamefont {Langen}, \citenamefont {Valtolina}, \citenamefont {Wang},\ and\ \citenamefont {Ye}}]{Langen2024Review}%
  \BibitemOpen
  \bibfield  {author} {\bibinfo {author} {\bibfnamefont {T.}~\bibnamefont {Langen}}, \bibinfo {author} {\bibfnamefont {G.}~\bibnamefont {Valtolina}}, \bibinfo {author} {\bibfnamefont {D.}~\bibnamefont {Wang}},\ and\ \bibinfo {author} {\bibfnamefont {J.}~\bibnamefont {Ye}},\ }\href {https://doi.org/10.1038/s41567-024-02423-1} {\bibfield  {journal} {\bibinfo  {journal} {Nature Physics}\ }\textbf {\bibinfo {volume} {20}},\ \bibinfo {pages} {702} (\bibinfo {year} {2024})}\BibitemShut {NoStop}%
\bibitem [{\citenamefont {Zhang}\ \emph {et~al.}(2023{\natexlab{b}})\citenamefont {Zhang}, \citenamefont {Yu}, \citenamefont {Conn}, \citenamefont {Hutzler},\ and\ \citenamefont {Cheng}}]{zhang2023relativistic}%
  \BibitemOpen
  \bibfield  {author} {\bibinfo {author} {\bibfnamefont {C.}~\bibnamefont {Zhang}}, \bibinfo {author} {\bibfnamefont {P.}~\bibnamefont {Yu}}, \bibinfo {author} {\bibfnamefont {C.~J.}\ \bibnamefont {Conn}}, \bibinfo {author} {\bibfnamefont {N.~R.}\ \bibnamefont {Hutzler}},\ and\ \bibinfo {author} {\bibfnamefont {L.}~\bibnamefont {Cheng}},\ }\href {https://pubs.rsc.org/en/content/articlehtml/2023/cp/d3cp04040b} {\bibfield  {journal} {\bibinfo  {journal} {Physical Chemistry Chemical Physics}\ }\textbf {\bibinfo {volume} {25}},\ \bibinfo {pages} {32613} (\bibinfo {year} {2023}{\natexlab{b}})}\BibitemShut {NoStop}%
\bibitem [{sup()}]{supmat}%
  \BibitemOpen
  \href@noop {} {}\bibinfo {note} {See supplementary materials.}\BibitemShut {Stop}%
\bibitem [{\citenamefont {Scielzo}\ \emph {et~al.}(2006)\citenamefont {Scielzo}, \citenamefont {Guest}, \citenamefont {Schulte}, \citenamefont {Ahmad}, \citenamefont {Bailey}, \citenamefont {Bowers}, \citenamefont {Holt}, \citenamefont {Lu}, \citenamefont {O’Connor},\ and\ \citenamefont {Potterveld}}]{scielzo2006measurement}%
  \BibitemOpen
  \bibfield  {author} {\bibinfo {author} {\bibfnamefont {N.~D.}\ \bibnamefont {Scielzo}}, \bibinfo {author} {\bibfnamefont {J.~R.}\ \bibnamefont {Guest}}, \bibinfo {author} {\bibfnamefont {E.~C.}\ \bibnamefont {Schulte}}, \bibinfo {author} {\bibfnamefont {I.}~\bibnamefont {Ahmad}}, \bibinfo {author} {\bibfnamefont {K.}~\bibnamefont {Bailey}}, \bibinfo {author} {\bibfnamefont {D.~L.}\ \bibnamefont {Bowers}}, \bibinfo {author} {\bibfnamefont {R.~J.}\ \bibnamefont {Holt}}, \bibinfo {author} {\bibfnamefont {Z.-T.}\ \bibnamefont {Lu}}, \bibinfo {author} {\bibfnamefont {T.~P.}\ \bibnamefont {O’Connor}},\ and\ \bibinfo {author} {\bibfnamefont {D.~H.}\ \bibnamefont {Potterveld}},\ }\href {https://doi.org/10.1103/PhysRevA.73.010501} {\bibfield  {journal} {\bibinfo  {journal} {Physical Review A}\ }\textbf {\bibinfo {volume} {73}},\ \bibinfo {pages} {010501} (\bibinfo {year} {2006})}\BibitemShut {NoStop}%
\bibitem [{\citenamefont {Pilgram}\ \emph {et~al.}(2021)\citenamefont {Pilgram}, \citenamefont {Jadbabaie}, \citenamefont {Zeng}, \citenamefont {Hutzler},\ and\ \citenamefont {Steimle}}]{Pilgram2021YbOHOdd}%
  \BibitemOpen
  \bibfield  {author} {\bibinfo {author} {\bibfnamefont {N.~H.}\ \bibnamefont {Pilgram}}, \bibinfo {author} {\bibfnamefont {A.}~\bibnamefont {Jadbabaie}}, \bibinfo {author} {\bibfnamefont {Y.}~\bibnamefont {Zeng}}, \bibinfo {author} {\bibfnamefont {N.~R.}\ \bibnamefont {Hutzler}},\ and\ \bibinfo {author} {\bibfnamefont {T.~C.}\ \bibnamefont {Steimle}},\ }\href {https://doi.org/10.1063/5.0055293} {\bibfield  {journal} {\bibinfo  {journal} {The Journal of Chemical Physics}\ }\textbf {\bibinfo {volume} {154}},\ \bibinfo {pages} {244309} (\bibinfo {year} {2021})}\BibitemShut {NoStop}%
\bibitem [{\citenamefont {Dzuba}\ and\ \citenamefont {Flambaum}(2006)}]{dzuba2006calculation}%
  \BibitemOpen
  \bibfield  {author} {\bibinfo {author} {\bibfnamefont {V.}~\bibnamefont {Dzuba}}\ and\ \bibinfo {author} {\bibfnamefont {V.}~\bibnamefont {Flambaum}},\ }\href {https://iopscience.iop.org/article/10.1088/0953-4075/40/1/021} {\bibfield  {journal} {\bibinfo  {journal} {Journal of Physics B}\ }\textbf {\bibinfo {volume} {40}},\ \bibinfo {pages} {227} (\bibinfo {year} {2006})}\BibitemShut {NoStop}%
\bibitem [{\citenamefont {Zhang}\ and\ \citenamefont {Cheng}(2022)}]{Zhang22}%
  \BibitemOpen
  \bibfield  {author} {\bibinfo {author} {\bibfnamefont {C.}~\bibnamefont {Zhang}}\ and\ \bibinfo {author} {\bibfnamefont {L.}~\bibnamefont {Cheng}},\ }\href {https://doi.org/10.1021/acs.jpca.2c02181} {\bibfield  {journal} {\bibinfo  {journal} {J. Phys. Chem. A}\ }\textbf {\bibinfo {volume} {126}},\ \bibinfo {pages} {4537} (\bibinfo {year} {2022})}\BibitemShut {NoStop}%
\bibitem [{\citenamefont {Asthana}\ \emph {et~al.}(2019)\citenamefont {Asthana}, \citenamefont {Liu},\ and\ \citenamefont {Cheng}}]{Asthana19}%
  \BibitemOpen
  \bibfield  {author} {\bibinfo {author} {\bibfnamefont {A.}~\bibnamefont {Asthana}}, \bibinfo {author} {\bibfnamefont {J.}~\bibnamefont {Liu}},\ and\ \bibinfo {author} {\bibfnamefont {L.}~\bibnamefont {Cheng}},\ }\href {https://pubs.aip.org/aip/jcp/article/150/7/074102/197572/Exact-two-component-equation-of-motion-coupled} {\bibfield  {journal} {\bibinfo  {journal} {J. Chem. Phys.}\ }\textbf {\bibinfo {volume} {150}},\ \bibinfo {pages} {074102} (\bibinfo {year} {2019})}\BibitemShut {NoStop}%
\bibitem [{\citenamefont {Ivanov}\ \emph {et~al.}(2019)\citenamefont {Ivanov}, \citenamefont {Bangerter},\ and\ \citenamefont {Krylov}}]{ivanov2019towards}%
  \BibitemOpen
  \bibfield  {author} {\bibinfo {author} {\bibfnamefont {M.~V.}\ \bibnamefont {Ivanov}}, \bibinfo {author} {\bibfnamefont {F.~H.}\ \bibnamefont {Bangerter}},\ and\ \bibinfo {author} {\bibfnamefont {A.~I.}\ \bibnamefont {Krylov}},\ }\href {https://pubs.rsc.org/en/content/articlelanding/2019/cp/c9cp03914g} {\bibfield  {journal} {\bibinfo  {journal} {Physical Chemistry Chemical Physics}\ }\textbf {\bibinfo {volume} {21}},\ \bibinfo {pages} {19447} (\bibinfo {year} {2019})}\BibitemShut {NoStop}%
\bibitem [{\citenamefont {Vilas}\ \emph {et~al.}(2023)\citenamefont {Vilas}, \citenamefont {Hallas}, \citenamefont {Anderegg}, \citenamefont {Robichaud}, \citenamefont {Zhang}, \citenamefont {Dawley}, \citenamefont {Cheng},\ and\ \citenamefont {Doyle}}]{Vilas2023BBR}%
  \BibitemOpen
  \bibfield  {author} {\bibinfo {author} {\bibfnamefont {N.~B.}\ \bibnamefont {Vilas}}, \bibinfo {author} {\bibfnamefont {C.}~\bibnamefont {Hallas}}, \bibinfo {author} {\bibfnamefont {L.}~\bibnamefont {Anderegg}}, \bibinfo {author} {\bibfnamefont {P.}~\bibnamefont {Robichaud}}, \bibinfo {author} {\bibfnamefont {C.}~\bibnamefont {Zhang}}, \bibinfo {author} {\bibfnamefont {S.}~\bibnamefont {Dawley}}, \bibinfo {author} {\bibfnamefont {L.}~\bibnamefont {Cheng}},\ and\ \bibinfo {author} {\bibfnamefont {J.~M.}\ \bibnamefont {Doyle}},\ }\href {https://doi.org/10.1103/PhysRevA.107.062802} {\bibfield  {journal} {\bibinfo  {journal} {Physical Review A}\ }\textbf {\bibinfo {volume} {107}},\ \bibinfo {pages} {062802} (\bibinfo {year} {2023})}\BibitemShut {NoStop}%
\bibitem [{\citenamefont {Kinsey-Nielsen}\ \emph {et~al.}(1986)\citenamefont {Kinsey-Nielsen}, \citenamefont {Brazier},\ and\ \citenamefont {Bernath}}]{kinsey1986rotational}%
  \BibitemOpen
  \bibfield  {author} {\bibinfo {author} {\bibfnamefont {S.}~\bibnamefont {Kinsey-Nielsen}}, \bibinfo {author} {\bibfnamefont {C.}~\bibnamefont {Brazier}},\ and\ \bibinfo {author} {\bibfnamefont {P.}~\bibnamefont {Bernath}},\ }\href {https://doi.org/10.1063/1.450566} {\bibfield  {journal} {\bibinfo  {journal} {The Journal of Chemical Physics}\ }\textbf {\bibinfo {volume} {84}},\ \bibinfo {pages} {698} (\bibinfo {year} {1986})}\BibitemShut {NoStop}%
\bibitem [{\citenamefont {Gustavsson}\ \emph {et~al.}(1991)\citenamefont {Gustavsson}, \citenamefont {Alcaraz}, \citenamefont {Berlande}, \citenamefont {Cuvellier}, \citenamefont {Mestdagh}, \citenamefont {Meynadier}, \citenamefont {De~Pujo}, \citenamefont {Sublemontier},\ and\ \citenamefont {Visticot}}]{gustavsson1991perturbations}%
  \BibitemOpen
  \bibfield  {author} {\bibinfo {author} {\bibfnamefont {T.}~\bibnamefont {Gustavsson}}, \bibinfo {author} {\bibfnamefont {C.}~\bibnamefont {Alcaraz}}, \bibinfo {author} {\bibfnamefont {J.}~\bibnamefont {Berlande}}, \bibinfo {author} {\bibfnamefont {J.}~\bibnamefont {Cuvellier}}, \bibinfo {author} {\bibfnamefont {J.-M.}\ \bibnamefont {Mestdagh}}, \bibinfo {author} {\bibfnamefont {P.}~\bibnamefont {Meynadier}}, \bibinfo {author} {\bibfnamefont {P.}~\bibnamefont {De~Pujo}}, \bibinfo {author} {\bibfnamefont {O.}~\bibnamefont {Sublemontier}},\ and\ \bibinfo {author} {\bibfnamefont {J.-P.}\ \bibnamefont {Visticot}},\ }\href {https://doi.org/10.1016/0022-2852(91)90364-G} {\bibfield  {journal} {\bibinfo  {journal} {Journal of Molecular Spectroscopy}\ }\textbf {\bibinfo {volume} {145}},\ \bibinfo {pages} {210} (\bibinfo {year} {1991})}\BibitemShut {NoStop}%
\bibitem [{\citenamefont {Patel}\ \emph {et~al.}(2025)\citenamefont {Patel}, \citenamefont {Howard}, \citenamefont {Steimle},\ and\ \citenamefont {Hutzler}}]{patel2025rapid}%
  \BibitemOpen
  \bibfield  {author} {\bibinfo {author} {\bibfnamefont {A.~N.}\ \bibnamefont {Patel}}, \bibinfo {author} {\bibfnamefont {M.~I.}\ \bibnamefont {Howard}}, \bibinfo {author} {\bibfnamefont {T.~C.}\ \bibnamefont {Steimle}},\ and\ \bibinfo {author} {\bibfnamefont {N.~R.}\ \bibnamefont {Hutzler}},\ }\href {https://arxiv.org/abs/2505.03650} {\bibfield  {journal} {\bibinfo  {journal} {arXiv:2505.03650}\ } (\bibinfo {year} {2025})}\BibitemShut {NoStop}%
\bibitem [{\citenamefont {Anderegg}\ \emph {et~al.}(2019)\citenamefont {Anderegg}, \citenamefont {Cheuk}, \citenamefont {Bao}, \citenamefont {Burchesky}, \citenamefont {Ketterle}, \citenamefont {Ni},\ and\ \citenamefont {Doyle}}]{anderegg2019optical}%
  \BibitemOpen
  \bibfield  {author} {\bibinfo {author} {\bibfnamefont {L.}~\bibnamefont {Anderegg}}, \bibinfo {author} {\bibfnamefont {L.~W.}\ \bibnamefont {Cheuk}}, \bibinfo {author} {\bibfnamefont {Y.}~\bibnamefont {Bao}}, \bibinfo {author} {\bibfnamefont {S.}~\bibnamefont {Burchesky}}, \bibinfo {author} {\bibfnamefont {W.}~\bibnamefont {Ketterle}}, \bibinfo {author} {\bibfnamefont {K.-K.}\ \bibnamefont {Ni}},\ and\ \bibinfo {author} {\bibfnamefont {J.~M.}\ \bibnamefont {Doyle}},\ }\href {https://doi.org/10.1126/science.aax1265} {\bibfield  {journal} {\bibinfo  {journal} {Science}\ }\textbf {\bibinfo {volume} {365}},\ \bibinfo {pages} {1156} (\bibinfo {year} {2019})}\BibitemShut {NoStop}%
\bibitem [{\citenamefont {Vilas}\ \emph {et~al.}(2024)\citenamefont {Vilas}, \citenamefont {Robichaud}, \citenamefont {Hallas}, \citenamefont {Li}, \citenamefont {Anderegg},\ and\ \citenamefont {Doyle}}]{vilas2024optical}%
  \BibitemOpen
  \bibfield  {author} {\bibinfo {author} {\bibfnamefont {N.~B.}\ \bibnamefont {Vilas}}, \bibinfo {author} {\bibfnamefont {P.}~\bibnamefont {Robichaud}}, \bibinfo {author} {\bibfnamefont {C.}~\bibnamefont {Hallas}}, \bibinfo {author} {\bibfnamefont {G.~K.}\ \bibnamefont {Li}}, \bibinfo {author} {\bibfnamefont {L.}~\bibnamefont {Anderegg}},\ and\ \bibinfo {author} {\bibfnamefont {J.~M.}\ \bibnamefont {Doyle}},\ }\href {https://doi.org/10.1038/s41586-024-07199-1} {\bibfield  {journal} {\bibinfo  {journal} {Nature}\ }\textbf {\bibinfo {volume} {628}},\ \bibinfo {pages} {282} (\bibinfo {year} {2024})}\BibitemShut {NoStop}%
\bibitem [{\citenamefont {Hudson}\ \emph {et~al.}(2011)\citenamefont {Hudson}, \citenamefont {Kara}, \citenamefont {Smallman}, \citenamefont {Sauer}, \citenamefont {Tarbutt},\ and\ \citenamefont {Hinds}}]{hudson2011improved}%
  \BibitemOpen
  \bibfield  {author} {\bibinfo {author} {\bibfnamefont {J.~J.}\ \bibnamefont {Hudson}}, \bibinfo {author} {\bibfnamefont {D.~M.}\ \bibnamefont {Kara}}, \bibinfo {author} {\bibfnamefont {I.}~\bibnamefont {Smallman}}, \bibinfo {author} {\bibfnamefont {B.~E.}\ \bibnamefont {Sauer}}, \bibinfo {author} {\bibfnamefont {M.~R.}\ \bibnamefont {Tarbutt}},\ and\ \bibinfo {author} {\bibfnamefont {E.~A.}\ \bibnamefont {Hinds}},\ }\href {https://doi.org/10.1038/nature10104} {\bibfield  {journal} {\bibinfo  {journal} {Nature}\ }\textbf {\bibinfo {volume} {473}},\ \bibinfo {pages} {493} (\bibinfo {year} {2011})}\BibitemShut {NoStop}%
\bibitem [{\citenamefont {Wilkins}\ \emph {et~al.}(2025)\citenamefont {Wilkins}, \citenamefont {Udrescu}, \citenamefont {Athanasakis-Kaklamanakis}, \citenamefont {Garcia~Ruiz}, \citenamefont {Au}, \citenamefont {Belo{\v{s}}evi{\'c}}, \citenamefont {Berger}, \citenamefont {Bissell}, \citenamefont {Breier}, \citenamefont {Brinson} \emph {et~al.}}]{wilkins2025observation}%
  \BibitemOpen
  \bibfield  {author} {\bibinfo {author} {\bibfnamefont {S.}~\bibnamefont {Wilkins}}, \bibinfo {author} {\bibfnamefont {S.}~\bibnamefont {Udrescu}}, \bibinfo {author} {\bibfnamefont {M.}~\bibnamefont {Athanasakis-Kaklamanakis}}, \bibinfo {author} {\bibfnamefont {R.}~\bibnamefont {Garcia~Ruiz}}, \bibinfo {author} {\bibfnamefont {M.}~\bibnamefont {Au}}, \bibinfo {author} {\bibfnamefont {I.}~\bibnamefont {Belo{\v{s}}evi{\'c}}}, \bibinfo {author} {\bibfnamefont {R.}~\bibnamefont {Berger}}, \bibinfo {author} {\bibfnamefont {M.}~\bibnamefont {Bissell}}, \bibinfo {author} {\bibfnamefont {A.}~\bibnamefont {Breier}}, \bibinfo {author} {\bibfnamefont {A.}~\bibnamefont {Brinson}}, \emph {et~al.},\ }\href@noop {} {\bibfield  {journal} {\bibinfo  {journal} {Science}\ }\textbf {\bibinfo {volume} {390}},\ \bibinfo {pages} {386} (\bibinfo {year} {2025})}\BibitemShut {NoStop}%
\bibitem [{\citenamefont {Hutzler}\ \emph {et~al.}(2026)\citenamefont {Hutzler}, \citenamefont {Conn},\ and\ \citenamefont {Yu}}]{hutzler_conn_yu_2026}%
  \BibitemOpen
  \bibfield  {author} {\bibinfo {author} {\bibfnamefont {N.~R.}\ \bibnamefont {Hutzler}}, \bibinfo {author} {\bibfnamefont {C.}~\bibnamefont {Conn}},\ and\ \bibinfo {author} {\bibfnamefont {P.}~\bibnamefont {Yu}},\ }\href {https://doi.org/10.22002/qc4at-7mn31} {\bibinfo {title} {Production and spectroscopy of cold radioactive molecules {[Data set]}}} (\bibinfo {year} {2026}),\ \bibinfo {note} {{CaltechDATA}. https://doi.org/10.22002/qc4at-7mn31}\BibitemShut {NoStop}%
\bibitem [{\citenamefont {Mampallil}\ and\ \citenamefont {Eral}(2018)}]{mampallil2018review}%
  \BibitemOpen
  \bibfield  {author} {\bibinfo {author} {\bibfnamefont {D.}~\bibnamefont {Mampallil}}\ and\ \bibinfo {author} {\bibfnamefont {H.~B.}\ \bibnamefont {Eral}},\ }\href {https://doi.org/10.1016/j.cis.2017.12.008} {\bibfield  {journal} {\bibinfo  {journal} {Advances in Colloid and Interface Science}\ }\textbf {\bibinfo {volume} {252}},\ \bibinfo {pages} {38} (\bibinfo {year} {2018})}\BibitemShut {NoStop}%
\bibitem [{\citenamefont {Kocjan}\ \emph {et~al.}(2008)\citenamefont {Kocjan}, \citenamefont {Krnel},\ and\ \citenamefont {Kosma{\v{c}}}}]{kocjan2008influence}%
  \BibitemOpen
  \bibfield  {author} {\bibinfo {author} {\bibfnamefont {A.}~\bibnamefont {Kocjan}}, \bibinfo {author} {\bibfnamefont {K.}~\bibnamefont {Krnel}},\ and\ \bibinfo {author} {\bibfnamefont {T.}~\bibnamefont {Kosma{\v{c}}}},\ }\href {https://doi.org/10.1016/j.jeurceramsoc.2007.09.012} {\bibfield  {journal} {\bibinfo  {journal} {Journal of the European Ceramic Society}\ }\textbf {\bibinfo {volume} {28}},\ \bibinfo {pages} {1003} (\bibinfo {year} {2008})}\BibitemShut {NoStop}%
\bibitem [{\citenamefont {Liu}\ \emph {et~al.}(2022)\citenamefont {Liu}, \citenamefont {Wang}, \citenamefont {Wright}, \citenamefont {Doppelbauer}, \citenamefont {Meijer}, \citenamefont {Truppe},\ and\ \citenamefont {P{\'e}rez-R{\'\i}os}}]{liu2022chemistry}%
  \BibitemOpen
  \bibfield  {author} {\bibinfo {author} {\bibfnamefont {X.}~\bibnamefont {Liu}}, \bibinfo {author} {\bibfnamefont {W.}~\bibnamefont {Wang}}, \bibinfo {author} {\bibfnamefont {S.}~\bibnamefont {Wright}}, \bibinfo {author} {\bibfnamefont {M.}~\bibnamefont {Doppelbauer}}, \bibinfo {author} {\bibfnamefont {G.}~\bibnamefont {Meijer}}, \bibinfo {author} {\bibfnamefont {S.}~\bibnamefont {Truppe}},\ and\ \bibinfo {author} {\bibfnamefont {J.}~\bibnamefont {P{\'e}rez-R{\'\i}os}},\ }\href {https://doi.org/10.1063/5.0098378} {\bibfield  {journal} {\bibinfo  {journal} {The Journal of Chemical Physics}\ }\textbf {\bibinfo {volume} {157}} (\bibinfo {year} {2022})}\BibitemShut {NoStop}%
\bibitem [{\citenamefont {Sharada}\ \emph {et~al.}(2014)\citenamefont {Sharada}, \citenamefont {Bell},\ and\ \citenamefont {Head-Gordon}}]{sharada2014finite}%
  \BibitemOpen
  \bibfield  {author} {\bibinfo {author} {\bibfnamefont {S.~M.}\ \bibnamefont {Sharada}}, \bibinfo {author} {\bibfnamefont {A.~T.}\ \bibnamefont {Bell}},\ and\ \bibinfo {author} {\bibfnamefont {M.}~\bibnamefont {Head-Gordon}},\ }\href {https://doi.org/10.1063/1.4871660} {\bibfield  {journal} {\bibinfo  {journal} {The Journal of Chemical Physics}\ }\textbf {\bibinfo {volume} {140}} (\bibinfo {year} {2014})}\BibitemShut {NoStop}%
\bibitem [{\citenamefont {Baker}(1986)}]{baker1986algorithm}%
  \BibitemOpen
  \bibfield  {author} {\bibinfo {author} {\bibfnamefont {J.}~\bibnamefont {Baker}},\ }\href {https://doi.org/10.1002/jcc.540070402} {\bibfield  {journal} {\bibinfo  {journal} {Journal of Computational Chemistry}\ }\textbf {\bibinfo {volume} {7}},\ \bibinfo {pages} {385} (\bibinfo {year} {1986})}\BibitemShut {NoStop}%
\bibitem [{\citenamefont {Epifanovsky}\ \emph {et~al.}(2021)\citenamefont {Epifanovsky}, \citenamefont {Gilbert}, \citenamefont {Feng}, \citenamefont {Lee}, \citenamefont {Mao}, \citenamefont {Mardirossian}, \citenamefont {Pokhilko}, \citenamefont {White}, \citenamefont {Coons}, \citenamefont {Dempwolff}, \citenamefont {Gan}, \citenamefont {Hait}, \citenamefont {Horn}, \citenamefont {Jacobson}, \citenamefont {Kaliman}, \citenamefont {Kussmann}, \citenamefont {Lange}, \citenamefont {Lao}, \citenamefont {Levine}, \citenamefont {Liu}, \citenamefont {McKenzie}, \citenamefont {Morrison}, \citenamefont {Nanda}, \citenamefont {Plasser}, \citenamefont {Rehn}, \citenamefont {Vidal}, \citenamefont {You}, \citenamefont {Zhu}, \citenamefont {Alam}, \citenamefont {Albrecht}, \citenamefont {Aldossary}, \citenamefont {Alguire}, \citenamefont {Andersen}, \citenamefont {Athavale}, \citenamefont {Barton}, \citenamefont {Begam}, \citenamefont {Behn}, \citenamefont {Bellonzi}, \citenamefont {Bernard}, \citenamefont {Berquist},
  \citenamefont {Burton}, \citenamefont {Carreras}, \citenamefont {Carter-Fenk}, \citenamefont {Chakraborty}, \citenamefont {Chien}, \citenamefont {Closser}, \citenamefont {Cofer-Shabica}, \citenamefont {Dasgupta}, \citenamefont {De~Wergifosse}, \citenamefont {Deng}, \citenamefont {Diedenhofen}, \citenamefont {Do}, \citenamefont {Ehlert}, \citenamefont {Fang}, \citenamefont {Fatehi}, \citenamefont {Feng}, \citenamefont {Friedhoff}, \citenamefont {Gayvert}, \citenamefont {Ge}, \citenamefont {Gidofalvi}, \citenamefont {Goldey}, \citenamefont {Gomes}, \citenamefont {Gonz{\'a}lez-Espinoza}, \citenamefont {Gulania}, \citenamefont {Gunina}, \citenamefont {Hanson-Heine}, \citenamefont {Harbach}, \citenamefont {Hauser}, \citenamefont {Herbst}, \citenamefont {Hern{\'a}ndez~Vera}, \citenamefont {Hodecker}, \citenamefont {Holden}, \citenamefont {Houck}, \citenamefont {Huang}, \citenamefont {Hui}, \citenamefont {Huynh}, \citenamefont {Ivanov}, \citenamefont {J{\'a}sz}, \citenamefont {Ji}, \citenamefont {Jiang},
  \citenamefont {Kaduk}, \citenamefont {K{\"a}hler}, \citenamefont {Khistyaev}, \citenamefont {Kim}, \citenamefont {Kis}, \citenamefont {Klunzinger}, \citenamefont {Koczor-Benda}, \citenamefont {Koh}, \citenamefont {Kosenkov}, \citenamefont {Koulias}, \citenamefont {Kowalczyk}, \citenamefont {Krauter}, \citenamefont {Kue}, \citenamefont {Kunitsa}, \citenamefont {Kus}, \citenamefont {Ladj{\'a}nszki}, \citenamefont {Landau}, \citenamefont {Lawler}, \citenamefont {Lefrancois}, \citenamefont {Lehtola}, \citenamefont {Li}, \citenamefont {Li}, \citenamefont {Liang}, \citenamefont {Liebenthal}, \citenamefont {Lin}, \citenamefont {Lin}, \citenamefont {Liu}, \citenamefont {Liu}, \citenamefont {Loipersberger}, \citenamefont {Luenser}, \citenamefont {Manjanath}, \citenamefont {Manohar}, \citenamefont {Mansoor}, \citenamefont {Manzer}, \citenamefont {Mao}, \citenamefont {Marenich}, \citenamefont {Markovich}, \citenamefont {Mason}, \citenamefont {Maurer}, \citenamefont {McLaughlin}, \citenamefont {Menger}, \citenamefont
  {Mewes}, \citenamefont {Mewes}, \citenamefont {Morgante}, \citenamefont {Mullinax}, \citenamefont {Oosterbaan}, \citenamefont {Paran}, \citenamefont {Paul}, \citenamefont {Paul}, \citenamefont {Pavo{\v{s}}evi{\'c}}, \citenamefont {Pei}, \citenamefont {Prager}, \citenamefont {Proynov}, \citenamefont {R{\'a}k}, \citenamefont {Ramos-Cordoba}, \citenamefont {Rana}, \citenamefont {Rask}, \citenamefont {Rettig}, \citenamefont {Richard}, \citenamefont {Rob}, \citenamefont {Rossomme}, \citenamefont {Scheele}, \citenamefont {Scheurer}, \citenamefont {Schneider}, \citenamefont {Sergueev}, \citenamefont {Sharada}, \citenamefont {Skomorowski}, \citenamefont {Small}, \citenamefont {Stein}, \citenamefont {Su}, \citenamefont {Sundstrom}, \citenamefont {Tao}, \citenamefont {Thirman}, \citenamefont {Tornai}, \citenamefont {Tsuchimochi}, \citenamefont {Tubman}, \citenamefont {Veccham}, \citenamefont {Vydrov}, \citenamefont {Wenzel}, \citenamefont {Witte}, \citenamefont {Yamada}, \citenamefont {Yao}, \citenamefont {Yeganeh},
  \citenamefont {Yost}, \citenamefont {Zech}, \citenamefont {Zhang}, \citenamefont {Zhang}, \citenamefont {Zhang}, \citenamefont {Zuev}, \citenamefont {Aspuru-Guzik}, \citenamefont {Bell}, \citenamefont {Besley}, \citenamefont {Bravaya}, \citenamefont {Brooks}, \citenamefont {Casanova}, \citenamefont {Chai}, \citenamefont {Coriani}, \citenamefont {Cramer}, \citenamefont {Cserey}, \citenamefont {DePrince}, \citenamefont {DiStasio}, \citenamefont {Dreuw}, \citenamefont {Dunietz}, \citenamefont {Furlani}, \citenamefont {Goddard}, \citenamefont {Hammes-Schiffer}, \citenamefont {Head-Gordon}, \citenamefont {Hehre}, \citenamefont {Hsu}, \citenamefont {Jagau}, \citenamefont {Jung}, \citenamefont {Klamt}, \citenamefont {Kong}, \citenamefont {Lambrecht}, \citenamefont {Liang}, \citenamefont {Mayhall}, \citenamefont {McCurdy}, \citenamefont {Neaton}, \citenamefont {Ochsenfeld}, \citenamefont {Parkhill}, \citenamefont {Peverati}, \citenamefont {Rassolov}, \citenamefont {Shao}, \citenamefont {Slipchenko}, \citenamefont
  {Stauch}, \citenamefont {Steele}, \citenamefont {Subotnik}, \citenamefont {Thom}, \citenamefont {Tkatchenko}, \citenamefont {Truhlar}, \citenamefont {Van~Voorhis}, \citenamefont {Wesolowski}, \citenamefont {Whaley}, \citenamefont {Woodcock}, \citenamefont {Zimmerman}, \citenamefont {Faraji}, \citenamefont {Gill}, \citenamefont {Head-Gordon}, \citenamefont {Herbert},\ and\ \citenamefont {Krylov}}]{epifanovsky2021software}%
  \BibitemOpen
  \bibfield  {author} {\bibinfo {author} {\bibfnamefont {E.}~\bibnamefont {Epifanovsky}}, \bibinfo {author} {\bibfnamefont {A.~T.~B.}\ \bibnamefont {Gilbert}}, \bibinfo {author} {\bibfnamefont {X.}~\bibnamefont {Feng}}, \bibinfo {author} {\bibfnamefont {J.}~\bibnamefont {Lee}}, \bibinfo {author} {\bibfnamefont {Y.}~\bibnamefont {Mao}}, \bibinfo {author} {\bibfnamefont {N.}~\bibnamefont {Mardirossian}}, \bibinfo {author} {\bibfnamefont {P.}~\bibnamefont {Pokhilko}}, \bibinfo {author} {\bibfnamefont {A.~F.}\ \bibnamefont {White}}, \bibinfo {author} {\bibfnamefont {M.~P.}\ \bibnamefont {Coons}}, \bibinfo {author} {\bibfnamefont {A.~L.}\ \bibnamefont {Dempwolff}}, \bibinfo {author} {\bibfnamefont {Z.}~\bibnamefont {Gan}}, \bibinfo {author} {\bibfnamefont {D.}~\bibnamefont {Hait}}, \bibinfo {author} {\bibfnamefont {P.~R.}\ \bibnamefont {Horn}}, \bibinfo {author} {\bibfnamefont {L.~D.}\ \bibnamefont {Jacobson}}, \bibinfo {author} {\bibfnamefont {I.}~\bibnamefont {Kaliman}}, \bibinfo {author} {\bibfnamefont
  {J.}~\bibnamefont {Kussmann}}, \bibinfo {author} {\bibfnamefont {A.~W.}\ \bibnamefont {Lange}}, \bibinfo {author} {\bibfnamefont {K.~U.}\ \bibnamefont {Lao}}, \bibinfo {author} {\bibfnamefont {D.~S.}\ \bibnamefont {Levine}}, \bibinfo {author} {\bibfnamefont {J.}~\bibnamefont {Liu}}, \bibinfo {author} {\bibfnamefont {S.~C.}\ \bibnamefont {McKenzie}}, \bibinfo {author} {\bibfnamefont {A.~F.}\ \bibnamefont {Morrison}}, \bibinfo {author} {\bibfnamefont {K.~D.}\ \bibnamefont {Nanda}}, \bibinfo {author} {\bibfnamefont {F.}~\bibnamefont {Plasser}}, \bibinfo {author} {\bibfnamefont {D.~R.}\ \bibnamefont {Rehn}}, \bibinfo {author} {\bibfnamefont {M.~L.}\ \bibnamefont {Vidal}}, \bibinfo {author} {\bibfnamefont {Z.-Q.}\ \bibnamefont {You}}, \bibinfo {author} {\bibfnamefont {Y.}~\bibnamefont {Zhu}}, \bibinfo {author} {\bibfnamefont {B.}~\bibnamefont {Alam}}, \bibinfo {author} {\bibfnamefont {B.~J.}\ \bibnamefont {Albrecht}}, \bibinfo {author} {\bibfnamefont {A.}~\bibnamefont {Aldossary}}, \bibinfo {author}
  {\bibfnamefont {E.}~\bibnamefont {Alguire}}, \bibinfo {author} {\bibfnamefont {J.~H.}\ \bibnamefont {Andersen}}, \bibinfo {author} {\bibfnamefont {V.}~\bibnamefont {Athavale}}, \bibinfo {author} {\bibfnamefont {D.}~\bibnamefont {Barton}}, \bibinfo {author} {\bibfnamefont {K.}~\bibnamefont {Begam}}, \bibinfo {author} {\bibfnamefont {A.}~\bibnamefont {Behn}}, \bibinfo {author} {\bibfnamefont {N.}~\bibnamefont {Bellonzi}}, \bibinfo {author} {\bibfnamefont {Y.~A.}\ \bibnamefont {Bernard}}, \bibinfo {author} {\bibfnamefont {E.~J.}\ \bibnamefont {Berquist}}, \bibinfo {author} {\bibfnamefont {H.~G.~A.}\ \bibnamefont {Burton}}, \bibinfo {author} {\bibfnamefont {A.}~\bibnamefont {Carreras}}, \bibinfo {author} {\bibfnamefont {K.}~\bibnamefont {Carter-Fenk}}, \bibinfo {author} {\bibfnamefont {R.}~\bibnamefont {Chakraborty}}, \bibinfo {author} {\bibfnamefont {A.~D.}\ \bibnamefont {Chien}}, \bibinfo {author} {\bibfnamefont {K.~D.}\ \bibnamefont {Closser}}, \bibinfo {author} {\bibfnamefont {V.}~\bibnamefont
  {Cofer-Shabica}}, \bibinfo {author} {\bibfnamefont {S.}~\bibnamefont {Dasgupta}}, \bibinfo {author} {\bibfnamefont {M.}~\bibnamefont {De~Wergifosse}}, \bibinfo {author} {\bibfnamefont {J.}~\bibnamefont {Deng}}, \bibinfo {author} {\bibfnamefont {M.}~\bibnamefont {Diedenhofen}}, \bibinfo {author} {\bibfnamefont {H.}~\bibnamefont {Do}}, \bibinfo {author} {\bibfnamefont {S.}~\bibnamefont {Ehlert}}, \bibinfo {author} {\bibfnamefont {P.-T.}\ \bibnamefont {Fang}}, \bibinfo {author} {\bibfnamefont {S.}~\bibnamefont {Fatehi}}, \bibinfo {author} {\bibfnamefont {Q.}~\bibnamefont {Feng}}, \bibinfo {author} {\bibfnamefont {T.}~\bibnamefont {Friedhoff}}, \bibinfo {author} {\bibfnamefont {J.}~\bibnamefont {Gayvert}}, \bibinfo {author} {\bibfnamefont {Q.}~\bibnamefont {Ge}}, \bibinfo {author} {\bibfnamefont {G.}~\bibnamefont {Gidofalvi}}, \bibinfo {author} {\bibfnamefont {M.}~\bibnamefont {Goldey}}, \bibinfo {author} {\bibfnamefont {J.}~\bibnamefont {Gomes}}, \bibinfo {author} {\bibfnamefont {C.~E.}\ \bibnamefont
  {Gonz{\'a}lez-Espinoza}}, \bibinfo {author} {\bibfnamefont {S.}~\bibnamefont {Gulania}}, \bibinfo {author} {\bibfnamefont {A.~O.}\ \bibnamefont {Gunina}}, \bibinfo {author} {\bibfnamefont {M.~W.~D.}\ \bibnamefont {Hanson-Heine}}, \bibinfo {author} {\bibfnamefont {P.~H.~P.}\ \bibnamefont {Harbach}}, \bibinfo {author} {\bibfnamefont {A.}~\bibnamefont {Hauser}}, \bibinfo {author} {\bibfnamefont {M.~F.}\ \bibnamefont {Herbst}}, \bibinfo {author} {\bibfnamefont {M.}~\bibnamefont {Hern{\'a}ndez~Vera}}, \bibinfo {author} {\bibfnamefont {M.}~\bibnamefont {Hodecker}}, \bibinfo {author} {\bibfnamefont {Z.~C.}\ \bibnamefont {Holden}}, \bibinfo {author} {\bibfnamefont {S.}~\bibnamefont {Houck}}, \bibinfo {author} {\bibfnamefont {X.}~\bibnamefont {Huang}}, \bibinfo {author} {\bibfnamefont {K.}~\bibnamefont {Hui}}, \bibinfo {author} {\bibfnamefont {B.~C.}\ \bibnamefont {Huynh}}, \bibinfo {author} {\bibfnamefont {M.}~\bibnamefont {Ivanov}}, \bibinfo {author} {\bibfnamefont {{\'A}.}~\bibnamefont {J{\'a}sz}}, \bibinfo
  {author} {\bibfnamefont {H.}~\bibnamefont {Ji}}, \bibinfo {author} {\bibfnamefont {H.}~\bibnamefont {Jiang}}, \bibinfo {author} {\bibfnamefont {B.}~\bibnamefont {Kaduk}}, \bibinfo {author} {\bibfnamefont {S.}~\bibnamefont {K{\"a}hler}}, \bibinfo {author} {\bibfnamefont {K.}~\bibnamefont {Khistyaev}}, \bibinfo {author} {\bibfnamefont {J.}~\bibnamefont {Kim}}, \bibinfo {author} {\bibfnamefont {G.}~\bibnamefont {Kis}}, \bibinfo {author} {\bibfnamefont {P.}~\bibnamefont {Klunzinger}}, \bibinfo {author} {\bibfnamefont {Z.}~\bibnamefont {Koczor-Benda}}, \bibinfo {author} {\bibfnamefont {J.~H.}\ \bibnamefont {Koh}}, \bibinfo {author} {\bibfnamefont {D.}~\bibnamefont {Kosenkov}}, \bibinfo {author} {\bibfnamefont {L.}~\bibnamefont {Koulias}}, \bibinfo {author} {\bibfnamefont {T.}~\bibnamefont {Kowalczyk}}, \bibinfo {author} {\bibfnamefont {C.~M.}\ \bibnamefont {Krauter}}, \bibinfo {author} {\bibfnamefont {K.}~\bibnamefont {Kue}}, \bibinfo {author} {\bibfnamefont {A.}~\bibnamefont {Kunitsa}}, \bibinfo {author}
  {\bibfnamefont {T.}~\bibnamefont {Kus}}, \bibinfo {author} {\bibfnamefont {I.}~\bibnamefont {Ladj{\'a}nszki}}, \bibinfo {author} {\bibfnamefont {A.}~\bibnamefont {Landau}}, \bibinfo {author} {\bibfnamefont {K.~V.}\ \bibnamefont {Lawler}}, \bibinfo {author} {\bibfnamefont {D.}~\bibnamefont {Lefrancois}}, \bibinfo {author} {\bibfnamefont {S.}~\bibnamefont {Lehtola}}, \bibinfo {author} {\bibfnamefont {R.~R.}\ \bibnamefont {Li}}, \bibinfo {author} {\bibfnamefont {Y.-P.}\ \bibnamefont {Li}}, \bibinfo {author} {\bibfnamefont {J.}~\bibnamefont {Liang}}, \bibinfo {author} {\bibfnamefont {M.}~\bibnamefont {Liebenthal}}, \bibinfo {author} {\bibfnamefont {H.-H.}\ \bibnamefont {Lin}}, \bibinfo {author} {\bibfnamefont {Y.-S.}\ \bibnamefont {Lin}}, \bibinfo {author} {\bibfnamefont {F.}~\bibnamefont {Liu}}, \bibinfo {author} {\bibfnamefont {K.-Y.}\ \bibnamefont {Liu}}, \bibinfo {author} {\bibfnamefont {M.}~\bibnamefont {Loipersberger}}, \bibinfo {author} {\bibfnamefont {A.}~\bibnamefont {Luenser}}, \bibinfo {author}
  {\bibfnamefont {A.}~\bibnamefont {Manjanath}}, \bibinfo {author} {\bibfnamefont {P.}~\bibnamefont {Manohar}}, \bibinfo {author} {\bibfnamefont {E.}~\bibnamefont {Mansoor}}, \bibinfo {author} {\bibfnamefont {S.~F.}\ \bibnamefont {Manzer}}, \bibinfo {author} {\bibfnamefont {S.-P.}\ \bibnamefont {Mao}}, \bibinfo {author} {\bibfnamefont {A.~V.}\ \bibnamefont {Marenich}}, \bibinfo {author} {\bibfnamefont {T.}~\bibnamefont {Markovich}}, \bibinfo {author} {\bibfnamefont {S.}~\bibnamefont {Mason}}, \bibinfo {author} {\bibfnamefont {S.~A.}\ \bibnamefont {Maurer}}, \bibinfo {author} {\bibfnamefont {P.~F.}\ \bibnamefont {McLaughlin}}, \bibinfo {author} {\bibfnamefont {M.~F. S.~J.}\ \bibnamefont {Menger}}, \bibinfo {author} {\bibfnamefont {J.-M.}\ \bibnamefont {Mewes}}, \bibinfo {author} {\bibfnamefont {S.~A.}\ \bibnamefont {Mewes}}, \bibinfo {author} {\bibfnamefont {P.}~\bibnamefont {Morgante}}, \bibinfo {author} {\bibfnamefont {J.~W.}\ \bibnamefont {Mullinax}}, \bibinfo {author} {\bibfnamefont {K.~J.}\ \bibnamefont
  {Oosterbaan}}, \bibinfo {author} {\bibfnamefont {G.}~\bibnamefont {Paran}}, \bibinfo {author} {\bibfnamefont {A.~C.}\ \bibnamefont {Paul}}, \bibinfo {author} {\bibfnamefont {S.~K.}\ \bibnamefont {Paul}}, \bibinfo {author} {\bibfnamefont {F.}~\bibnamefont {Pavo{\v{s}}evi{\'c}}}, \bibinfo {author} {\bibfnamefont {Z.}~\bibnamefont {Pei}}, \bibinfo {author} {\bibfnamefont {S.}~\bibnamefont {Prager}}, \bibinfo {author} {\bibfnamefont {E.~I.}\ \bibnamefont {Proynov}}, \bibinfo {author} {\bibfnamefont {{\'A}.}~\bibnamefont {R{\'a}k}}, \bibinfo {author} {\bibfnamefont {E.}~\bibnamefont {Ramos-Cordoba}}, \bibinfo {author} {\bibfnamefont {B.}~\bibnamefont {Rana}}, \bibinfo {author} {\bibfnamefont {A.~E.}\ \bibnamefont {Rask}}, \bibinfo {author} {\bibfnamefont {A.}~\bibnamefont {Rettig}}, \bibinfo {author} {\bibfnamefont {R.~M.}\ \bibnamefont {Richard}}, \bibinfo {author} {\bibfnamefont {F.}~\bibnamefont {Rob}}, \bibinfo {author} {\bibfnamefont {E.}~\bibnamefont {Rossomme}}, \bibinfo {author} {\bibfnamefont
  {T.}~\bibnamefont {Scheele}}, \bibinfo {author} {\bibfnamefont {M.}~\bibnamefont {Scheurer}}, \bibinfo {author} {\bibfnamefont {M.}~\bibnamefont {Schneider}}, \bibinfo {author} {\bibfnamefont {N.}~\bibnamefont {Sergueev}}, \bibinfo {author} {\bibfnamefont {S.~M.}\ \bibnamefont {Sharada}}, \bibinfo {author} {\bibfnamefont {W.}~\bibnamefont {Skomorowski}}, \bibinfo {author} {\bibfnamefont {D.~W.}\ \bibnamefont {Small}}, \bibinfo {author} {\bibfnamefont {C.~J.}\ \bibnamefont {Stein}}, \bibinfo {author} {\bibfnamefont {Y.-C.}\ \bibnamefont {Su}}, \bibinfo {author} {\bibfnamefont {E.~J.}\ \bibnamefont {Sundstrom}}, \bibinfo {author} {\bibfnamefont {Z.}~\bibnamefont {Tao}}, \bibinfo {author} {\bibfnamefont {J.}~\bibnamefont {Thirman}}, \bibinfo {author} {\bibfnamefont {G.~J.}\ \bibnamefont {Tornai}}, \bibinfo {author} {\bibfnamefont {T.}~\bibnamefont {Tsuchimochi}}, \bibinfo {author} {\bibfnamefont {N.~M.}\ \bibnamefont {Tubman}}, \bibinfo {author} {\bibfnamefont {S.~P.}\ \bibnamefont {Veccham}}, \bibinfo
  {author} {\bibfnamefont {O.}~\bibnamefont {Vydrov}}, \bibinfo {author} {\bibfnamefont {J.}~\bibnamefont {Wenzel}}, \bibinfo {author} {\bibfnamefont {J.}~\bibnamefont {Witte}}, \bibinfo {author} {\bibfnamefont {A.}~\bibnamefont {Yamada}}, \bibinfo {author} {\bibfnamefont {K.}~\bibnamefont {Yao}}, \bibinfo {author} {\bibfnamefont {S.}~\bibnamefont {Yeganeh}}, \bibinfo {author} {\bibfnamefont {S.~R.}\ \bibnamefont {Yost}}, \bibinfo {author} {\bibfnamefont {A.}~\bibnamefont {Zech}}, \bibinfo {author} {\bibfnamefont {I.~Y.}\ \bibnamefont {Zhang}}, \bibinfo {author} {\bibfnamefont {X.}~\bibnamefont {Zhang}}, \bibinfo {author} {\bibfnamefont {Y.}~\bibnamefont {Zhang}}, \bibinfo {author} {\bibfnamefont {D.}~\bibnamefont {Zuev}}, \bibinfo {author} {\bibfnamefont {A.}~\bibnamefont {Aspuru-Guzik}}, \bibinfo {author} {\bibfnamefont {A.~T.}\ \bibnamefont {Bell}}, \bibinfo {author} {\bibfnamefont {N.~A.}\ \bibnamefont {Besley}}, \bibinfo {author} {\bibfnamefont {K.~B.}\ \bibnamefont {Bravaya}}, \bibinfo {author}
  {\bibfnamefont {B.~R.}\ \bibnamefont {Brooks}}, \bibinfo {author} {\bibfnamefont {D.}~\bibnamefont {Casanova}}, \bibinfo {author} {\bibfnamefont {J.-D.}\ \bibnamefont {Chai}}, \bibinfo {author} {\bibfnamefont {S.}~\bibnamefont {Coriani}}, \bibinfo {author} {\bibfnamefont {C.~J.}\ \bibnamefont {Cramer}}, \bibinfo {author} {\bibfnamefont {G.}~\bibnamefont {Cserey}}, \bibinfo {author} {\bibfnamefont {A.~E.}\ \bibnamefont {DePrince}}, \bibinfo {author} {\bibfnamefont {R.~A.}\ \bibnamefont {DiStasio}}, \bibinfo {author} {\bibfnamefont {A.}~\bibnamefont {Dreuw}}, \bibinfo {author} {\bibfnamefont {B.~D.}\ \bibnamefont {Dunietz}}, \bibinfo {author} {\bibfnamefont {T.~R.}\ \bibnamefont {Furlani}}, \bibinfo {author} {\bibfnamefont {W.~A.}\ \bibnamefont {Goddard}}, \bibinfo {author} {\bibfnamefont {S.}~\bibnamefont {Hammes-Schiffer}}, \bibinfo {author} {\bibfnamefont {T.}~\bibnamefont {Head-Gordon}}, \bibinfo {author} {\bibfnamefont {W.~J.}\ \bibnamefont {Hehre}}, \bibinfo {author} {\bibfnamefont {C.-P.}\ \bibnamefont
  {Hsu}}, \bibinfo {author} {\bibfnamefont {T.-C.}\ \bibnamefont {Jagau}}, \bibinfo {author} {\bibfnamefont {Y.}~\bibnamefont {Jung}}, \bibinfo {author} {\bibfnamefont {A.}~\bibnamefont {Klamt}}, \bibinfo {author} {\bibfnamefont {J.}~\bibnamefont {Kong}}, \bibinfo {author} {\bibfnamefont {D.~S.}\ \bibnamefont {Lambrecht}}, \bibinfo {author} {\bibfnamefont {W.}~\bibnamefont {Liang}}, \bibinfo {author} {\bibfnamefont {N.~J.}\ \bibnamefont {Mayhall}}, \bibinfo {author} {\bibfnamefont {C.~W.}\ \bibnamefont {McCurdy}}, \bibinfo {author} {\bibfnamefont {J.~B.}\ \bibnamefont {Neaton}}, \bibinfo {author} {\bibfnamefont {C.}~\bibnamefont {Ochsenfeld}}, \bibinfo {author} {\bibfnamefont {J.~A.}\ \bibnamefont {Parkhill}}, \bibinfo {author} {\bibfnamefont {R.}~\bibnamefont {Peverati}}, \bibinfo {author} {\bibfnamefont {V.~A.}\ \bibnamefont {Rassolov}}, \bibinfo {author} {\bibfnamefont {Y.}~\bibnamefont {Shao}}, \bibinfo {author} {\bibfnamefont {L.~V.}\ \bibnamefont {Slipchenko}}, \bibinfo {author} {\bibfnamefont
  {T.}~\bibnamefont {Stauch}}, \bibinfo {author} {\bibfnamefont {R.~P.}\ \bibnamefont {Steele}}, \bibinfo {author} {\bibfnamefont {J.~E.}\ \bibnamefont {Subotnik}}, \bibinfo {author} {\bibfnamefont {A.~J.~W.}\ \bibnamefont {Thom}}, \bibinfo {author} {\bibfnamefont {A.}~\bibnamefont {Tkatchenko}}, \bibinfo {author} {\bibfnamefont {D.~G.}\ \bibnamefont {Truhlar}}, \bibinfo {author} {\bibfnamefont {T.}~\bibnamefont {Van~Voorhis}}, \bibinfo {author} {\bibfnamefont {T.~A.}\ \bibnamefont {Wesolowski}}, \bibinfo {author} {\bibfnamefont {K.~B.}\ \bibnamefont {Whaley}}, \bibinfo {author} {\bibfnamefont {H.~L.}\ \bibnamefont {Woodcock}}, \bibinfo {author} {\bibfnamefont {P.~M.}\ \bibnamefont {Zimmerman}}, \bibinfo {author} {\bibfnamefont {S.}~\bibnamefont {Faraji}}, \bibinfo {author} {\bibfnamefont {P.~M.~W.}\ \bibnamefont {Gill}}, \bibinfo {author} {\bibfnamefont {M.}~\bibnamefont {Head-Gordon}}, \bibinfo {author} {\bibfnamefont {J.~M.}\ \bibnamefont {Herbert}},\ and\ \bibinfo {author} {\bibfnamefont {A.~I.}\
  \bibnamefont {Krylov}},\ }\href {https://doi.org/10.1063/5.0055522} {\bibfield  {journal} {\bibinfo  {journal} {The Journal of Chemical Physics}\ }\textbf {\bibinfo {volume} {155}} (\bibinfo {year} {2021})}\BibitemShut {NoStop}%
\bibitem [{\citenamefont {Yanai}\ \emph {et~al.}(2004)\citenamefont {Yanai}, \citenamefont {Tew},\ and\ \citenamefont {Handy}}]{yanai2004new}%
  \BibitemOpen
  \bibfield  {author} {\bibinfo {author} {\bibfnamefont {T.}~\bibnamefont {Yanai}}, \bibinfo {author} {\bibfnamefont {D.~P.}\ \bibnamefont {Tew}},\ and\ \bibinfo {author} {\bibfnamefont {N.~C.}\ \bibnamefont {Handy}},\ }\href {https://doi.org/10.1016/j.cplett.2004.06.011} {\bibfield  {journal} {\bibinfo  {journal} {Chemical Physics Letters}\ }\textbf {\bibinfo {volume} {393}},\ \bibinfo {pages} {51} (\bibinfo {year} {2004})}\BibitemShut {NoStop}%
\bibitem [{\citenamefont {Dunning~Jr}(1989)}]{dunning1989gaussian}%
  \BibitemOpen
  \bibfield  {author} {\bibinfo {author} {\bibfnamefont {T.~H.}\ \bibnamefont {Dunning~Jr}},\ }\href {https://doi.org/10.1063/1.456153} {\bibfield  {journal} {\bibinfo  {journal} {The Journal of Chemical Physics}\ }\textbf {\bibinfo {volume} {90}},\ \bibinfo {pages} {1007} (\bibinfo {year} {1989})}\BibitemShut {NoStop}%
\bibitem [{\citenamefont {Hill}\ and\ \citenamefont {Peterson}(2017)}]{hill2017gaussian}%
  \BibitemOpen
  \bibfield  {author} {\bibinfo {author} {\bibfnamefont {J.~G.}\ \bibnamefont {Hill}}\ and\ \bibinfo {author} {\bibfnamefont {K.~A.}\ \bibnamefont {Peterson}},\ }\href {https://doi.org/10.1063/1.5010587} {\bibfield  {journal} {\bibinfo  {journal} {The Journal of Chemical Physics}\ }\textbf {\bibinfo {volume} {147}} (\bibinfo {year} {2017})}\BibitemShut {NoStop}%
\bibitem [{\citenamefont {Lim}\ \emph {et~al.}(2006)\citenamefont {Lim}, \citenamefont {Stoll},\ and\ \citenamefont {Schwerdtfeger}}]{lim2006relativistic}%
  \BibitemOpen
  \bibfield  {author} {\bibinfo {author} {\bibfnamefont {I.~S.}\ \bibnamefont {Lim}}, \bibinfo {author} {\bibfnamefont {H.}~\bibnamefont {Stoll}},\ and\ \bibinfo {author} {\bibfnamefont {P.}~\bibnamefont {Schwerdtfeger}},\ }\href {https://doi.org/10.1063/1.2148945} {\bibfield  {journal} {\bibinfo  {journal} {The Journal of Chemical Physics}\ }\textbf {\bibinfo {volume} {124}} (\bibinfo {year} {2006})}\BibitemShut {NoStop}%
\bibitem [{\citenamefont {Weigend}\ and\ \citenamefont {Ahlrichs}(2005)}]{weigend2005balanced}%
  \BibitemOpen
  \bibfield  {author} {\bibinfo {author} {\bibfnamefont {F.}~\bibnamefont {Weigend}}\ and\ \bibinfo {author} {\bibfnamefont {R.}~\bibnamefont {Ahlrichs}},\ }\href {https://doi.org/10.1039/B508541A} {\bibfield  {journal} {\bibinfo  {journal} {Physical Chemistry Chemical Physics}\ }\textbf {\bibinfo {volume} {7}},\ \bibinfo {pages} {3297} (\bibinfo {year} {2005})}\BibitemShut {NoStop}%
\bibitem [{\citenamefont {Dolg}\ \emph {et~al.}(1989)\citenamefont {Dolg}, \citenamefont {Stoll},\ and\ \citenamefont {Preuss}}]{dolg1989energy}%
  \BibitemOpen
  \bibfield  {author} {\bibinfo {author} {\bibfnamefont {M.}~\bibnamefont {Dolg}}, \bibinfo {author} {\bibfnamefont {H.}~\bibnamefont {Stoll}},\ and\ \bibinfo {author} {\bibfnamefont {H.}~\bibnamefont {Preuss}},\ }\href {https://doi.org/10.1063/1.456066} {\bibfield  {journal} {\bibinfo  {journal} {The Journal of Chemical Physics}\ }\textbf {\bibinfo {volume} {90}},\ \bibinfo {pages} {1730} (\bibinfo {year} {1989})}\BibitemShut {NoStop}%
\bibitem [{\citenamefont {Hanwell}\ \emph {et~al.}(2012)\citenamefont {Hanwell}, \citenamefont {Curtis}, \citenamefont {Lonie}, \citenamefont {Vandermeersch}, \citenamefont {Zurek},\ and\ \citenamefont {Hutchison}}]{hanwell2012avogadro}%
  \BibitemOpen
  \bibfield  {author} {\bibinfo {author} {\bibfnamefont {M.~D.}\ \bibnamefont {Hanwell}}, \bibinfo {author} {\bibfnamefont {D.~E.}\ \bibnamefont {Curtis}}, \bibinfo {author} {\bibfnamefont {D.~C.}\ \bibnamefont {Lonie}}, \bibinfo {author} {\bibfnamefont {T.}~\bibnamefont {Vandermeersch}}, \bibinfo {author} {\bibfnamefont {E.}~\bibnamefont {Zurek}},\ and\ \bibinfo {author} {\bibfnamefont {G.~R.}\ \bibnamefont {Hutchison}},\ }\href {https://jcheminf.biomedcentral.com/articles/10.1186/1758-2946-4-17} {\bibfield  {journal} {\bibinfo  {journal} {Journal of Cheminformatics}\ }\textbf {\bibinfo {volume} {4}},\ \bibinfo {pages} {17} (\bibinfo {year} {2012})}\BibitemShut {NoStop}%
\bibitem [{\citenamefont {Steimle}\ \emph {et~al.}(2019)\citenamefont {Steimle}, \citenamefont {Linton}, \citenamefont {Mengesha}, \citenamefont {Bai},\ and\ \citenamefont {Le}}]{Steimle2019YbOH}%
  \BibitemOpen
  \bibfield  {author} {\bibinfo {author} {\bibfnamefont {T.~C.}\ \bibnamefont {Steimle}}, \bibinfo {author} {\bibfnamefont {C.}~\bibnamefont {Linton}}, \bibinfo {author} {\bibfnamefont {E.~T.}\ \bibnamefont {Mengesha}}, \bibinfo {author} {\bibfnamefont {X.}~\bibnamefont {Bai}},\ and\ \bibinfo {author} {\bibfnamefont {A.~T.}\ \bibnamefont {Le}},\ }\href {https://doi.org/10.1103/PhysRevA.100.052509} {\bibfield  {journal} {\bibinfo  {journal} {Physical Review A}\ }\textbf {\bibinfo {volume} {100}},\ \bibinfo {pages} {052509} (\bibinfo {year} {2019})}\BibitemShut {NoStop}%
\bibitem [{\citenamefont {Lefebvre-Brion}\ and\ \citenamefont {Field}(2004)}]{Lefebvre2004Book}%
  \BibitemOpen
  \bibfield  {author} {\bibinfo {author} {\bibfnamefont {H.}~\bibnamefont {Lefebvre-Brion}}\ and\ \bibinfo {author} {\bibfnamefont {R.~W.}\ \bibnamefont {Field}},\ }\href {https://www.sciencedirect.com/book/9780124414556/the-spectra-and-dynamics-of-diatomic-molecules} {\emph {\bibinfo {title} {{The Spectra and Dynamics of Diatomic Molecules}}}},\ \bibinfo {edition} {2nd}\ ed.\ (\bibinfo  {publisher} {Academic Press},\ \bibinfo {year} {2004})\BibitemShut {NoStop}%
\bibitem [{\citenamefont {Brown}\ and\ \citenamefont {Carrington}(2003)}]{brown2003rotational}%
  \BibitemOpen
  \bibfield  {author} {\bibinfo {author} {\bibfnamefont {J.~M.}\ \bibnamefont {Brown}}\ and\ \bibinfo {author} {\bibfnamefont {A.}~\bibnamefont {Carrington}},\ }\href {https://doi.org/10.1017/CBO9780511814808} {\emph {\bibinfo {title} {Rotational Spectroscopy of Diatomic Molecules}}}\ (\bibinfo  {publisher} {Cambridge University Press},\ \bibinfo {year} {2003})\BibitemShut {NoStop}%
\bibitem [{\citenamefont {Osika}\ \emph {et~al.}(2024)\citenamefont {Osika}, \citenamefont {Sharashkin}, \citenamefont {Pitsevich},\ and\ \citenamefont {Shundalau}}]{Osika2024RaOH}%
  \BibitemOpen
  \bibfield  {author} {\bibinfo {author} {\bibfnamefont {Y.}~\bibnamefont {Osika}}, \bibinfo {author} {\bibfnamefont {S.}~\bibnamefont {Sharashkin}}, \bibinfo {author} {\bibfnamefont {G.}~\bibnamefont {Pitsevich}},\ and\ \bibinfo {author} {\bibfnamefont {M.}~\bibnamefont {Shundalau}},\ }\href {https://doi.org/10.1016/j.jqsrt.2023.108852} {\bibfield  {journal} {\bibinfo  {journal} {Journal of Quantitative Spectroscopy and Radiative Transfer}\ }\textbf {\bibinfo {volume} {314}},\ \bibinfo {pages} {108852} (\bibinfo {year} {2024})}\BibitemShut {NoStop}%
\bibitem [{\citenamefont {Zaitsevskii}\ \emph {et~al.}(2022)\citenamefont {Zaitsevskii}, \citenamefont {Skripnikov}, \citenamefont {Mosyagin}, \citenamefont {Isaev}, \citenamefont {Berger}, \citenamefont {Breier},\ and\ \citenamefont {Giesen}}]{Zaitsevskii2022RaF}%
  \BibitemOpen
  \bibfield  {author} {\bibinfo {author} {\bibfnamefont {A.}~\bibnamefont {Zaitsevskii}}, \bibinfo {author} {\bibfnamefont {L.~V.}\ \bibnamefont {Skripnikov}}, \bibinfo {author} {\bibfnamefont {N.~S.}\ \bibnamefont {Mosyagin}}, \bibinfo {author} {\bibfnamefont {T.}~\bibnamefont {Isaev}}, \bibinfo {author} {\bibfnamefont {R.}~\bibnamefont {Berger}}, \bibinfo {author} {\bibfnamefont {A.~A.}\ \bibnamefont {Breier}},\ and\ \bibinfo {author} {\bibfnamefont {T.~F.}\ \bibnamefont {Giesen}},\ }\href {https://doi.org/10.1063/5.0079618} {\bibfield  {journal} {\bibinfo  {journal} {The Journal of Chemical Physics}\ }\textbf {\bibinfo {volume} {156}},\ \bibinfo {pages} {044306} (\bibinfo {year} {2022})}\BibitemShut {NoStop}%
\bibitem [{\citenamefont {Stanton}\ \emph {et~al.}()\citenamefont {Stanton}, \citenamefont {Gauss}, \citenamefont {Cheng}, \citenamefont {Harding}, \citenamefont {Matthews},\ and\ \citenamefont {Szalay}}]{CFOURfull}%
  \BibitemOpen
  \bibfield  {author} {\bibinfo {author} {\bibfnamefont {J.~F.}\ \bibnamefont {Stanton}}, \bibinfo {author} {\bibfnamefont {J.}~\bibnamefont {Gauss}}, \bibinfo {author} {\bibfnamefont {L.}~\bibnamefont {Cheng}}, \bibinfo {author} {\bibfnamefont {M.~E.}\ \bibnamefont {Harding}}, \bibinfo {author} {\bibfnamefont {D.~A.}\ \bibnamefont {Matthews}},\ and\ \bibinfo {author} {\bibfnamefont {P.~G.}\ \bibnamefont {Szalay}},\ }\href@noop {} {\bibinfo {title} {{CFOUR, Coupled-Cluster techniques for Computational Chemistry, a quantum-chemical program package}}},\ \bibinfo {note} {{W}ith contributions from {A}. {A}sthana, {A}.{A}. {A}uer, {R}.{J}. {B}artlett, {U}. {B}enedikt, {C}. {B}erger, {D}.{E}. {B}ernholdt, {S}. {B}laschke, {Y}. {J}. {B}omble, {S}. {B}urger, {O}. {C}hristiansen, {D}. {D}atta, {F}. {E}ngel, {R}. {F}aber, {J}. {G}reiner, {M}. {H}eckert, {O}. {H}eun, {M}. Hilgenberg, {C}. {H}uber, {T}.-{C}. {J}agau, {D}. {J}onsson, {J}. {J}us{\'e}lius, {T}. Kirsch, {M}.-{P}. {K}itsaras, {K}. {K}lein, {G}.{M}. {K}opper,
  {W}.{J}. {L}auderdale, {F}. {L}ipparini, {J}. {L}iu, {T}. {M}etzroth, {L.} {M}onzel, {L}.{A}. {M}{\"u}ck, {D}.{P}. {O}'{N}eill, {T}. {N}ottoli, {J}. {O}swald, {D}.{R}. {P}rice, {E}. {P}rochnow, {C}. {P}uzzarini, {K}. {R}uud, {F}. {S}chiffmann, {W}. {S}chwalbach, {C}. {S}immons, {S}. {S}topkowicz, {A}. {T}ajti, {T.} Uhl\'{i}{\v{r}ov\'{a}, {J}. {V}{\'a}zquez, {F}. {W}ang, {J}.{D}. {W}atts, {P.} Yerg{\"u}n. {C}. {Z}hang, {X}. {Z}heng, and the integral packages {MOLECULE} ({J}. {A}lml{\"o}f and {P}.{R}. {T}aylor), {PROPS} ({P}.{R}. {T}aylor), {ABACUS} ({T}. {H}elgaker, {H}.{J}. {A}a. {J}ensen, {P}. {J}{\o}rgensen, and {J}. {O}lsen), and {ECP} routines by {A}. {V}. {M}itin and {C}. van {W}{\"u}llen. {F}or the current version, see http://www.cfour.de.}}\BibitemShut {Stop}%
\bibitem [{\citenamefont {Matthews}\ \emph {et~al.}(2020)\citenamefont {Matthews}, \citenamefont {Cheng}, \citenamefont {Harding}, \citenamefont {Lipparini}, \citenamefont {Stopkowicz}, \citenamefont {Jagau}, \citenamefont {Szalay}, \citenamefont {Gauss},\ and\ \citenamefont {Stanton}}]{Matthews20a}%
  \BibitemOpen
  \bibfield  {author} {\bibinfo {author} {\bibfnamefont {D.~A.}\ \bibnamefont {Matthews}}, \bibinfo {author} {\bibfnamefont {L.}~\bibnamefont {Cheng}}, \bibinfo {author} {\bibfnamefont {M.~E.}\ \bibnamefont {Harding}}, \bibinfo {author} {\bibfnamefont {F.}~\bibnamefont {Lipparini}}, \bibinfo {author} {\bibfnamefont {S.}~\bibnamefont {Stopkowicz}}, \bibinfo {author} {\bibfnamefont {T.-C.}\ \bibnamefont {Jagau}}, \bibinfo {author} {\bibfnamefont {P.~G.}\ \bibnamefont {Szalay}}, \bibinfo {author} {\bibfnamefont {J.}~\bibnamefont {Gauss}},\ and\ \bibinfo {author} {\bibfnamefont {J.~F.}\ \bibnamefont {Stanton}},\ }\href {https://doi.org/10.1063/5.0004837} {\bibfield  {journal} {\bibinfo  {journal} {J. Chem. Phys.}\ }\textbf {\bibinfo {volume} {152}},\ \bibinfo {pages} {214108} (\bibinfo {year} {2020})}\BibitemShut {NoStop}%
\bibitem [{\citenamefont {Stanton}\ and\ \citenamefont {Bartlett}(1993)}]{Stanton93a}%
  \BibitemOpen
  \bibfield  {author} {\bibinfo {author} {\bibfnamefont {J.~F.}\ \bibnamefont {Stanton}}\ and\ \bibinfo {author} {\bibfnamefont {R.~J.}\ \bibnamefont {Bartlett}},\ }\href {https://pubs.aip.org/aip/jcp/article/98/9/7029/857758/The-equation-of-motion-coupled-cluster-method-A} {\bibfield  {journal} {\bibinfo  {journal} {J. Chem. Phys.}\ }\textbf {\bibinfo {volume} {98}},\ \bibinfo {pages} {7029} (\bibinfo {year} {1993})}\BibitemShut {NoStop}%
\bibitem [{\citenamefont {Nooijen}\ and\ \citenamefont {Bartlett}(1995)}]{Nooijen95}%
  \BibitemOpen
  \bibfield  {author} {\bibinfo {author} {\bibfnamefont {M.}~\bibnamefont {Nooijen}}\ and\ \bibinfo {author} {\bibfnamefont {R.~J.}\ \bibnamefont {Bartlett}},\ }\href {https://pubs.aip.org/aip/jcp/article/102/9/3629/481399/Equation-of-motion-coupled-cluster-method-for} {\bibfield  {journal} {\bibinfo  {journal} {J. Chem. Phys.}\ }\textbf {\bibinfo {volume} {102}},\ \bibinfo {pages} {3629} (\bibinfo {year} {1995})}\BibitemShut {NoStop}%
\bibitem [{\citenamefont {Dyall}(1997)}]{Dyall97}%
  \BibitemOpen
  \bibfield  {author} {\bibinfo {author} {\bibfnamefont {K.~G.}\ \bibnamefont {Dyall}},\ }\href {https://pubs.aip.org/aip/jcp/article/106/23/9618/293591/Interfacing-relativistic-and-nonrelativistic} {\bibfield  {journal} {\bibinfo  {journal} {J. Chem. Phys.}\ }\textbf {\bibinfo {volume} {106}},\ \bibinfo {pages} {9618} (\bibinfo {year} {1997})}\BibitemShut {NoStop}%
\bibitem [{\citenamefont {Kutzelnigg}\ and\ \citenamefont {Liu}(2005)}]{Kutzelnigg05}%
  \BibitemOpen
  \bibfield  {author} {\bibinfo {author} {\bibfnamefont {W.}~\bibnamefont {Kutzelnigg}}\ and\ \bibinfo {author} {\bibfnamefont {W.}~\bibnamefont {Liu}},\ }\href {https://pubs.aip.org/aip/jcp/article/123/24/241102/349274/Quasirelativistic-theory-equivalent-to-fully} {\bibfield  {journal} {\bibinfo  {journal} {J. Chem. Phys.}\ }\textbf {\bibinfo {volume} {123}},\ \bibinfo {pages} {241102} (\bibinfo {year} {2005})}\BibitemShut {NoStop}%
\bibitem [{\citenamefont {Ilia{\v{s}}}\ and\ \citenamefont {Saue}(2007)}]{Ilias07}%
  \BibitemOpen
  \bibfield  {author} {\bibinfo {author} {\bibfnamefont {M.}~\bibnamefont {Ilia{\v{s}}}}\ and\ \bibinfo {author} {\bibfnamefont {T.}~\bibnamefont {Saue}},\ }\href {https://pubs.aip.org/aip/jcp/article/126/6/064102/953046} {\bibfield  {journal} {\bibinfo  {journal} {J. Chem. Phys.}\ }\textbf {\bibinfo {volume} {126}},\ \bibinfo {pages} {064102} (\bibinfo {year} {2007})}\BibitemShut {NoStop}%
\bibitem [{\citenamefont {Liu}\ and\ \citenamefont {Peng}(2009)}]{Liu09}%
  \BibitemOpen
  \bibfield  {author} {\bibinfo {author} {\bibfnamefont {W.}~\bibnamefont {Liu}}\ and\ \bibinfo {author} {\bibfnamefont {D.}~\bibnamefont {Peng}},\ }\href {https://pubs.aip.org/aip/jcp/article/131/3/031104/71694/Exact-two-component-Hamiltonians-revisited} {\bibfield  {journal} {\bibinfo  {journal} {J. Chem. Phys.}\ }\textbf {\bibinfo {volume} {131}},\ \bibinfo {pages} {031104} (\bibinfo {year} {2009})}\BibitemShut {NoStop}%
\bibitem [{\citenamefont {He{\ss}}\ \emph {et~al.}(1996)\citenamefont {He{\ss}}, \citenamefont {Marian}, \citenamefont {Wahlgren},\ and\ \citenamefont {Gropen}}]{Hess96a}%
  \BibitemOpen
  \bibfield  {author} {\bibinfo {author} {\bibfnamefont {B.~A.}\ \bibnamefont {He{\ss}}}, \bibinfo {author} {\bibfnamefont {C.~M.}\ \bibnamefont {Marian}}, \bibinfo {author} {\bibfnamefont {U.}~\bibnamefont {Wahlgren}},\ and\ \bibinfo {author} {\bibfnamefont {O.}~\bibnamefont {Gropen}},\ }\href {https://www.sciencedirect.com/science/article/pii/0009261496001194} {\bibfield  {journal} {\bibinfo  {journal} {Chem. Phys. Lett.}\ }\textbf {\bibinfo {volume} {251}},\ \bibinfo {pages} {365} (\bibinfo {year} {1996})}\BibitemShut {NoStop}%
\bibitem [{\citenamefont {Liu}\ and\ \citenamefont {Cheng}(2018)}]{Liu18}%
  \BibitemOpen
  \bibfield  {author} {\bibinfo {author} {\bibfnamefont {J.}~\bibnamefont {Liu}}\ and\ \bibinfo {author} {\bibfnamefont {L.}~\bibnamefont {Cheng}},\ }\href {https://pubs.aip.org/aip/jcp/article/148/14/144108/196372} {\bibfield  {journal} {\bibinfo  {journal} {J. Chem. Phys.}\ }\textbf {\bibinfo {volume} {148}},\ \bibinfo {pages} {144108} (\bibinfo {year} {2018})}\BibitemShut {NoStop}%
\bibitem [{\citenamefont {Dyall}(2001)}]{Dyall01}%
  \BibitemOpen
  \bibfield  {author} {\bibinfo {author} {\bibfnamefont {K.~G.}\ \bibnamefont {Dyall}},\ }\href {https://pubs.aip.org/aip/jcp/article/115/20/9136/442213/Interfacing-relativistic-and-nonrelativistic} {\bibfield  {journal} {\bibinfo  {journal} {J. Chem. Phys.}\ }\textbf {\bibinfo {volume} {115}},\ \bibinfo {pages} {9136} (\bibinfo {year} {2001})}\BibitemShut {NoStop}%
\bibitem [{\citenamefont {Cheng}\ and\ \citenamefont {Gauss}(2011)}]{Cheng11b}%
  \BibitemOpen
  \bibfield  {author} {\bibinfo {author} {\bibfnamefont {L.}~\bibnamefont {Cheng}}\ and\ \bibinfo {author} {\bibfnamefont {J.}~\bibnamefont {Gauss}},\ }\href {https://pubs.aip.org/aip/jcp/article/135/8/084114/72411/Analytic-energy-gradients-for-the-spin-free-exact} {\bibfield  {journal} {\bibinfo  {journal} {J. Chem. Phys.}\ }\textbf {\bibinfo {volume} {135}},\ \bibinfo {pages} {084114} (\bibinfo {year} {2011})}\BibitemShut {NoStop}%
\bibitem [{\citenamefont {Zhang}\ \emph {et~al.}(2023{\natexlab{c}})\citenamefont {Zhang}, \citenamefont {Zheng}, \citenamefont {Liu}, \citenamefont {Asthana},\ and\ \citenamefont {Cheng}}]{Zhang23a}%
  \BibitemOpen
  \bibfield  {author} {\bibinfo {author} {\bibfnamefont {C.}~\bibnamefont {Zhang}}, \bibinfo {author} {\bibfnamefont {X.}~\bibnamefont {Zheng}}, \bibinfo {author} {\bibfnamefont {J.}~\bibnamefont {Liu}}, \bibinfo {author} {\bibfnamefont {A.}~\bibnamefont {Asthana}},\ and\ \bibinfo {author} {\bibfnamefont {L.}~\bibnamefont {Cheng}},\ }\href {https://doi.org/10.1063/5.0175041} {\bibfield  {journal} {\bibinfo  {journal} {J. Chem. Phys.}\ }\textbf {\bibinfo {volume} {159}},\ \bibinfo {pages} {244113} (\bibinfo {year} {2023}{\natexlab{c}})}\BibitemShut {NoStop}%
\bibitem [{\citenamefont {Skoff}\ \emph {et~al.}(2011)\citenamefont {Skoff}, \citenamefont {Hendricks}, \citenamefont {Sinclair}, \citenamefont {Hudson}, \citenamefont {Segal}, \citenamefont {Sauer}, \citenamefont {Hinds},\ and\ \citenamefont {Tarbutt}}]{skoff2011diffusion}%
  \BibitemOpen
  \bibfield  {author} {\bibinfo {author} {\bibfnamefont {S.}~\bibnamefont {Skoff}}, \bibinfo {author} {\bibfnamefont {R.}~\bibnamefont {Hendricks}}, \bibinfo {author} {\bibfnamefont {C.}~\bibnamefont {Sinclair}}, \bibinfo {author} {\bibfnamefont {J.}~\bibnamefont {Hudson}}, \bibinfo {author} {\bibfnamefont {D.}~\bibnamefont {Segal}}, \bibinfo {author} {\bibfnamefont {B.}~\bibnamefont {Sauer}}, \bibinfo {author} {\bibfnamefont {E.}~\bibnamefont {Hinds}},\ and\ \bibinfo {author} {\bibfnamefont {M.}~\bibnamefont {Tarbutt}},\ }\href {https://doi.org/10.1103/PhysRevA.83.023418} {\bibfield  {journal} {\bibinfo  {journal} {Physical Review A}\ }\textbf {\bibinfo {volume} {83}},\ \bibinfo {pages} {023418} (\bibinfo {year} {2011})}\BibitemShut {NoStop}%
\bibitem [{\citenamefont {Mengesha}\ \emph {et~al.}(2020)\citenamefont {Mengesha}, \citenamefont {Le}, \citenamefont {Steimle}, \citenamefont {Cheng}, \citenamefont {Zhang}, \citenamefont {Augenbraun}, \citenamefont {Lasner},\ and\ \citenamefont {Doyle}}]{Mengesha2020YbOHBranching}%
  \BibitemOpen
  \bibfield  {author} {\bibinfo {author} {\bibfnamefont {E.~T.}\ \bibnamefont {Mengesha}}, \bibinfo {author} {\bibfnamefont {A.~T.}\ \bibnamefont {Le}}, \bibinfo {author} {\bibfnamefont {T.~C.}\ \bibnamefont {Steimle}}, \bibinfo {author} {\bibfnamefont {L.}~\bibnamefont {Cheng}}, \bibinfo {author} {\bibfnamefont {C.}~\bibnamefont {Zhang}}, \bibinfo {author} {\bibfnamefont {B.~L.}\ \bibnamefont {Augenbraun}}, \bibinfo {author} {\bibfnamefont {Z.}~\bibnamefont {Lasner}},\ and\ \bibinfo {author} {\bibfnamefont {J.}~\bibnamefont {Doyle}},\ }\href {https://doi.org/10.1021/acs.jpca.0c00850} {\bibfield  {journal} {\bibinfo  {journal} {The Journal of Physical Chemistry A}\ }\textbf {\bibinfo {volume} {124}},\ \bibinfo {pages} {3135} (\bibinfo {year} {2020})}\BibitemShut {NoStop}%
\bibitem [{\citenamefont {Kozyryev}\ \emph {et~al.}(2015)\citenamefont {Kozyryev}, \citenamefont {Baum}, \citenamefont {Matsuda}, \citenamefont {Olson}, \citenamefont {Hemmerling},\ and\ \citenamefont {Doyle}}]{Kozyryev2015Quench}%
  \BibitemOpen
  \bibfield  {author} {\bibinfo {author} {\bibfnamefont {I.}~\bibnamefont {Kozyryev}}, \bibinfo {author} {\bibfnamefont {L.}~\bibnamefont {Baum}}, \bibinfo {author} {\bibfnamefont {K.}~\bibnamefont {Matsuda}}, \bibinfo {author} {\bibfnamefont {P.}~\bibnamefont {Olson}}, \bibinfo {author} {\bibfnamefont {B.}~\bibnamefont {Hemmerling}},\ and\ \bibinfo {author} {\bibfnamefont {J.~M.}\ \bibnamefont {Doyle}},\ }\href {https://doi.org/10.1088/1367-2630/17/4/045003} {\bibfield  {journal} {\bibinfo  {journal} {New Journal of Physics}\ }\textbf {\bibinfo {volume} {17}},\ \bibinfo {pages} {045003} (\bibinfo {year} {2015})}\BibitemShut {NoStop}%
\end{thebibliography}%

\newpage

\onecolumngrid
\setcounter{section}{0}
\setcounter{equation}{0}
\setcounter{figure}{0}
\setcounter{table}{0}
\setcounter{page}{1}
\makeatletter
\renewcommand{\theequation}{S\arabic{equation}}
\renewcommand{\thefigure}{S\arabic{figure}}
\renewcommand{\thetable}{S\arabic{table}}
\renewcommand{\thesection}{S\arabic{section}}

\newpage

\FloatBarrier

\section{Materials and methods}

\subsection{Target preparation}

\subsubsection{Radium target}

\begin{figure}
    \centering
    \includegraphics[width=\linewidth]{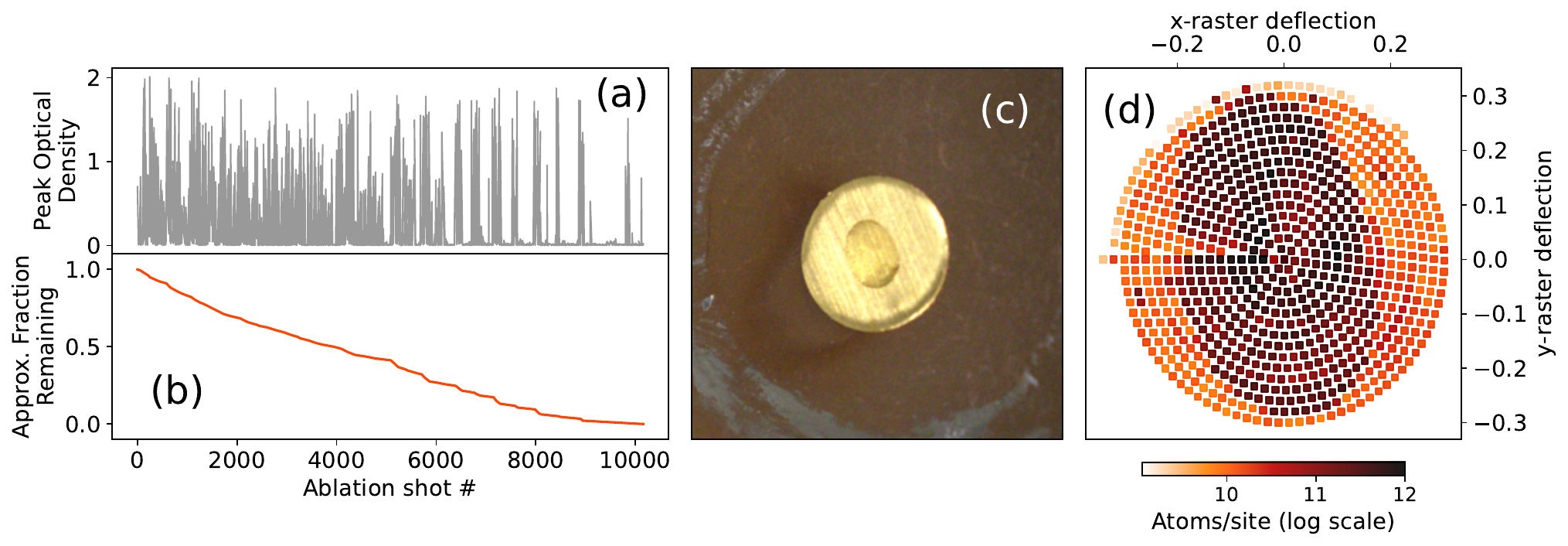}
    \caption{Drop casted target ablation record from 50 $\mu$Ci of radium nitrate salt deposited onto gold foil. Dissolving xylitol into the solution yields a  viscous syrup that evaporates into a translucent gel depicted in panel (c). Atomic ablation yields off of the target can be inferred by absorption on the $^1P_1-^1S_0$ Ra transition, which indicates good shot-to-shot consistency over the spatial extent of the target, as shown in panels (b) and (d). Panel (a) depicts optical density per shot (OD = $-\text{ln}(T)$ for transmission $T$) with respect to ablation shot number, and panel (b) depicts the estimated mass fraction of the target that has been used based on rate of depletion per shot. Approximately 10$^4$ shots are taken over $\sim10^3$ unique ablation raster sites per target. Atomic absorption data here is taken in the presence of strong optical pumping from the $^3P_1-^1S_0$ light which partially depletes the $^1S_0$ ground state population, thus reducing the total atoms detected. Atom numbers in this figure should be interpreted for relative shot-to-shot comparison, but are not indicative of total yields. See section \ref{sec:numbers} for details on atom number number estimation in the absence of optical pumping. }\label{fig:Ra_target}
\end{figure}

Radium-226 ablation targets are made by evaporation of an aqueous radium salt solution onto an inert substrate, a process referred to as ``drop casting." Both radium chloride (RaCl$_2$) and radium nitrate (Ra(NO$_3$)$_2$) were employed in this work; for initial testing and RaOH survey spectroscopy, 10 $\mu$Ci radium chloride solutions supplied by Eckert \& Ziegler were used. Subsequent targets were made from radium nitrate, which was acquired as 1 mCi of dried salt from the U.S. Department of Energy National Isotope Development Center and Oak Ridge National Laboratory. Upon receipt, we dissolved and aliquoted this salt into smaller activities for target preparation and medium-term storage. 

Drop casting solutions contain 10--50 $\mu$Ci of Ra-226 in 100 $\mu$L of 0.1 M acid (HCl for radium chloride, HNO$_3$ for radium nitrate). The weak acid helps prevent radium ions from sticking to the glass container while in solution. Shortly before drop casting, we dissolve a small amount ($\sim$ 0.1 mg, 1-10 mM final concentration) of the sugar alcohol xylitol into the solution, which facilitates a uniform deposition of radium during evaporation and allows for more consistent and repeatable targets by disrupting the formation of salt crystals and ``coffee rings''~\cite{mampallil2018review}.  The consistency and atomic distribution for a typical radium target with xylitol is depicted in Fig.~\ref{fig:Ra_target}.

From the radium solution, 2 $\mu$L-sized drops are pipetted onto a heated target plate and allowed to fully evaporate until the entire sample has been deposited. The target plate consists of a copper C10100 substrate with a 0.1-mm-thick, 5/16-inch-diameter piece of pure gold foil epoxied on the center using Stycast 2850FT Black with catalyst 24LV.  The radium solution is drop casted onto this gold foil.  Initial attempts to drop cast the solution onto bare or gold-electroplated copper resulted in unexpected and unreliable chemical reactions, which led to inconsistent targets.  We attribute this reactivity primarily to radiolysis since these effects were not present in barium salt solutions used for testing, even at higher acid concentrations and temperatures.

With 50 $\mu$Ci of radium-226 (221 nmol $\approx$ 10$^{17}$ atoms), a single target will produce a usable number of atoms for 10$^3$--10$^4$ ablation shots (Fig.~\ref{fig:Ra_target}). Target production parameters described above are chosen to maximize the number of shots that produce large molecule signals, and can be adjusted to get an even larger number of atomic ablation shots at the cost of smaller molecule fluorescence. Because the evaporated material forms a thin layer, all of the desired atoms at a particular ablation laser focus spot will be depleted after around 10 to 100 shots. Motorized mirror mounts are used to automatically raster the ablation laser over the bounds of the target, shown in Fig.~\ref{fig:Ra_target}d. Once the target has been completely depleted, the \red{vacuum-sealed cell} is disposed of and a new one is prepared for installation.  We have tested our protocol with barium solutions containing milligrams of barium and are still able to consistently create atoms and molecules; we therefore anticipate that this protocol could be used for larger quantities and activities.

\subsubsection{Reagent co-targets}

Co-ablation enables flexibility in choosing the ligand and thus the resulting molecule without opening or changing the experimental apparatus or interfering with the production of the more complex radium ablation target. The process is largely chemically agnostic, showing comparable efficiency across different metal-reagent combinations.
The reagent target, or ``co-target,'' is composed of a powder mixture consisting of a reagent (e.g., Al(OH)\textsubscript{3}, CuF\textsubscript{2}, or Al(OD)\textsubscript{3}) and 5\% by mass of polyethylene glycol (PEG) to act as a binder. These powders are generally commercially available, except for Al(OD)\textsubscript{3}, which we synthesize in-house as described below.
The powder mixture is hydraulically pressed in an 8-mm-diameter die under a pressure of 1~{GPa} for approximately 30 minutes, forming a pellet with a typical mass of $\sim$~0.2~g. The approximate densities of the respective pressed powder targets are 2.9~g/cm\textsuperscript{3} for CuF\textsubscript{2}, 2.0~g/cm\textsuperscript{3} for Al(OH)\textsubscript{3}, and 1.9~g/cm\textsuperscript{3} for Al(OD)\textsubscript{3}. After pressing, the circular pellets are cut to the desired size with a razor blade and epoxied to the copper target plate containing the deposited radium using Stycast 2850FT Black with catalyst 24LV. We tried several different targets containing various forms of hydroxides and fluorides and found that the performance between them was similar, but that these reagents were most cost-effective and safest to handle. 

\begin{figure}
    \centering
    \includegraphics[width=0.8\linewidth]{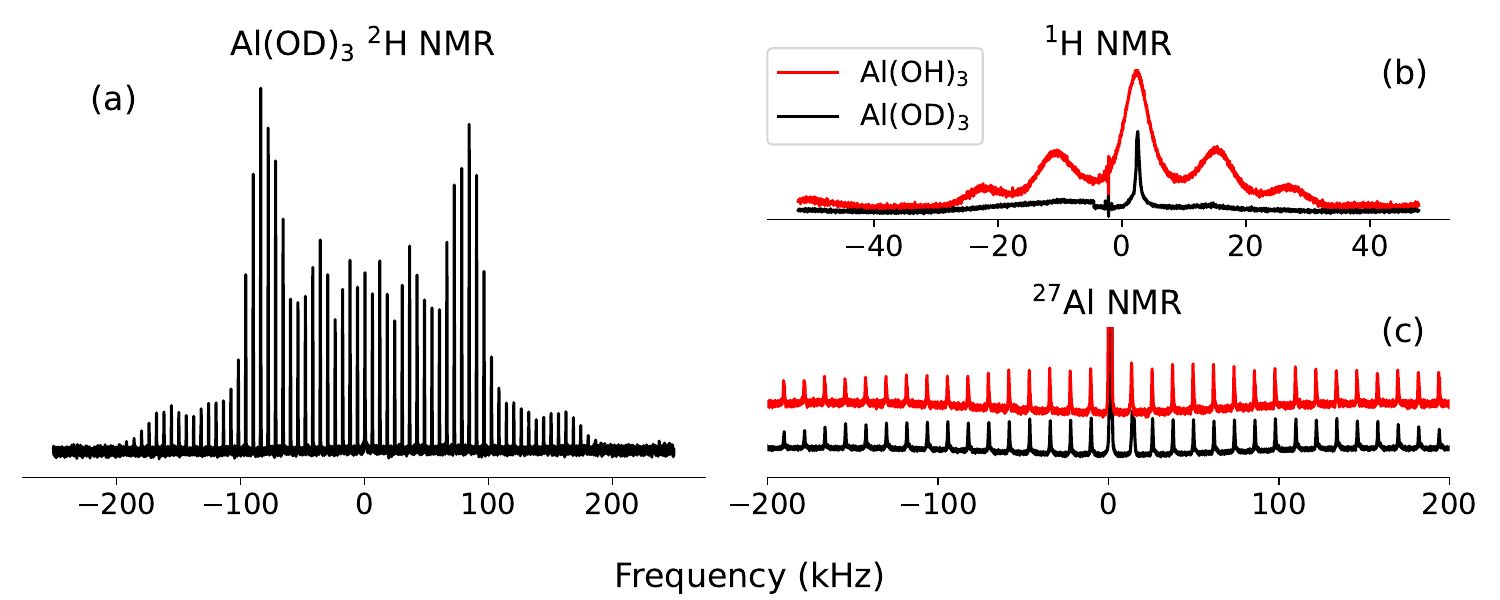}
    \caption{(a) $^2$H magic-angle spinning NMR spectrum confirming synthesis of aluminum deuteroxide powder. The two peaks near $\pm 84$ kHz form a Pake doublet corresponding to the $I=-1\rightarrow0$ and $I=0\rightarrow1$ transitions in deuterium, which are split into narrow sidebands separated by the magic-angle spinning frequency of 6 kHz. (b) $^1$H NMR of synthesized Al(OD)$_3$ against commercial Al(OH)$_3$. The integrated signals indicate that the synthesized sample is roughly 5\% H instead of D, likely owing to uncontrolled ambient humidity during synthesis. (c) $^{27}$Al magic-angle spinning NMR of both samples (offset for visibility), demonstrating structural agreement between the two. This data was taken at the Caltech Solid State NMR Facility.}
    \label{fig:Full_NMR_plot}
\end{figure}

The deuterated analogue of our hydroxide co-target, aluminum deuteroxide (Al(OD)$_3$) is not commercially available. We synthesized aluminum deuteroxide via the decomposition of aluminum nitride in heavy water, $\text{AlN} \ + \ 3\text{D}_2\text{O} \ \rightarrow \ \text{Al(OD)}_3 \ + \ \text{ND}_3$.  Following the corresponding procedure for aluminum hydroxide in~\cite{kocjan2008influence}, we added 3.675 g of AlN to 100 ml of D$_2$O and stirred it at 40 \textdegree C for 48 hours. After the solution cooled, we poured off the majority of the ND$_3$ supernatant and dried the precipitate in a vacuum oven for 24 hours at 50 \textdegree C. The final mass was 6.13 g; assuming a full conversion of AlN $\rightarrow $ Al(OD)$_3$, this corresponds to a chemical yield of 85\%. To confirm the presence of deuterium, we performed $^2$H NMR on our sample, which shows the expected structure for a deuterium-containing molecule (Fig.~\ref{fig:Full_NMR_plot}a). We also compared our sample against a commercially available Al(OH)$_3$ powder using $^1$H and $^{27}$Al NMR to verify that our product was largely Al(OD)$_3$, shown in figs.~\ref{fig:Full_NMR_plot}b,c.

\subsection{Ablation Production and Cooling}\label{sec:ablation_cooling}

Molecules are produced in a cryogenic buffer gas~\cite{hutzler2012buffer}. Helium is initially loaded into a known room-temperature calibration volume until a desired pressure is reached, after which it is added to the cell to achieve an approximate density of $\sim(10^{15} -10^{16})$ atoms/cm$^3$. During operation, the instantaneous helium pressure inside the cell can deviate from simple ideal gas law scalings due to the effects of helium desorption from the pressed powder reagent targets. The operational buffer gas quantity is determined empirically by optimizing on a molecular fluorescence signal for a known species, such as YbOH. While the cryosystem can achieve a base temperature of 4.2~K, heaters are used to hold the cell at 7~K during routine operation. At this temperature, we observed more intense and longer-lived molecule fluorescence signals, which we attribute to the desorption of helium off of the co-targets following ablation. \red{To reduce background light scatter from lasers and ablation processes, the fluorescence detection region of the cell is painted via airbrush with a low-outgassing reflection control coating (MH2200, currently manufactured by HII Advanced Materials and Coating Laboratory).}

The ablation beams originate from separate pulsed Nd:YAG lasers, which emit at 532 nm with a pulse duration of $\approx$~10~ns and beam diameter of $\approx$~1~cm.  A 200 mm focal length lens external to the cryostat focuses the beams onto the target plate. Pulse energies of the two lasers are optimized for molecule production and may vary depending on the target, but are generally 5 mJ for the radium drop cast target and 20 mJ for the pressed powder reagent. The co-target is ablated 10 ms before the radium, filling the cell volume with a reagent cloud and desorbing helium to aid with thermalization and diffusion. Many delay timings were tested, and we observed similar molecule production efficiency when delaying the Ra ablation anywhere from 5 to 50 ms after the co-target.

As the targets are ablated away, the lasers must be moved to hit new regions. For the powder reagent this only needs to happen after a few hundred ablation shots, and is done by periodically steering a mirror mount by hand. Spots on the drop cast radium target need to be refreshed much more frequently, from every ten to hundred ablation shots. In order to maximize usage of the drop cast radium material, motorized mirror mounts are used to raster the ablation beam over the spatial extent of the drop cast target, shown in Fig.~\ref{fig:Ra_target}c. Because of the trace quantities involved, we limit the experimental duty cycle to roughly 1 Hz to enable human intervention and real-time adjustment of experimental parameters, although much higher data taking rates are possible. 

\subsection{Chemical Production and Enhancement}

\begin{figure}
  \begin{minipage}[c]{0.57\textwidth}
    \includegraphics[width=0.8\linewidth]{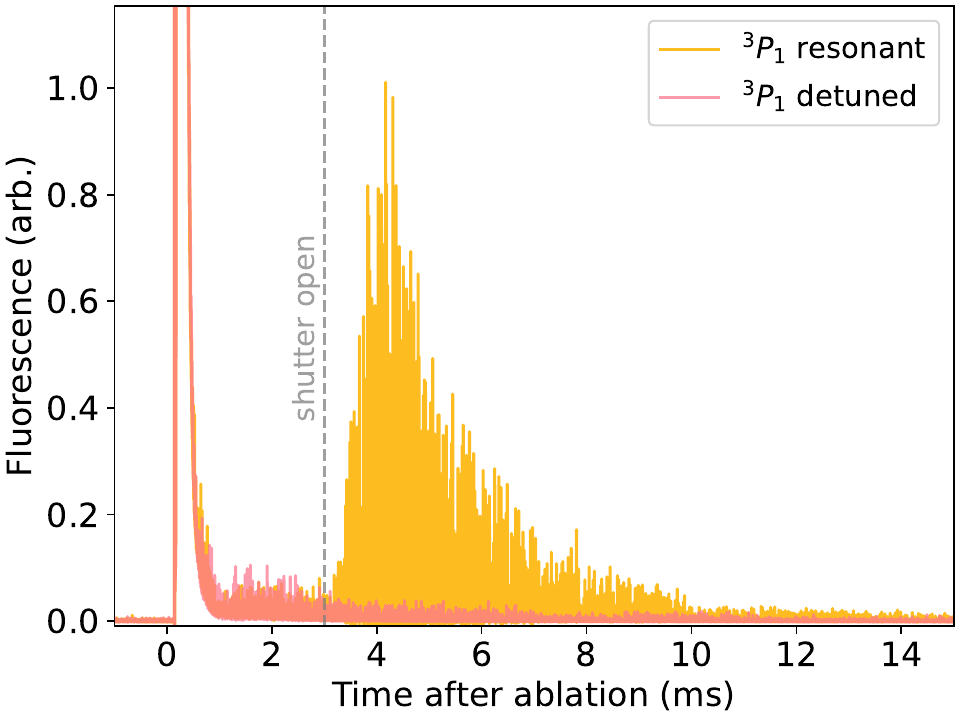}
  \end{minipage}\hfill
  \begin{minipage}[c]{0.4\textwidth}
    \caption{Demonstration of chemical enhancement on a representative RaF fluorescence signal. Measurements were taken similarly to \ref{fig:RaOD}d on the RaF $C^2\Sigma_{1/2}^+ (v' = 0) - X^2\Sigma_{1/2}^+ (v'' = 1)$ transition using pulsed dye excitation while ablating a pressed powder CuF$_2$ co-target. A laser shutter blocking the enhancement light is opened 3 ms after ablation and results in a 20-fold increase in molecule production when resonant with the $^3P_1\leftarrow ^1S_0$ transition in $^{226}$Ra, an effect not observed in experiments producing alkaline earth(-like) monofluorides using SF$_6$ and NF$_3$ reagents~\cite{liu2022chemistry}.}
    \label{fig:RaF_enh}
  \end{minipage}
\end{figure}

As discussed in the main text, we enhance our molecule yields via laser excitation of radium to the metastable $^3P_1$ state at 714~nm. Experiments that produce alkaline earth monofluorides generally use fluoride-containing gases, such as SF$_6$,  CF$_4$, or NF$_3$, as the reagent. For this reaction, chemical enhancement has been shown to often reduce the total molecule signal, as the ground-state reaction is exothermic and further excitation favors the formation of difluorides~\cite{liu2022chemistry}. With our powder target reagent, however, we observe that enhancement does in fact benefit molecule production to a similar extent as for hydroxides, shown in Fig.~\ref{fig:RaF_enh}.  We have also observed chemical enhancement of BaF in the same apparatus, and expect that this approach could be used for other alkaline earth fluorides as well.

In order to understand the energetics of the chemical reaction between excited Ra atoms and molecules produced by ablation, we performed electron density-based quantum chemical calculations of the reaction profile, similar to those conducted for YbOH~\cite{jadbabaie2020enhanced}.  As an example reaction, we considered the formation of RaOH via Ra insertion into H$_2$O, which is a prototypical although not exclusive byproduct of ablation.
To obtain the equilibrium energies of reaction minima and transition states on the singlet Ra($^1$S)+H$_2$O and triplet Ra($^3$P)+H$_2$O potential energy surfaces, critical point searches were performed on the reaction geometry using numerical Hessian~\cite{sharada2014finite} and eigenvector-following methods~\cite{baker1986algorithm}. These computations were performed in Q-Chem 5.3~\cite{epifanovsky2021software} using the spin-unrestricted UCAM-B3LYP density functional \cite{yanai2004new} and the diffuse, correlation-consistent aug-cc-pVTZ basis set for O and H atoms~\cite{dunning1989gaussian}, and the correlation-consistent cc-pVQZ-PP basis set~\cite{hill2017gaussian} with the Stuttgart ECP78MDF effective core potential~\cite{lim2006relativistic} for the Ra atom.  The calculation approach was validated by reproducing the reaction Yb+H$_2$O~\cite{jadbabaie2020enhanced} using the UCAM-B3LYP functional, the aug-cc-pVTZ basis set for O and H atoms, and the segmented def2-TZVP basis set~\cite{weigend2005balanced} with the ECP28MWB pseudopotential~\cite{dolg1989energy} for the Yb atom.

\begin{figure}
    \centering
    \includegraphics[width=0.75\linewidth]{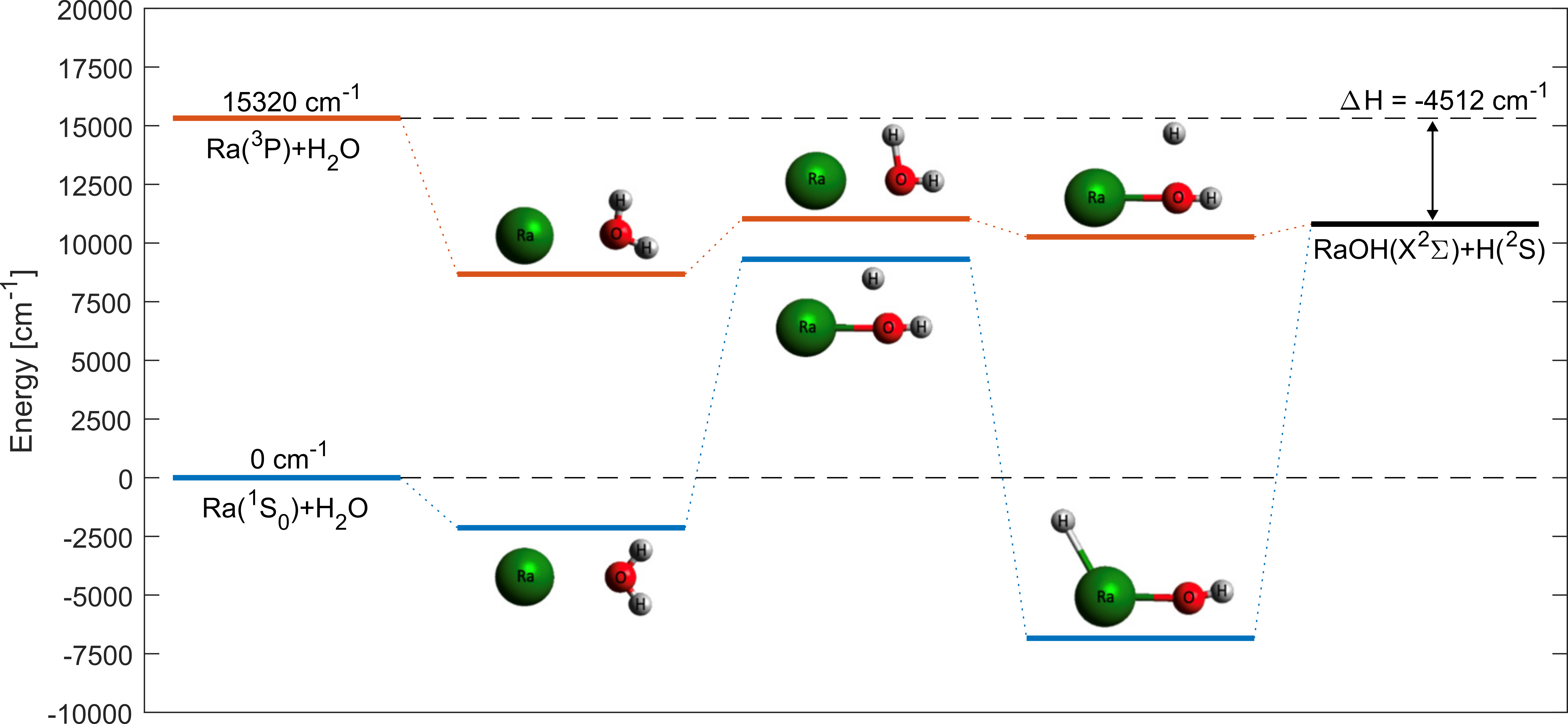}
    \caption{The critical points on the potential energy surface of Ra+H$_2$O. The molecular models, generated using Avogadro 1.9 \cite{hanwell2012avogadro}, represent
the geometries at those critical points. The energies are from calculations described in the text.}
    \label{fig:sm-reaction}
\end{figure}

The results for the RaOH formation pathways are summarized in Fig.~\ref{fig:sm-reaction}.  The reaction between ground state Ra and H$_2$O is endothermic, requiring energy to yield products, while the excited Ra reaction is exothermic and has a submerged transition state barrier. Thus, the Ra($^3$P)+H$_2$O reaction should have a large reaction rate and is preferable for the synthesis of RaOH.

\FloatBarrier
\section{Spectral Data}

\begin{figure}
    \centering
    \includegraphics[width=\linewidth]{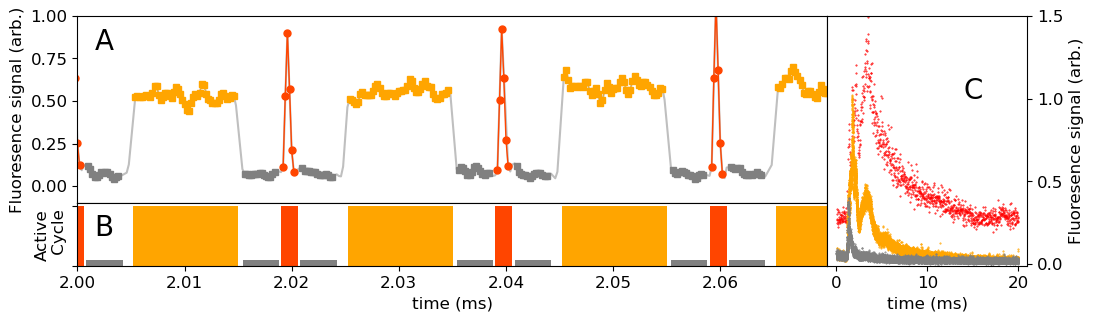}
    \caption{\red{Time-modulated fluorescence detection scheme utilized for high-resolution data acquisition. Panels (a) and (b) depict a narrow time window for a single ablation shot, demonstrating the raw data recorded on the photomultiplier tube in (a), with the laser duty cycle represented diagrammatically in (b). In this scheme, a scanning ``spectroscopy" CW laser (yellow) is interspersed with fixed-frequency ``normalization" pulsed dye excitation (red), the latter of which is used for monitoring molecular production fluctuations across ablation shots. Within each pulse there is dead time where no laser is on (gray), which can be used to recover the non-resonant glow backgrounds due to ablation and cryoplasma formation. The entire chopping sequence is performed at 50 kHz. Binning the data over the entire acquisition window (panel (c)) produces near-concurrent traces for the different sources of fluorescence.}}
    \label{fig:cycle}
\end{figure}

\begin{figure}
    \centering
    \includegraphics[width=1\linewidth]{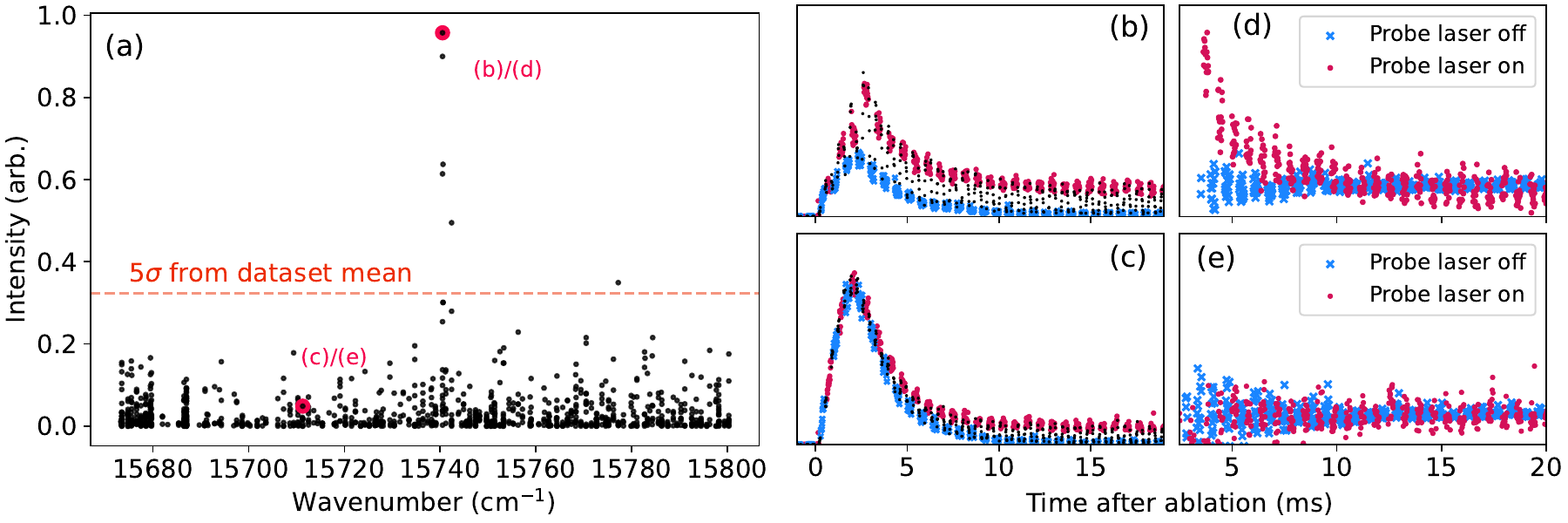}
    \caption{Low-resolution survey scan data showing the initial identification of RaOH. More than 100 cm$^{-1}$  was scanned with linewidth $\Delta\nu\sim1$ cm$^{-1}$ around the predicted transition energy of 15749 cm$^{-1}$~\cite{zhang2023relativistic}. The entire scan, which took one afternoon to perform, is shown in (a), highlighting the region of high fluorescence intensity near 15740 cm$^{-1}$. Raw data for highlighted  points at 15740.5 and 15711.3 cm$^{-1}$ are shown in (b) and (c), respectively, showing the early-time fluorescence due to nonresonant glow. Magenta (blue) points are when the laser is unblocked (blocked) through the cell, and black is the switching region between the two. The modulation frequency is only 1.5 kHz as this data was taken using a mechanical chopper wheel instead of an acousto-optic modulator as in later data. Traces (d)/(e) show the background-subtracted signal of (b)/(c), emphasizing the excess fluorescence present at early time when the laser is resonant and passing through the cell in (b).}
    \label{fig:RaOH_lowres}
\end{figure}

As discussed in the main text, RaOH fluorescence spectra were collected at iterative levels of resolution, from coarse $\Delta \nu\sim$30~GHz survey resolution, followed by $\Delta \nu\sim2.5$ GHz resolution, and ultimately $\Delta\nu\sim$~100~MHz Doppler-limited resolution. \red{Laser amplitude modulation is used to distinguish molecular fluorescence from non-resonant scattered light backgrounds inside the cell. Additionally, dual time-multiplexed lasers are used for the high-resolution data acqusition to provide simultaneous fixed-frequency monitoring for production normalization and frequency scanning for spectroscopy, as depicted in Figure \ref{fig:cycle}.} The $\sim$30~GHz scan data is shown in Fig.~\ref{fig:RaOH_lowres}, while the others are shown in the main text.  The high resolution lines and fits are shown in Table~\ref{tab:sm-raoxlines}. 

\begin{figure}
    \centering
    \includegraphics[width=1\linewidth]{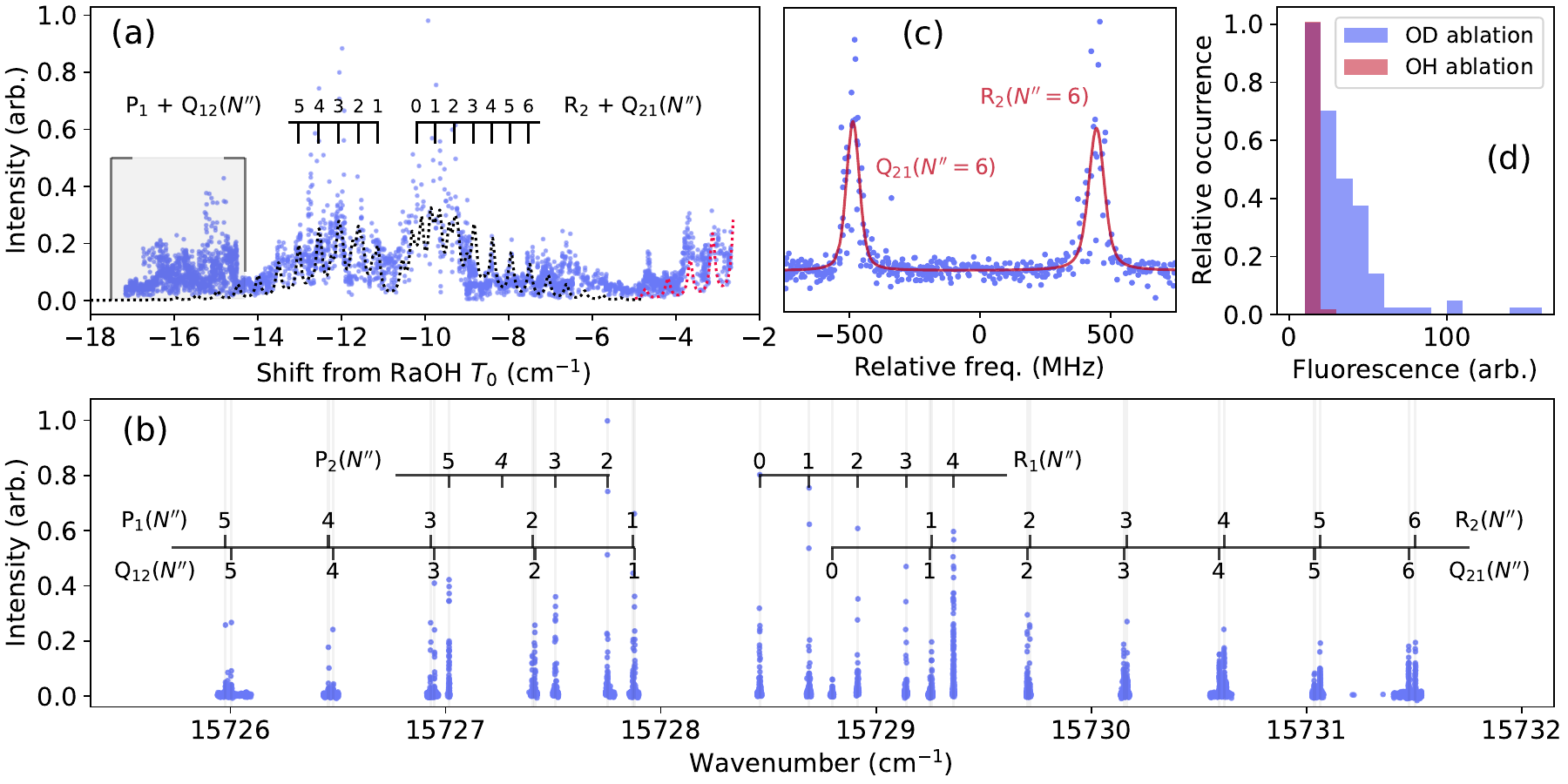}
    \caption{Laser-induced fluorescence data for the deuterated isotopologue RaOD. (a) Medium-resolution ($\sim$~0.07 cm$^{-1}$) pulsed dye spectrum identifying the band head and coarse features relative to the RaOH $\tilde{C}-\tilde{X}$ origin $T_0$. Black dotted lines are the predicted spectrum at the recorded linewidth and temperature of 7 K. Red dotted lines are the prediction for RaOH, which we observe because of hydroxide residue in our target materials. The grayed region to the left is believed to be the $\tilde{C}(100)-\tilde{X}(100)$ vibrational overtone in RaOD, which is expected to be redshifted from the main line by $\sim 4$ cm$^{-1}$. (b) High-resolution spectrum of the low-$N$ region, identifying the 31 lines used in the Hamiltonian fit. The $P_2(N^{\prime\prime}=4)$ line was not scanned. The $R$-branch $N^{\prime\prime}=6$ doublet from this spectrum is enlarged in (c), overlaid by the Voigt fits that were used to extract line centers. (d) Histograms of RaOD fluorescence signals while ablating hydroxide and deuteroxide co-targets. A strong line in the RaOD spectrum was probed as the reagent ablation laser was switched between the Al(OD)$_3$ and Al(OH)$_3$ pellets, resulting in a fluorescence signal only when vaporizing the deuterated target.}
    
    \label{fig:RaOD}
\end{figure}

Survey spectroscopy on the RaOD $\tilde{C}^2\Sigma^+_{1/2}-\tilde{X}^2\Sigma^+_{1/2}$ origin band utilized only the $\Delta \nu \sim 2.5$~GHz linewidth pulsed dye laser, as the isotope-shifted band head could be predicted much more precisely given our prior observation of the same transition in RaOH. Initial scans covering 15719 cm$^{-1}$ to 15741 cm$^{-1}$ identified a strong fluorescence signal at $\sim$ 15729 cm$^{-1}$, redshifted from RaOH by $\sim9$ cm$^{-1}$. A finer-grained scan around this region is shown in Fig.~\ref{fig:RaOD}a. Following this, a narrow-linewidth laser was used to resolve the low-$N$ lines in high-resolution, identifying a total of 32 spectral features. These features were detected and fit using the same methods for RaOH described above and are collected in Table \ref{tab:sm-raoxlines}. The extracted $B_0$ and $\gamma$ constants listed in Table \ref{tab:raoh_params} scale as expected for the replacement of H$\rightarrow $D. Because our wavemeter system was not referenced to an absolute frequency standard, a global uncertainty of $\pm 60$ MHz applies to all line centers, limited by the calibration error of the helium-neon reference laser. This uncertainty should not affect the relative separation between lines as this dataset was taken over a short period of time.

To further identify these lines with RaOD, points in this region were probed while switching the ablation between the Al(OH)$_3$ and Al(OD)$_3$ co-targets. In the region near the band head, the signal depends entirely on ablation of the deuterated co-target (Fig.~\ref{fig:RaOD}d). Because the Al(OD)$_3$ target contains hydroxide groups from the PEG binder, the correlation decreases towards the blue side of this scan (from 15734 cm$^{-1}$ onwards), owing to the presence of overlapping RaOH lines and vibrational overtones. Molecular fluorescence also relies on resonant $^3P_1$ excitation of Ra across the entire scan region, indicating the dependence on Ra-driven chemistry for molecule production (Fig.~\ref{fig:fig2}c).

\begin{figure}
    \begin{minipage}[c]{0.5\textwidth}
    \includegraphics[width=\textwidth]{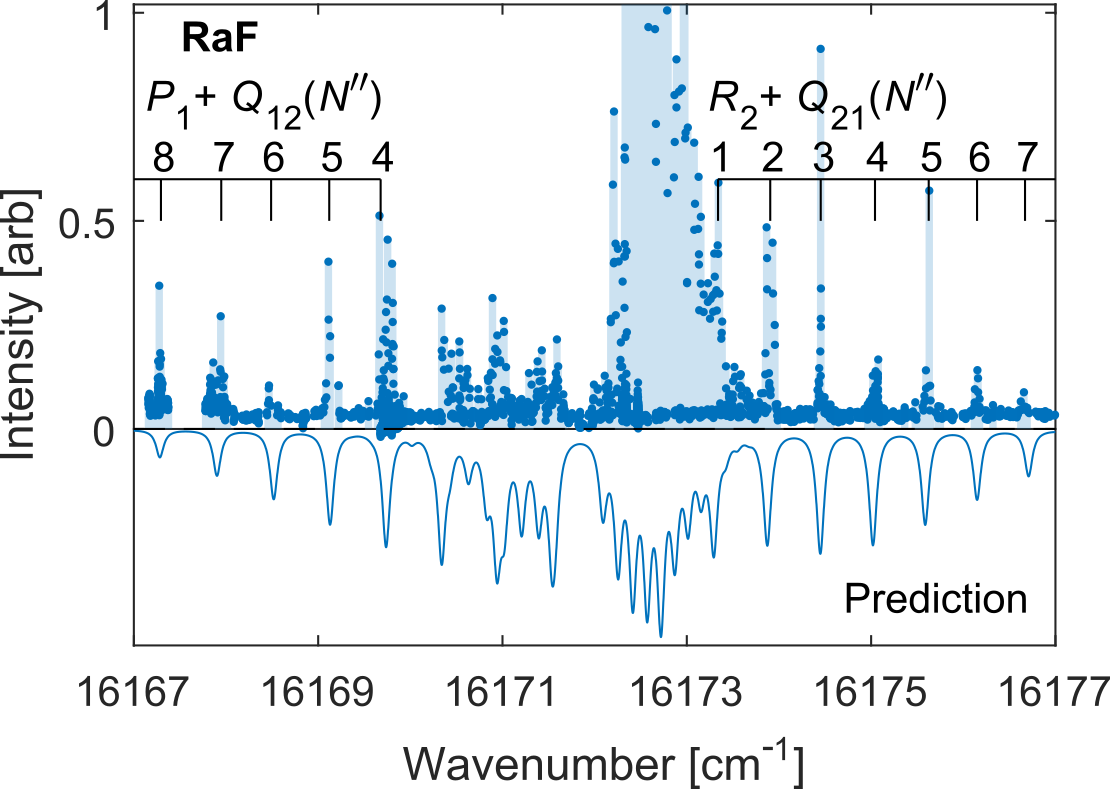}
    \end{minipage}
    \hfill
    \begin{minipage}[c]{0.45\textwidth}
    \captionof{figure}{Fluorescence spectrum without normalization of the RaF $C^2\Sigma^+_{1/2}(v' = 0) - X^2\Sigma^+_{1/2} (v''= 1)$ vibronic band system, recorded via pulsed dye excitation at $\Delta\nu\sim 2.5$ GHz laser linewidth. The top panel contains background-subtracted fluorescence spectra, identifying resolved lines from the $P_1$ and $R_2$ branch progressions. The bottom panel shows simulated band features based on fit parameters (see Table \ref{tab:raf_params}), laser linewidth, and rotational temperature $T=7$~K.}
    \label{fig:SM_RaF_midres}
    \end{minipage}
\end{figure}

The spectrum of the RaF $C^2\Sigma_{1/2}^+ (v' = 0)-X^2\Sigma_{1/2}^+ (v'' = 1)$ vibronic band system was recorded at medium-resolution ($\Delta\nu\sim 2.5$) GHz pulsed dye laser linewidth. Fluorescence detection is performed on the strong $(v' = 0) - (v'' = 0)$ decay near 602 nm with optical filtering. As depicted in Fig.~\ref{fig:fig3}A of the main text and Fig.~\ref{fig:SM_RaF_midres}, a total of 12 rotationally resolved features clear of the band head are recorded at this linewidth. The observations were fitted to a Voigt profile whose centers and standard errors are listed in Table \ref{tab:RaF_linelist}. 

At this resolution, ground state spin-rotation is not resolved, and therefore $P_1 + Q_{12}$ as well as the $R_1 + Q_{21}$ progression pairs are not distinguishable from each other in the spectra. However, ground state rotational parameters, including spin-rotation, for the $X^2\Sigma^+ (v'' = 1)$ manifold of RaF were previously determined in~\cite{udrescu2024precision}. For the band analysis in the present work, $X^2\Sigma^+ $ lower state parameters are held fixed to the literature values of~\cite{udrescu2024precision} in a Hamiltonian model of the $C^2\Sigma^+ (v' = 0)-X^2\Sigma^+ (v'' = 1)$ band. From this, we determine updated fine structure parameters for the $C^2\Sigma^+ (v' = 0)$ excited state manifold, which are listed in Table \ref{tab:raf_params} of the main text. 

\begin{figure}
  \begin{minipage}[c]{0.37\textwidth}
  \vspace{3mm}
    \begin{tabular}{l|l}
\hline
\hline
 Branches & Observation\\
\hline
$P_1 + Q_{12}(N'' = 8)$ & 16167.2938(2)\\
$P_1 + Q_{12}(N'' = 7)$  & 16167.9479(6)\\
$P_1 + Q_{12}(N'' = 6)$  & 16168.4890(1) \\
$P_1 + Q_{12}(N'' = 5)$  & 16169.1198(5) \\
$P_1 + Q_{12}(N'' = 4)$  & 16169.6773(2) \\
\hline
\multicolumn{2}{c}{\textit{band head}}\\
\hline
$R_2 + Q_{21}(N'' = 1)$ & 16173.3392(3) \\
$R_2 + Q_{21}(N'' = 2)$  & 16173.9048(8) \\
$R_2 + Q_{21}(N'' = 3)$  & 16174.4530(5) \\
$R_2 + Q_{21}(N'' = 4)$  & 16175.0417(4) \\
$R_2 + Q_{21}(N'' = 5)$  & 16175.6246(1) \\
$R_2 + Q_{21}(N'' = 6)$  & 16176.1491(3) \\
$R_2 + Q_{21}(N'' = 7)$  & 16176.6699(2) \\
\hline
\hline
\end{tabular}
  \end{minipage}
  \begin{minipage}[c]{0.4\textwidth}
\vspace{3mm}\captionof{table}{\label{tab:RaF_linelist} Medium-resolution ($\Delta\nu \sim 2.5$ GHz) rotationally resolved features on the RaF $C^2\Sigma^+_{1/2}(0)-X^2\Sigma^+_{1/2}(1)$ band system resolved with pulsed dye excitation. The ground state spin-rotation splitting is below the laser linewidth, and two branch designators are therefore assigned to each observation. As with the high-resolution line list, feature centers are determined from Voigt profile fits to the fluorescence data. Parentheses reflect the standard error of the fit which is used for error weighting in Hamiltonian parameter determination. All values are in cm$^{-1}$.}
  \end{minipage}
\end{figure}

\section{Spectral Fitting}
\label{sec:effective_ham}

Two analyses of the high-resolution spectra were performed. There was no evidence of local perturbations or proton/deuteron magnetic hyperfine splitting. The data set was restricted to the lowest rotational levels and did not require the inclusion of centrifugal distortion terms. In the initial analysis, the energy levels for the $\tilde{X}^2\Sigma^+$ and $\tilde{C}^2\Sigma^+$ states were modeled using an effective Hamiltonian approach that included the rotational and spin-rotation terms in space-fixed convention:
\begin{align}
    \hat{H}_\text{eff} = \hat{H}_\text{rot} + \hat{H}_{SR} = B\hat{N}^2 + \gamma \hat{N}\cdot \hat{S},\label{eq:effham}
\end{align}
where $N=J-S$. The energy level pattern of the $\tilde{X}^2\Sigma^+$ is that of a molecule near the Hund's case (b) limit, with the closely spaced rotational levels split by a small spin-rotation splitting, while the pattern of the $\tilde{C}^2\Sigma^+$ state is nearer to the Hund's case (c) limit. For generalizability, we work in the Hund's case (a) basis $|\Lambda,  S, \Sigma, J, \Omega\rangle$ for both states. The matrix elements in case (a) for the effective Hamiltonian terms (eq. \ref{eq:effham}) are: 
\begin{align}
    \langle \Lambda,  S, \Sigma, J, \Omega|&\hat{H}_{\text{rot}}|\Lambda,  S, \Sigma', J, \Omega'\rangle =B_0\bigg[\delta_{\Sigma, \Sigma'} \delta_{\Omega, \Omega'} (J(J+1)+S(S+1)-2\Omega\Sigma)\notag\\
    &-2(-1)^{J-\Omega+S-\Sigma}\sum_{q}\threeJ{J}{-\Omega}{1}{q}{J}{\Omega'}\threeJ{S}{-\Sigma}{1}{q}{S}{\Sigma'}\sqrt{J(J+1)(2J+1)S(S+1)(2S+1)}\bigg].\\
    \langle \Lambda,  S, \Sigma, J, \Omega|&\hat{H}_{SR}|\Lambda,  S, \Sigma', J, \Omega'\rangle = \gamma\bigg[\delta_{\Sigma, \Sigma'} \delta_{\Omega, \Omega'}(\Omega\Sigma - S(S+1))\notag\\ &+ (-1)^{J-\Omega+S-\Sigma}\sum_{q}\threeJ{J}{-\Omega}{1}{q}{J}{\Omega'}\threeJ{S}{-\Sigma}{1}{q}{S}{\Sigma'}\sqrt{J(J+1)(2J+1)S(S+1)(2S+1)}\bigg].\label{eq:spinrot}
\end{align}

\begin{table}[]
\begin{tabular}{r|cc|cc|cc}
\hline\hline 
& \multicolumn{2}{c|}{\textbf{RaOH}} & \multicolumn{2}{c|}{\textbf{RaOD}} & \multicolumn{2}{c}{\textbf{RaF}} \\
      & Eff. Ham.$^a$     & Explicit   (3x3)$^b$  & Eff. Ham.$^a$     & Explicit   (3x3)$^b$ & Eff. Ham.$^a$     & Explicit   (3x3)$^c$ \\ \hline
$B(\tilde{C}^2\Sigma^+)$  & 0.192810(23) &  0.191552(19) & 0.173750(21) &0.169692(22) & 0.18890(19) & 0.18652(57) \\
$\gamma(\tilde{C}^2\Sigma^+)$ & --0.25488(14)  & -- & --0.224787(97) & -- & --0.4132(30) & --   \\
$M_1$    & --             & 0.22697(10) & -- & 0.204299(54) & -- & 0.34047(18) \\
$E(\tilde{C}^2\Sigma^+)$ & -- & 15338.47958(20) & -- & 15325.54202(17) & -- & 15747.993(23) \\
$B(\tilde{X}^2\Sigma^+)$  & 0.193944(19)    & 0.193929(24) & 0.175333(19) &0.175392(17) & 0.19092$^d$ & 0.19092$^d$ \\
$\gamma(\tilde{X}^2\Sigma^+)$ & 0.00504(12) & \red{0.00496(15)} & 0.004759(56)& 0.00470(7) & 0.00581$^d$ & 0.00581$^d$ \\
$T_{0}$   & 15739.42204(21) & -- & 15728.22464(24) & -- & 16171.9194(79)$^\dagger$ & -- \\
Std. of fit   & 0.0012        & 0.0009  & 0.0005 & 0.0006 & 0.041 & 0.040     \\
\hline\hline 
\end{tabular}
\caption{Determined spectroscopic parameters for the $\tilde{C}^2\Sigma^+(0,0,0) - \tilde{X}^2\Sigma^+(0,0,0)$ band of RaOH/D  and the $C^2\Sigma^+(v' = 0) - X^2\Sigma^+(v'' = 1)$ band of RaF in wavenumbers (cm$^{-1}$). Values in parentheses denote the $1\sigma$ statistical error on the estimated parameters from least squares fitting. a)	Obtained using the Hamiltonian given in Eq. (\ref{eq:effham}).  b)	Obtained using energies by  Eqs. (\ref{eq:sm-dpme}) : $A$=1450~cm$^{-1}$, $M_2$=1025~cm$^{-1}$, and $E(\tilde{A}^2\Pi) =13845$~cm$^{-1}$. $E(\tilde{C}^2\Sigma^+)$ and $E(\tilde{A}^2\Pi)$ are the energies of the interacting excited states in the absence of the spin-orbit interaction.
c)	Obtained using energies by  Eqs. (\ref{eq:sm-dpme}) : $A$=1450~cm$^{-1}$, $M_2$=1025~cm$^{-1}$, and $E(\tilde{A}^2\Pi) =14420$~cm$^{-1}$.  d) Fixed at values determined by~\cite{udrescu2024precision}. ($\dagger$) referenced relative to $X^2\Sigma^+(v'' = 1) - C^2\Sigma^+(v' = 0)$ transition.
\label{tab:sm-raoxfit}}
\end{table}

\begin{table}[]
\begin{tabular}{clrr|clrr}
\hline\hline \multicolumn{8}{c}{\textbf{RaOH}}  \\
Assign.$^a$ & Obs.$^b$   & $\Delta\nu^c$     & $\Delta\nu^d$     & Assign.$^a$ & Obs.$^b$   & $\Delta\nu^c$     & $\Delta\nu^d$      \\ \hline
$^PP_{11}(N'')$    &         &         &         & $^PP_{22}(N'')$  &         &         &         \\ 
(1) & 39.03143(4) & --0.0002 & --0.0011 & (2) & 38.9078(2)  & 0.0014  & 0.0008  \\
(2) & 38.51250(8) & 0.0010  & 0.0001  & (3) & 38.64436(3) & 0.0004  & 0.0000  \\
(3) & 37.9902(2)  & 0.0011  & 0.0003  & (4) & 38.37921(1) & 0.0000  & 0.0001  \\
(4) & 37.4653(1)  & 0.0008  & 0.0005  & (5) & 38.1116(1)  & --0.0006 & --0.0002 \\ \hline
$^PQ_{12}(N'')$    &         &         &         & $^RQ_{21}(N'')$ &         &         &         \\ 
(1) & 39.03924(2) & 0.0001  & --0.0007 & (0) & 40.06471(5) & 0.0022  & 0.0016  \\
(2) & 38.52575(9) & 0.0016  & 0.0010  & (1) & 40.57113(5) & 0.0003  & --0.0001 \\
(3) & 38.0086(2)  & 0.0018  & 0.0013  & (2) & 41.07680(3) & 0.0000  & 0.0000  \\
(4) & 37.48898(9) & 0.0018  & 0.0018  & (3) & 41.5787(1)  & --0.0018 & --0.0016 \\ \hline
$^RR_{11}(N'')$    &         &         &         & $^RR_{22}(N'')$  &         &         &         \\ 
(0) & 39.6815(2)  & 0.0013  & 0.0006  & (1) & 40.57882(3)  & 0.0005  & 0.0002  \\
(1) & 39.9346(2)  & 0.0009  & 0.0004  & (2) & 41.08966(2)  & 0.0003  & 0.0005  \\
(2) & 40.1851(1)  & 0.0003  & 0.0003  & (3) & 41.59675(10) & --0.0014 & --0.0009 \\
(3) & 40.4329(1)  & --0.0007 & --0.0006 & (4) & 42.1041(2)   & --0.0006 & 0.0010  \\
(4) & 40.68031(6) & 0.0001  & 0.0013  & (5) &  --            &   --      &   --      \\
(5) & 40.92336(9) & --0.0012 & --0.0000 & (6) & 43.10720(2)  & --0.0037 & --0.0002 \\ \hline\hline
\multicolumn{8}{c}{\textbf{RaOD}}  \\
Assign.$^a$ & Obs.$^b$   & $\Delta\nu^c$     & $\Delta\nu^d$     & Assign.$^a$ & Obs.$^b$   & $\Delta\nu^c$     & $\Delta\nu^d$      \\ \hline
$^PP_{11}(N'')$    &         &         &         & $^PP_{22}(N'')$  &         &         &         \\ 
(1)      & 27.871(1)  & --0.0003 & --0.0005 & (2)      & 27.752(2) & --0.0004 & 0.0005  \\
(2)      & 27.403(2)  & 0.0000  & 0.0001  & (3)      & 27.511(8) & 0.0009  & 0.0020  \\
(3)      & 26.931(3)  & --0.0002 & --0.0002 & (5)      & 27.016(4) & 0.0005  & 0.0014  \\
(4)      & 26.456(4)  & --0.0005 & --0.0001 & (9)      & 25.988(2) & --0.0016 & --0.0001 \\
(5)      & 25.978(1)  & 0.0002  & --0.0002 &          &           &         &         \\ \hline
$^PQ_{12}(N'')$    &         &         &         & $^RQ_{21}(N'')$ &         &         &         \\ 
(1)      & 27.8780(9) & --0.0007 & --0.0006 & (0)      & 28.797(1) & 0.0001  & 0.0002  \\
(2)      & 27.415(2)  & --0.0001 & 0.0003  & (1)      & 29.251(5) & 0.0001  & --0.0002 \\
(3)      & 26.948(3)  & 0.0006  & 0.0004  & (2)      & 29.703(1) & 0.0005  & 0.0007  \\
(4)      & 26.478(2)  & 0.0002  & 0.0008  & (3)      & 30.151(9) & 0.0007  & 0.0005  \\
(5)      & 26.004(2)  & 0.0000  & 0.0000  & (4)      & 30.595(6) & 0.0001  & --0.0001 \\
         &            &         &         & (5)      & 31.037(2) & --0.0001 & --0.0002 \\
         &            &         &         & (6)      & 31.475(3) & --0.0009 & --0.0001 \\ \hline
$^RR_{11}(N'')$    &         &         &         & $^RR_{22}(N'')$  &         &         &         \\ 
(0)      & 28.460(2)  & --0.0002 & 0.0000  & (1)      & 29.259(1) & 0.0003  & 0.0008  \\
(1)      & 28.689(3)  & --0.0001 & --0.0008 & (2)      & 29.714(2) & --0.0002 & --0.0001 \\
(2)      & 28.915(2)  & --0.0004 & --0.0013 & (3)      & 30.166(1) & --0.0010 & --0.0009 \\
(3)      & 29.140(2)  & 0.0007  & 0.0001  & (4)      & 30.617(5) & 0.0003  & 0.0008  \\
(4)      & 29.360(1)  & 0.0008  & 0.0003  & (5)      & 31.063(6) & 0.0000  & --0.0001 \\
         &            &         &         & (6)      & 31.506(5) & --0.0007 & 0.0004 \\ \hline
\end{tabular}
\caption{Observed and predicted transition wavenumbers for the $\tilde{C}^2\Sigma^+(0,0,0) - \tilde{X}^2\Sigma^+(0,0,0)$ band of RaOH and RaOD in wavenumbers (cm$^{-1}$). 
	a) Line assignment: $^{\Delta N}\Delta J_{F'F''}(N'')$. 
	b) Observed transition wavenumber -- 15700 cm$^{-1}$. Parentheses indicate the $1\sigma$ parameter error on the line center determined from Voigt profile fit. Absolute calibration error on individual lines is $\pm$ 60 MHz, defined by wavemeter reference uncertainty. 
	c) The observed-calculated using optimized parameter from the effective Hamiltonian model.
	d) The observed-calculated using optimized parameter from the $\tilde{C}^2\Sigma^+/\tilde{A}^2\Pi$  mixed state perturbation model.
    \label{tab:sm-raoxlines}    
}
\end{table}

The most convenient representation of both the $\tilde{X}^2\Sigma$ and $\tilde{C}^2\Sigma$ states is in the parity-conserving Hund's case (a) basis, which is diagonal in $\hat{H}_\text{eff}$:

\begin{equation}
\Psi^\pm = \frac{1}{\sqrt{2}} \left[ |\eta,\Lambda;S, \Sigma; J, \Omega \rangle \pm (-1)^{J-S} |\eta, -\Lambda; S, -\Sigma; J, -\Omega \rangle \right]
\end{equation}
where the ``$\pm$'' refers to the plus and minus parity, and $\eta$ refers to all other quantum numbers. $\hat{H}_\text{eff}$ is an eigenoperator of these basis functions with plus parity energies, $E^{(+)}(J)$, and minus parity energies, $E^{(-)}(J)$, given by:
\begin{equation}
\red{E^{(\pm)}(J) = B\left( J(J+1) + \frac{1}{4} \right) \pm (-1)^{J-1/2} B\left( -J - \frac{1}{2} \right) - \gamma\left( -\frac{1}{2} \pm (-1)^{J-1/2} \left[ \frac{J + \frac{1}{2}}{2} \right] \right)\label{eq:sm-epm}.}
\end{equation}

In the effective Hamiltonian analysis, weighted linear least squares fits of the 27 measured RaOH transition wavenumbers and the 32 measured RaOD transition wavenumbers of Table~\ref{tab:sm-raoxlines} were performed against $\hat{H}_\text{eff}$. The weights were taken as the inverse of the square of the standard error of the lineshape fits, which are also given in the same tables. A similar analysis was performed for the RaF medium resolution lines in Table \ref{tab:RaF_linelist}, fixing the ground state parameters to values determined in \cite{udrescu2024precision} as described earlier. For RaOH/D, the optimized $B$ and $\gamma$ parameters for the $\tilde{X}^2\Sigma^+$ and $\tilde{C}^2\Sigma^+$ states, the origin, $T_{0}$, and associated errors are presented in Table~\ref{tab:raoh_params} of the main text and Table~\ref{tab:sm-raoxfit}. The standard deviation of the fits of 0.0012 cm$^{-1}$ (RaOH) and 0.0005 cm$^{-1}$ (RaOD) are commensurate with estimated measurement uncertainty of approximately 0.001 cm$^{-1}$ and there are no systematic trends in the difference between the observed and predicted transition wavenumber which are given in Table~\ref{tab:sm-raoxlines}. Fig.~\ref{fig:confidence_intervals} depicts the two-dimensional statistical confidence regions up to $2\sigma$ for estimated parameters computed via $F$-test, which confirms that the effective model is well-behaved and the estimated parameters converge to global minima.

\begin{figure}
\hspace{1cm}
  \begin{minipage}[b]{.55\linewidth}
    \centering
    \includegraphics[width=0.8\textwidth]{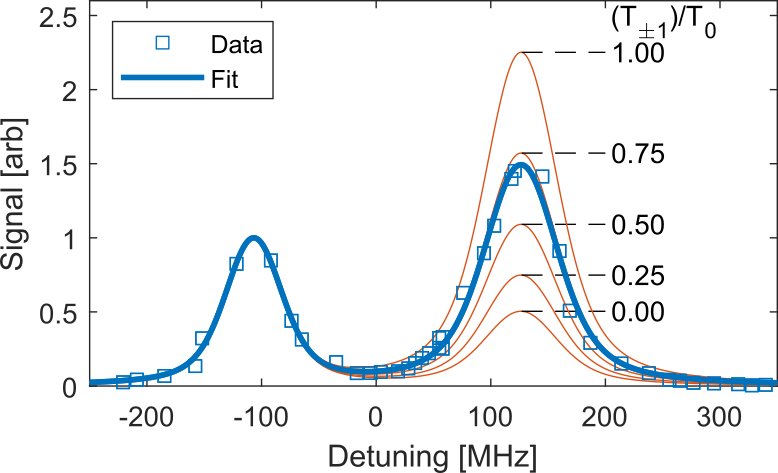}
    \captionof{figure}{Relative heights of the $^PP_{11}(1)$ and $^PQ_{12}(1)$ lines as a function of the TDM ratio $T_{\pm 1}/T_0$.}
    \label{fig:deperturbation}
  \end{minipage}\hfill
  \begin{minipage}[b]{.25\linewidth}
    \centering
    \begin{tabular}{c|c}
TDM Ratio& Intensity Ratio  \\
$T_{\pm 1}/T_0$ & $^PQ_{12}(1)$/$^PP_{11}(1)$   \\ \hline
0.00    & 0.50  \\
0.25  & 0.75    \\
0.50   & 1.1   \\
0.75  & 1.6    \\
1.00  & 2.3    \\
Expt. & 1.5
\end{tabular}
\vspace{5mm}
\captionof{table}{Table of expected intensity ratio of $^PQ_{12}(1)$ and $^PP_{11}(1)$ lines as a function of TDM ratio.\label{tab:dp_intensities}}
  \end{minipage}
  \hspace{1cm}
\end{figure}

\begin{figure}
    \centering
    \includegraphics[width=0.85\linewidth]{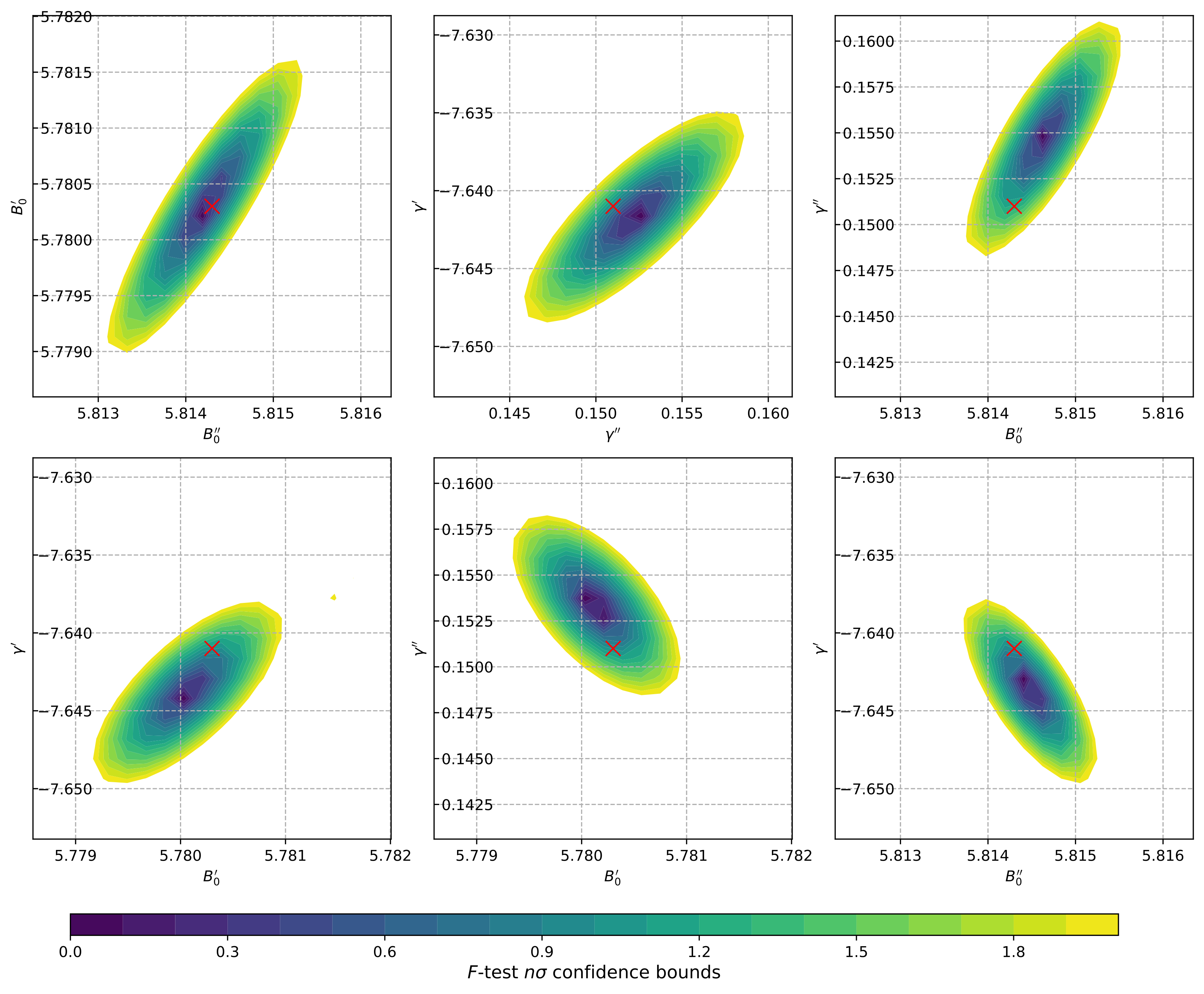}
    \caption{Two-dimensional $F$-test statistical confidence regions for selected parameters in the RaOH $\tilde{C}^2\Sigma^+_{1/2}-\tilde{X}^2\Sigma^+_{1/2}$ high-resolution effective Hamiltonian fit. Probabilistic regions up to $2\sigma$ levels are plotted. The red cross denotes the optimized nonlinear least squares parameters (see Table \ref{tab:raoh_params} in main text).}
    \label{fig:confidence_intervals}
\end{figure}

The eigenvectors and energies from this effective Hamiltonian model were used to predict the spectrum of RaOH/D. The intensities were computed by the square of the amplitude of the Hund's case (a) transition moment matrix element, scaled by the ground state rotational occupation described by a Boltzmann factor. Although the predicted spectra were in overall good agreement with the observations, there were noticeable discrepancies in the predicted relative intensities. For example, the RaOH satellite $^PQ_{12}(1)$ ($\nu = 15739.0392$ cm$^{-1}$) line is predicted to be approximately a factor of two less intense than the $^PP_{11}(1)$ ($\nu = 15739.0314$ cm$^{-1}$) line whereas it is observed to be a factor of around 1.5 times more intense (see Fig.~\ref{fig:deperturbation}). One reason for the apparent inconsistency is the assumption that the excited state is a pure $^2\Sigma^+$ electronic state. As evidenced by the large and negative spin-rotation fit parameter, $\gamma = -0.25488(14)$ cm$^{-1}$, which gives rise to large spin-rotation splitting ($\rho$-doubling), the excited $|\Omega| = 1/2$ state is in fact a $^2\Sigma^+ / ^2\Pi_{1/2}$ admixture, analogous to other related molecules such as YbOH~\cite{Steimle2019YbOH} and BaOH~\cite{kinsey1986rotational}.

Motivated by this observation, the second analysis modeled the energies and wavefunctions of the ``$\tilde{C}^2\Sigma^+$'' state by explicitly including interaction with the yet-to-be-detected $\tilde{A}^2\Pi_{1/2}$ state. This analysis aimed to improve the prediction of the relative intensities, thereby gaining insight into the spin-orbit-induced excited state mixing. The $\tilde{A}^2\Pi_{1/2}$ state is predicted~\cite{zhang2023relativistic} to be 3031 cm$^{-1}$ below the $\tilde{C}^2\Sigma^+$ state. In the zeroth order picture, the levels of the $\tilde{A}^2\Pi_{1/2}$ state are systematically shifted to lower energy and those of the $\tilde{C}^2\Sigma^+$ state shifted higher by spin-orbit and spin-electronic interactions~\cite{Lefebvre2004Book}. The rate of the shifting of the positive and negative parity levels of both the $\tilde{A}^2\Pi_{1/2}$ and $\tilde{C}^2\Sigma^+$ states are slightly different, giving rise to $\Lambda$-doubling in the $\tilde{A}^2\Pi_{1/2}$ state and $\rho$-doubling in the $\tilde{C}^2\Sigma^+$ state. In this second approach, rotational energies were obtained by diagonalizing a $3 \times 3$ matrix constructed in a parity conserving Hund's case (a) function for the $^2\Sigma^+$, $^2\Pi_{1/2}$, and $^2\Pi_{3/2}$ electronic states. The complete set of matrix elements are tabulated by Brown and Carrington~\cite[8.5.2(d)]{brown2003rotational} and those relevant to the present analysis are reproduced here for convenience:
\begin{equation}
\begin{aligned}
\langle ^2\Pi_{3/2}(\pm)|\hat{H}|^2\Pi_{3/2}(\pm) \rangle &= E(^2\Pi) + \frac{1}{2}A + B[J(J + 1) - 7/4],\\
\langle ^2\Pi_{1/2}(\pm)|\hat{H}|^2\Pi_{1/2}(\pm) \rangle &= E(^2\Pi) - \frac{1}{2}A + B[J(J + 1) + 1/4],\\
\langle ^2\Sigma_{1/2}(\pm)|\hat{H}|^2\Sigma_{1/2}(\pm) \rangle &= E(^2\Sigma) + B[J(J + 1) + 1/4] \pm B(-1)^{J-1/2}(J + 1/2),\\
\langle ^2\Pi_{1/2}(\pm)|\hat{H}|^2\Sigma_{1/2}(\pm) \rangle &= M_2 \mp M_1(-1)^{J-1/2}(J + 1/2).\label{eq:sm-dpme}
\end{aligned}
\end{equation}
In eqs.~\ref{eq:sm-dpme}, $M_1 = \langle ^2\Pi |B L^+| ^2\Sigma \rangle$ and \red{$M_2 = \langle ^2\Pi| (A/2 + B)L^+| ^2\Sigma \rangle$} are the spin-electronic and spin-orbit mixing matrix elements, $E(^2\Pi)$ and $E(^2\Sigma)$ are the unperturbed energies, $A$ is the spin-orbit parameter, and $B$ is the rotational parameter. There is currently insufficient information about the excited states of RaOH to perform a global fit to determine all the parameters in eqs.~\ref{eq:sm-dpme}. In fitting the observed spectra of RaOH and RaOD to this model, the $M_1$, $B$, and $E(^2\Sigma)$ parameters of the $\tilde{C}^2\Sigma^+$/ $\tilde{A}^2\Pi$ mixed states were optimized and $E(^2\Pi)$, $A$, and $M_2$ parameters were fixed to estimated values. The unperturbed energy of the $\tilde{A}^2\Pi$ state, $E(^2\Pi)$, was constrained to 13845 cm$^{-1}$, which is the value that reproduced the predicted~\cite{zhang2023relativistic} splitting of 3031 cm$^{-1}$ between the $\tilde{A}^2\Pi_{1/2}$ and $\tilde{C}^2\Sigma^+$ states of RaOH. Estimates for $A$ and $M_2$ for the analysis were obtained by using the experimentally measured energies~\cite{Osika2024RaOH,GarciaRuiz2020RaF} of the $\tilde{A}^2\Pi_{1/2}$ (13288 cm$^{-1}$), $\tilde{A}^2\Pi_{3/2}$ (15355 cm$^{-1}$), and $\tilde{C}^2\Sigma^+$ (16620 cm$^{-1}$) states of isoelectronic RaF. If it is assumed that the $\tilde{A}^2\Pi$ and $\tilde{C}^2\Sigma^+$ of RaF and RaOH arise from an electronic configuration with a lone unpaired electron in a 7p orbital, then:
\begin{equation}
\langle ^2\Pi | L^+ | ^2\Sigma \rangle \approx \langle 5p_{\pm 1} | L^+ | 5p_0 \rangle = \sqrt{2} \Rightarrow M_2 \approx \frac{\sqrt{2}}{2} A.
\end{equation}
With this assumption, the relative energies of three RaF states can be approximately reproduced using the $3 \times 3$ matrix approach with spin-orbit parameter $A = 1450$ cm$^{-1}$ and $^2\Pi-^2\Sigma$ energy splitting $\Delta E = 1400$ cm$^{-1}$. A non-linear weighted least-squares fit of the 27 observed transition wavenumbers of RaOH and 32 observed transition wavenumbers of RaOD given in Table~\ref{tab:sm-raoxlines} were performed to determine optimized $B$ and $\gamma$ values for the $\tilde{X}^2\Sigma^+$, as well as $B$, $M_1$, and $E(^2\Sigma)$ for the $\tilde{C}^2\Sigma^+$/ $\tilde{A}^2\Pi$ mixed states. The optimized parameters and associated errors are given in Table~\ref{tab:sm-raoxfit}. The standard deviations of the fits are 0.009 cm$^{-1}$ (RaOH) and 0.007 cm$^{-1}$ (RaOD) which are commensurate with the estimated measurement uncertainties. The difference between the observed and predicted transition wavenumbers, which are given in Table~\ref{tab:sm-raoxlines}, are nearly identical to those for the effective Hamiltonian model.

The $\tilde{C}^2\Sigma^+$/ $\tilde{A}^2\Pi$ state mixing is significant. The excited state eigenvectors for energy levels associated with the observed transitions are approximately 87\% $^2\Sigma^+$ and 13\% $^2\Pi$ character, which is consistent with the \textit{ab initio} prediction~\cite{Zaitsevskii2022RaF} of the composition of the excited state of RaF. The eigenvectors from the mixed $\tilde{C}^2\Sigma^+$ and $\tilde{A}^2\Pi$ state analysis were used to predict the spectra. The Hund's case (a) transition moment matrix elements are:
\begin{align}
    \langle \Lambda,  S, \Sigma, J, \Omega|T^1_q(\mu)|\Lambda',  S, \Sigma, J', \Omega'\rangle = (-1)^{J-\Omega}\sqrt{(2J+1)(2J'+1)}\threeJ{J}{-\Omega}{1}{q}{J'}{\Omega'}\big\langle\Lambda\big|T^1_q(\mu)\big|\Lambda'\big\rangle.\label{eq:tdm}
\end{align}
The $q = 0$ components of the transition moment operator are responsible for the parallel ($\Delta\Lambda = 0$) $^2\Sigma^+ \leftrightarrow ^2\Sigma^+$ transition and the $q = \pm 1$ components are resposibe for the perpendicular ($\Delta\Lambda = \pm 1$) $^2\Pi \leftrightarrow ^2\Sigma^+$ transitions. The effective Hamiltonian assumes a pure $^2\Sigma^+$ character and thus excludes the $^2\Pi \leftrightarrow ^2\Sigma^+$ component. As can be seen from the predicted $^PQ_{12}(1)$ and $^PP_{11}(1)$ lines (see Fig.~\ref{fig:deperturbation}) the relative intensities are reproduced when $T^1_{\pm 1}(\mu) \approx 0.7 \times T^1_0(\mu)$. 

\red{Due to the limited number of electronic states that have been characterized, the scope of the present de-pertubation analysis is restricted to the $A^2\Pi_{1/2}$ and $C^2\Sigma^+_{1/2}$ excited states. While the mixing between these two states is expected to be the dominant interaction, additional excited states, in particular the $^2\Pi_{3/2}$ and $^2\Delta_{3/2}$ manifolds, are expected to also interact with the $C^2\Sigma^+_{1/2}$  state via spin-orbit coupling and alter the final state admixtures. The inclusion of these states -- which have not yet been spectroscopically characterized -- in a future multi-state de-pertubation model could further improve the predictions of the observed rotational intensity anomalies.}

\subsection{Spin-rotation coupling}

The electron spin-rotation interaction contributes a bilinear term $\gamma N\cdot S$ to the molecular energy. 
For the $\tilde{X}^2\Sigma^+$ electronic ground states of RaOH/D, 
$\gamma$ corresponds to the splitting of the rotational ground states of the $\Omega=\pm 1/2$ manifold
due to the $L$-uncoupling interaction $BL_{\pm}$. Specifically, the $L$-uncoupling interaction mixes
the $\Omega=\pm 1/2$ components and induces a splitting of the value $\gamma$ between
$\frac{1}{\sqrt{2}}(|\Omega=1/2\rangle+|\Omega=-1/2\rangle)$ and $\frac{1}{\sqrt{2}}(|\Omega=1/2\rangle-|\Omega=-1/2\rangle)$. For \textit{ab initio} electronic structure calculations of $\gamma$, we include a small magnetic field to evenly mix the $\Omega=\pm 1/2$ components and obtain the parity states $\frac{1}{\sqrt{2}}(|\Omega=1/2\rangle\pm|\Omega=-1/2\rangle)$.
The $L$-uncoupling interaction is then added as a finite field to obtain the energy splitting.
These calculations were performed using the CFOUR program package \cite{CFOURfull, Matthews20a}. 
The wave functions were calculated using the electron attachment version of equation-of-motion coupled-cluster singles and doubles (EOMEA-CCSD) method \cite{Stanton93a,Nooijen95}. 
The relativistic exact two-component (X2C) \cite{Dyall97,Kutzelnigg05,Ilias07,Liu09} approach with atomic mean-field (AMF) \cite{Hess96a} integrals based on the Dirac-Coulomb-Breit Hamiltonian [the X2CAMF(DCB) scheme] \cite{Liu18,Zhang22} was used to account for relativistic effects. 
The X2CAMF scheme features variational treatments of scalar-relativistic and spin-orbit coupling effects in the spinor representation.

As described in the main text and Table \ref{tab:raoh_params}, we obtain good agreement between the X2C-CC calculations and measurements for the $\tilde{X}^2\Sigma^+$ ground state spin rotation parameters of RaF and RaOD/H. The X2CAMF-EOMEA-CCSD value of 165 MHz for the $\tilde{X}^2\Sigma^+$ state of RaOH agrees well with the measured value of 151(3) MHz. 
The corresponding X2CAMF-EOMEA-CCSD value for RaOD is 149 MHz, in good agreement with the measured value of 143(2) MHz. Similarly, the X2CAMF-EOMEA-CCSD value of 187 MHz for RaF is also in good agreement with the measured ground state literature value of 175 MHz \cite{udrescu2024precision}. 

Agreement between calculation and measurement for the excited $\tilde{C}^2\Sigma^+$ states of RaF and RaOH is more mixed. The X2CAMF-EOMEA-CCSD computed value of $-13443$ MHz for RaF is in reasonable agreement with the measured value of $-12388$ MHz.
On the other hand, the computed values of $-13574$ MHz for RaOH and $-12342$ MHz for RaOD are in qualitative agreement with the measured values of $-7641(4)$ MHz and $-6739(3)$ MHz, but exhibit substantial overestimation of the absolute magnitudes. 
The measured value for RaOH is around half of that for RaF, while the computed values for RaOH is similar to that of RaF. 

These results suggest that certain differences between the RaOH and RaF electronic structure have not been fully captured by the EOMEA-CCSD calculations. An EOMEA-CCSD calculation uses the closed-shell ground state of the cation as the reference. The calculation accurately takes into account the contributions from the electron-attached states differing from the reference state in a single electron attachment. However, contributions from configurations differing from the reference by more than a single attachment are accounted for less accurately. One possibility is that the spin-rotation constants of the $\tilde{C}^2\Sigma$ state in RaOH/RaOD have more contributions from charge-transfer excited states than in RaF. The lowest charge-transfer excited state with $\Omega=1/2$ in RaOH or RaOD, originating from a transition from an oxygen 2p orbital to Ra 7s orbital, is around 5.3 eV above the ground state. They are substantially lower than the charge-transfer excited states in RaF with an excitation from fluorine 2p orbital to Ra 7s orbital, which is around 6.8 eV above the ground state. Alternatively, the values for the spin-rotation constants of the $\tilde{C}^2\Sigma$ state could be affected by coupling with nearby vibronic states, for example as in BaOH/BaOD \cite{kinsey1986rotational, gustavsson1991perturbations} where a factor of two difference between spin-rotation constants between the $\tilde{B}^2\Sigma^+$(000) and $\tilde{B}^2\Sigma^+$(100) states was observed. 

In addition to the finite field calculations, we have also performed an estimate of the spin rotation parameters for the $\tilde{C}^2\Sigma$ states via perturbation theory and obtain qualitatively consistent results. For this computation, we assume that the $\gamma N\cdot S$ effective interactions arise primarily as a consequence of the second order couplings described in eq. \ref{eqn:2ndorder}.
Assuming a single perturber, the excited spin rotation constant draws leading second-order contributions from the $L$-uncoupling and microscopic spin-orbit terms of the rotational Hamiltonian: 
\begin{equation}
    \gamma^{(2)}_{SR} = 2|\langle v''|v'\rangle|^2\langle ^2\Sigma^+_{-1/2}| B L_-|^2\Pi_{1/2}\rangle
    \times\frac{ \langle ^2\Pi_{1/2} | \sum_i a_i l_i^+ s_i^-|^2\Sigma^+_{1/2}\rangle}{E(^2\Pi_{1/2})-E(^2\Sigma^+_{1/2})}, \label{eqn:2ndorder}
\end{equation}
where $|\langle v''|v'\rangle|^2$ is the Franck-Condon factor between the $\tilde{A}^2\Pi_{1/2}$ and excited $\tilde{C}^2\Sigma^+_{1/2}$ manifolds.
Taking the standard approximations, $\langle ^2\Pi | \sum_i a_i l_i^+ s_i^-|^2\Sigma^+\rangle\approx A\sqrt{l(l+1)}/2$,  $\langle ^2\Sigma^+| B L_-|^2\Pi\rangle \approx B_{\Sigma} \sqrt{l(l+1)}$, $\langle v''|v '\rangle = 1$, and $l = 1$, we obtain:
\begin{align}
    \gamma^{(2)}_{SR}&\approx \frac{2l(l+1)(A/2) B_\Sigma}{E_\Pi - E_\Sigma}=-\frac{2A B_\Sigma}{E_\Pi - E_\Sigma}.\label{eqn:sr_est}
\end{align}
Table \ref{tab:sr_values} shows a list of inputs used to compute the value of $\gamma$ reported in the main text using eq. \ref{eqn:sr_est}.

Alternatively, this relationship can be used to perform a coarse, semi-empirical estimate for the location of the yet-to-be-observed $\tilde{A}^2\Pi_{1/2}$ electronic manifold in RaOH, which is anticipated to support highly diagonal optical cycling transitions~\cite{zhang2023intensity}. Due to the expected similarity in ligand field effects, we can take $A_{\Sigma,\Pi}\sim [E(^2\Pi_{3/2})-E(^2\Pi_{1/2})]/2\sim 1000$ cm$^{-1}$ from observed RaF splittings~\cite{athanasakis2025electron}. Combined with fine structure parameters of the RaOH $\tilde{C}^2\Sigma^+$ state, this simplified perturber model implies the location of the $\tilde{A}^2\Pi_{1/2}$ state $\sim 3000$ cm$^{-1}$ below the $\tilde{C}^2\Sigma^+$ origin, consistent with reference \textit{ab initio} calculations~\cite{zhang2023relativistic}.

\begin{table}[]
    \centering
    \begin{tabular}{c|cccc|c}
         & $A$ (cm$^{-1}$) & $B_{\Sigma}$ (MHz) & $E_\Pi$ (cm$^{-1}$) & $E_\Sigma$ (cm$^{-1}$) & $\gamma$ (MHz) \\
         \hline
         RaOH($\tilde{C}$) & 1412 & 5783 & 13778 & 15673 & --8618 \\
         RaOD($\tilde{C}$) & 1412 & 5226 & 13778& 15673 & --7787 \\
    \end{tabular}
    \caption{Values used for the perturbative estimate (see eq. \ref{eqn:sr_est}) of the spin rotation parameters ($\gamma$) in the $\tilde{X}$ and $\tilde{C}$ states of RaOH/D reported in the main text. Energies $E_\Pi$ are taken as the calculated spin-orbit free equilibrium energy of the $^2\Pi$ electronic configuration,  $E_\Sigma$ is taken as the equilibrium energies of the $\tilde{X}^2\Sigma$ and $\tilde{C}^2\Sigma$ states for the respective manifolds calculated. $A$ is the spin-orbit parameter, which is obtained from EOM-X2C-CC calculations.}\label{tab:sr_values}
\end{table}

\subsection{Bond length and geometry}

The measured high-resolution parameters also enable direct determination of molecular geometries. Consistent with isoelectronic MOH systems and electronic structure calculation, the spectral pattern for RaOH/D indicates a $C_{\infty v}$ linear geometry. We compute vibrationally averaged bond lengths ($r_0$) using the RaOH/D rotational constant ratios and the effective Hamiltonian parameters listed in Table \ref{tab:sm-raoxfit}. Ground $\tilde{X}$ state parameters are determined via isotope substitution calculation, where we assume identical geometry between RaOH/D and neglect differential zero-point vibrational energy (ZPE) effects. This is accomplished by solving eqs. \ref{eq:H_B} and \ref{eq:D_B} for $r_{0(\text{Ra-O})}$ and $r_{0(\text{O-H})}$:
\begin{align}
    B_{\text{RaOH}} = \frac{h}{8\pi^2 c}\frac{m_\text{Ra}+m_\text{O}+m_\text{H}}{m_\text{Ra}m_\text{O}r_{0(\text{Ra-O})}^2+m_\text{O}m_\text{H}r_{0(\text{O-H})}^2+m_\text{Ra}m_\text{H}[r_{0(\text{Ra-O})}+r_{0(\text{O-H})}]^2},\label{eq:H_B}\\
    B_{\text{RaOD}} = \frac{h}{8\pi^2 c}\frac{m_\text{Ra}+m_\text{O}+m_\text{D}}{m_\text{Ra}m_\text{O}r_{0(\text{Ra-O})}^2+m_\text{O}m_\text{D}r_{0(\text{O-H})}^2+m_\text{Ra}m_\text{D}[r_{0(\text{Ra-O})}+r_{0(\text{O-H})}]^2}.\label{eq:D_B}
\end{align}
In the excitation to the $\tilde{C}$ state, we observe that the RaOD rotational constant shifts by a larger amount compared to RaOH, likely a consequence of changes to vibronic couplings induced by ZPE shifts. We therefore treat the RaOH and RaOD excited state geometry calculations separately and only compute the shift to the Ra-O bond length from the excited state rotational constant, holding O-H fixed to the ground state.

Table \ref{tab:bondlength} lists a summary of ground and excited state bond lengths calculated from experiment and their comparison to values computed from relativistic EOM-CC theory. The scalar-relativistic EOMEA-CCSD calculations of structural parameters have used systematically enlarged TZ, QZ, and 5Z basis sets as constructed in Ref. \cite{zhang2023relativistic} to treat basis-set effects. The scalar-relativistic effects have been taken into account using the spin-free exact two-component theory in its one-electron variant (the SFX2C-1e scheme) \cite{Dyall01,Liu09,Cheng11b}.
The spin-orbit corrections have been included as the differences between the X2CAMF and SFX2C-1e results obtained using the TZ basis. The computations of equilibrium structures have been greatly facilitated by the implementation of analytic SFX2C-1e and X2CAMF-EOM-CCSD gradients \cite{Cheng11b,Zhang23a}. Detailed results are summarized in Table \ref{tab:geocomp}.

\begin{table}[]
    \begin{minipage}[c]{0.47\textwidth}
    \begin{tabular}{c|cc|cc}
         & \multicolumn{2}{c|}{Experiment} & \multicolumn{2}{c}{Theory}\\
         \cline{2-5}
         & $r_0$(Ra-O) & $r_0$(O-H/D) & $r_a$(Ra-O) & $r_a$(O-H/D)\\
         \hline
         RaOH($\tilde{X}$) & \multirow{2}{*}{2.2791(1)} & \multirow{2}{*}{0.9278(2)} & 2.2795 & 0.9213 \\
         RaOD($\tilde{X}$) & & & 2.2780 & 0.9304 \\ \hline
         RaOH($\tilde{C}$) & 2.2861(2) & (fixed) & 2.2861 & 0.9280 \\ 
         RaOD($\tilde{C}$) & 2.2902(2) & (fixed) & 2.2846 & 0.9352
    \end{tabular}
    \caption{Experimental values for vibrationally averaged bond lengths ($r_0$) of RaOH/D determined via isotope substitution and comparison to theory values from EOM-X2C-CC calculation. All data is given in angstroms (\AA). Parentheses indicate $1\sigma$ statistical uncertainties propogated from the least-squares effective Hamiltonian fit. Systematic uncertainties due to neglecting differential zero-point vibrational energy shifts between RaOH and RaOD are expected at similar levels ($\sim 10 ^{-3} - 10^{-4}$ \AA). For calculation of the excited $\tilde{C}$ state bond lengths, $r_0$(O-H/D) is held fixed as described in the text. The theory values for vibrationally averaged bond lengths $r_a$ were obtained by augmenting the computed equilibrium bond lengths with vibrational corrections calculated at the second-order vibrational perturbation theory.} \label{tab:bondlength}
    \end{minipage}\hfill
    \begin{minipage}[c]{0.47\textwidth}
    \begin{tabular}{c|cc|cc}
         & \multicolumn{2}{c|}{$\tilde{X}^2\Sigma$} & \multicolumn{2}{c}{$\tilde{C}^2\Sigma$}\\
         \cline{2-5}
         & $r_e$(Ra-O) & $r_e$(O-H/D) & $r_e$(Ra-O) & $r_e$(O-H/D)\\
         \hline
         TZ & 2.2915 & 0.9533 & 2.2951 & 0.9532 \\
         QZ & 2.2820 & 0.9519 & 2.2881 & 0.9519 \\
         5Z & 2.2787 & 0.9515 & 2.2866 & 0.9516 \\
  $\infty$Z & 2.2752 & 0.9511 & 2.2851 & 0.9512 \\
       +$\Delta$SOC & 2.2703 & 0.9510 & 2.2767 & 0.9512 \\
    \end{tabular}
    \caption{The EOMEA-CCSD/TZ, QZ, and 5Z bond lengths (in {\AA}) for RaOH were calculated using the SFX2C1e scheme to treat scalar-relativistic effects. The spin-orbit coupling corrections ($\Delta$SOC)
    were obtained as the difference between the X2CAMF and SFX2C-1e-EOMEA-CCSD/TZ results.
    These computed equilibrium structures correspond to equilibrium rotational constants of 5845 MHz, 5813 MHz, 5273 MHz, and 5245 MHz for the $X^2\Sigma$ state of RaOH, $C^2\Sigma$ state of RaOH, $X^2\Sigma$ state of RaOD, and $C^2\Sigma$ state of RaOD. Augmentation of these $B_e$ values with vibrational corrections obtained from second-order vibrational perturbation theory calculations give the theoretical $B_0$ values in Table \ref{tab:raoh_params}. } \label{tab:geocomp}
    \end{minipage}
\end{table}

\begin{table}[]
    \centering
      \begin{tabular}{cccccc}
      State & TZ & QZ & 5Z & $\infty$Z & +$\Delta$SOC \\ \hline
      $X^2\Sigma$ & 2.2631 & 2.2501 & 2.2455 & 2.2407 & 2.2360 \\
      $C^2\Sigma$  & 2.2832  & 2.2728  & 2.2693 & 2.2655 & 2.2568  \\
    \end{tabular}
    \caption{The EOMEA-CCSD/TZ, QZ, and 5Z bond lengths (in {\AA}) for RaF were calculated using the SFX2C1e scheme to treat scalar-relativistic effects. 
    The spin-orbit coupling corrections ($\Delta$SOC)
    were obtained as the difference between the X2CAMF and SFX2C-1e-EOMEA-CCSD/TZ results.
    These computed equilibrium structures correspond to equilibrium rotational constants of 5768 MHz and 5662 MHz for the $X^2\Sigma$ state and the $C^2\Sigma$ state. Augmentation of these $B_e$ values with vibrational corrections obtained from second-order vibrational perturbation theory calculations give the theoretical $B_0$ values of 5723 MHz for $v=1$ in the $X^2\Sigma$ state
    and 5647 MHz for $v=0$ in the $C^2\Sigma$ state. } \label{tab:rafgeocomp}
\end{table}

\section{Number estimates}
\label{sec:numbers}
\subsection{Atom number density}
\begin{figure}
    \centering
    \includegraphics[width=0.9\linewidth]{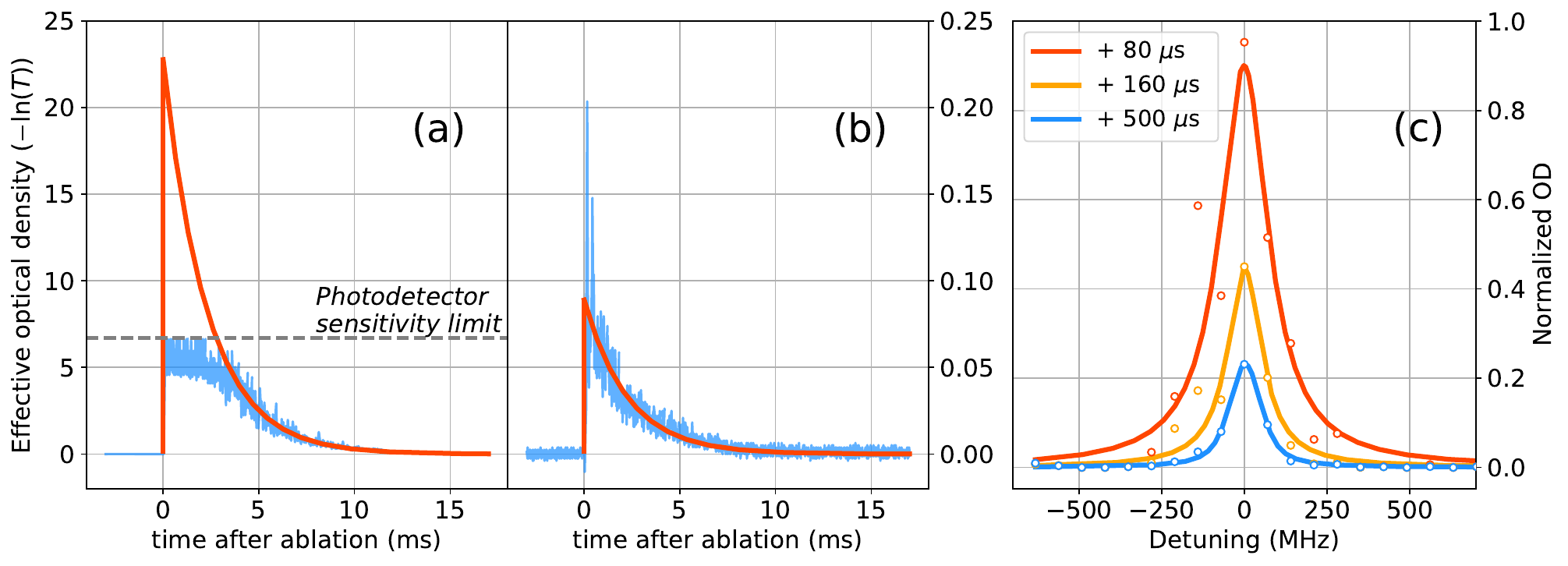}
    \caption{Ra optical densities (OD $=-$ln($T$)) within a single ablation pulse, monitored via laser absorption on the strong $^1P_1-^1S_0$ transition without the presence of the $^3P_1-^1S_0$ laser. Panel (a) shows a typical optical density trace (in blue) when ablating a new spot on the drop casted target. Due to high Ra density, the absorption channel is opaque for several milliseconds and the OD asymptotes to a finite measurement limit set by the smallest possible $T$ distinguishable from zero on the photodetectors ($\sim 1:10^5$). Subsequent shots on the same spot yield lower peak OD, which can be measured below the opacity limit, as shown in panel (b). Note the two orders of magnitude difference in density scales between panel (a) and (b). Data traces in both panels are fitted to a functional decay form (superimposed in red), which is described in the text. For panel (a), this enables the estimation via interpolation of peak densities. Panel (c) depicts OD as a function of detuning from the $^1P_1-^1S_0$ resonance and time after ablation of the drop casted target. The linewidth, which is dominated by Doppler broadening, settles to $\Gamma_D\sim 2\pi\times 100$ MHz after thermalizing with the cryogenic buffer gas. This occurs within 0.5 ms of the initial ablation pulse on the drop casted target.}
    \label{fig:atom_number}
\end{figure}
Following ablation of the drop casted target, Ra optical densities (OD) are monitored in real-time via atomic absorption on the $^1S_0-{}^1P_1$ transition at 483 nm. The relationship between transmission $(T)$ and absorber density $n$ is given by the Beer-Lambert Law:
\begin{align}
    n(t) = -\frac{\text{ln}[T(t)]}{\sigma_\text{abs} L} = \frac{\text{OD}}{\sigma_\text{abs} L}\label{beerslaw}
\end{align}
where $\sigma_\text{abs}$ is the absorption cross section of the $^1S_0-{}^1P_1$ transition and the absorption length is set by the geometry of the buffer gas cell $L = 89$ mm. For an absorption line where the Dopper width ($\Gamma_D$) is much larger than the radiative linewidth ($\Gamma_\text{rad}$), the cross section $\sigma_\text{abs}$ is
\begin{align}
    \sigma_\text{abs} = \frac{\tilde{g}\lambda^2}{4\sqrt{2 \pi}}\frac{\Gamma_\text{rad}}{\Gamma_D}
\end{align}
where in the present case $\Gamma_\text{rad} \sim 2\pi\times 26$ MHz, $\lambda = 483$ nm, and the degeneracy fraction $\tilde{g} = (2J'+1)/(2J''+1)$. Post-thermalization, the width settles to $\Gamma_D\sim 2\pi\times 100$ MHz, as depicted in panel (c) of Fig.~\ref{fig:atom_number}, set mostly by Doppler broadening.

In the apparatus, ablation energies and spot sizes are configured such that majority of material on the drop casted Ra target is removed during the first few ablation shots at each new raster position (see Fig.~\ref{fig:Ra_target}). These initial ablation shots produce peak atomic densities that are sufficiently high to be strongly opaque to the probe laser for several milliseconds (see Fig.~\ref{fig:atom_number}), with transmission below the minimum resolution ($\sim$ one part in $10^5$) of the photodetector system. To interpolate peak OD beyond this detection limit, we fit a simple exponential decay, $n(t) = A\cdot \text{Exp}(-k \cdot (t-t_0))$, to weaker absorption traces taken from repeated shots on the same ablation site where the laser probe is not strongly blocked by the atomic absorption, as shown in panel (b) of Fig.~\ref{fig:atom_number}. The decay constant $k$, which is directly related to the atom diffusion time and should be the same between ablation pulses, is held fixed for a subsequent decay fit of the high-density absorption traces that are initially opaque. 

By interpolation, we can estimate the peak optical densities during initial shots on the ablation raster to be $\sim 23$, as shown in panel (a) of Fig.~\ref{fig:atom_number}. Given an initial Doppler width of $\sim 2\pi\times 400$ MHz, this gives a peak atom density in the range of $\sim 0.5-1\times10^{11}$ atoms cm$^{-3}$. To extend this to a total atom number, we assume that post-ablation the atoms rapidly fill the cell ballistically before diffusing to the walls. This model is consistent with the approximate static helium density ($\sim 5\times 10^{15}$ cm$^{-3}$) inside the cell and prior diffusion studies of closed buffer gas cells \cite{skoff2011diffusion}. Given a total cell volume of 48 cm$^{3}$, we can therefore infer a total atom yield on the order of $\sim 10^{12}$ Ra atoms vaporized per ablation shot.

\subsection{Molecule number estimate}
\begin{figure}
    \centering
    \includegraphics[width=0.51\linewidth]{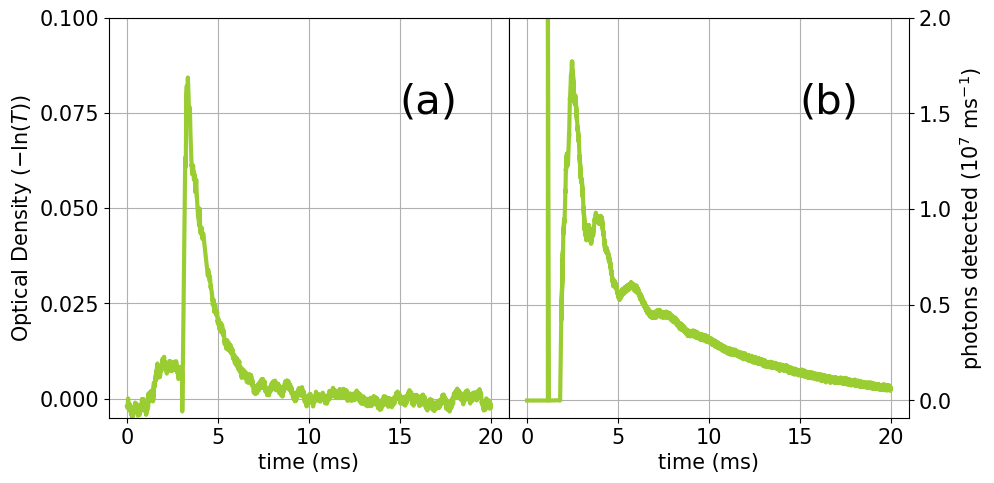} 
    \includegraphics[width=0.33\linewidth]{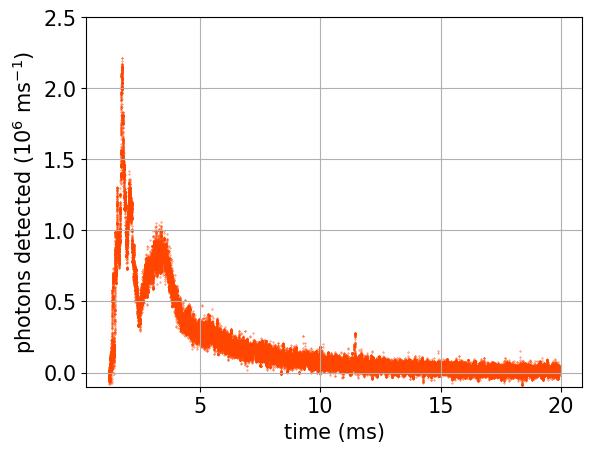}
    \caption{Left: $^{174}$YbOH optical density (a) compared against estimated molecule photons detected per unit time (b). Detection is performed by driving the $\tilde{X}^2\Sigma^+-\tilde{A}^2\Pi_{1/2}$ $^PQ_{12}(N''=1)$ rotational transition.  Right: Estimated photons detected per unit time, obtained from demodulated and background-subtracted RaOH single-frequency fluorescence trace. Detection is performed on the $\tilde{X}^2\Sigma^+-\tilde{C}^2\Sigma^+$ $R_2(N''=4)$ rotational transition.}
    \label{fig:xoh_number}
\end{figure}

The interpolation of molecule number from laser-induced fluorescence signals requires understanding the light collection efficiencies of the optical system and expected photons per emitter. We estimate these factors quasi-empirically by calibrating to signals generated with dense clouds of stable isotope molecules, where fluorescence signals can be directly referenced to density measurements performed via laser absorption. Fig.~\ref{fig:xoh_number} depicts $^{174}$YbOH absorption (a) and fluorescence (b) traces recorded on the $\tilde{X}^2\Sigma^+-\tilde{A}^2\Pi_{1/2}$ $^PQ_{12}(N''=1)$ rotational transition at 17323.5669 cm$^{-1}$ under identical ablation and production conditions. 

The total photon flux on the fluorescence detectors can be readily calculated from the PMT quantum efficiencies and gain factors. The Hamamatsu H13543-20 PMT module used for fluorescence detection has a cathode radiant sensitivity of 70 and 65 mA/W for the 595 nm and 660 nm wavelengths used for YbOH and RaOH off-diagonal detection, respectively. Following corrections for PMT gain and filter transmission efficiencies, we estimate a peak detection rate of at $1.6\times 10^7$ photons ms$^{-1}$ and $2\times 10^6$ photons ms$^{-1}$ and a total photon flux of $6.9\times 10^7$ and $ 3.6\times 10^6$ integrated over the 20 ms pulse for YbOH and RaOH, respectively, as depicted in Fig.~\ref{fig:xoh_number}. 

Extrapolating the detected photon number to molecule number requires considering multiple rate effects on the internal state population of the molecule: the excitation rate from the probe laser, the decay rates (with branching ratios) from the excited state, the diffusion time of the molecules out of the probe laser beam, as well as rotational remixing in the ground state that is induced by helium collisions in the buffer gas environment.
In the limit where the optical excitation rate and rotational remixing rate are fast relative to the diffusion time, the expected number of photons emitted in the detected wavelength range per molecule is $\sim 1$, keeping in mind that a photon will only be detected if the molecule decays into a vibrationally excited state which can therefore scatter no more photons.  Excited vibrational states will relax back down to the ground vibrational state on a timescale too slow to be relevant~\cite{hutzler2012buffer}.

Rotational remixing is less trivial.  If a molecule decays into a rotational state not addressed by the laser it will stop scattering photons, but collisional remixing of rotational states is not negligible on these timescales.  The probability that a molecule will decay down to a different rotational state in the ground vibrational state is $D\approx 1-(1-F)R$, where $F$ is the probably of decaying into a different vibrational state and $R$ is the rotational branching ratio (H\"{o}nl-London factor).  For YbOH and RaOH we have $F\ll 1$, and while $R$ is branch dependent it is typically $\approx1/3$.  Thus we typically have a rotational state change every few photon scatters, and  the photon scattering rate is limited by $\sim\Gamma_\text{remix}$, the rate of rotational remixing, in the limit where this rate is much slower than the excitation rate for a molecule resonantly interacting with the driving laser.  Therefore, the rate of detecting photons, which must arise from vibrational state changing decays, is $\sim \Gamma_\text{remix}F$.
If this timescale is less than the diffusion timescale $\tau_\text{diffusion}$, that is, the length of time needed for the molecules to diffuse to the cell walls where they are lost, then there are multiple opportunities to detect a photon.

Applying some specific numbers, we take $F_\text{YbOH}\sim 9.11\%$~\cite{Mengesha2020YbOHBranching} and $F_\text{RaOH}\sim 1.23\%$~\cite{zhang2023relativistic} for the branching ratios to off-diagonal vibrational detection channels. 
Given a steady-state helium density of $\sim 5 \times 10^{15}$ cm$^{-3}$, a molecule-helium cross section of $\sigma \sim 5 \times 10^{-14}$ cm$^{2}$~\cite{Kozyryev2015Quench}, and relative velocity $v = \sqrt{8 k_b T/\pi \mu}~\sim 200$ m/s, we estimate the mean time between diffusive helium-molecule collisions to be $\tau_c = [\bar{r}\sigma v]^{-1} \sim 0.2$ $\mu$s. We typically expect 10 - 100 collisions before a rotational state change~\cite{hutzler2012buffer}, implying a mean time of $\sim (2 - 20)$~$\mu$s before collisionally induced rotational state changes and a rotational remixing rate on the order of $\Gamma_\text{remix}\sim(5-50)\times10^3$ s$^{-1}$. This implies that the expected photon number per molecule $\langle n\rangle$ approaches 1 as $\tau_\text{remix} \sim 0.01 - 0.01$ ms for YbOH detection and $\tau_\text{remix} \sim 1 - 10$ ms for RaOH detection. These values are comparable to or smaller than the measured molecular diffusion $\tau_\text{diffusion}$ through the probe laser, which suggests that ratio of detected molecules to detected photons in Fig.~\ref{fig:xoh_number} is close to unity. 

Ray tracing optics simulation of the cell and detector geometry indicates an optical collection efficiency of $\sim 1\%$ from a point source at the focal point of the cell lens to the PMT detector. For the YbOH data, this indicates a total number of $7\times 10^9$ molecules addressed by the fluorescence probe. The beam waist radius is $\sim 1.5$ mm at the focal plane, which traces a cylindrical cross section section that is $\sim 40$x smaller from the 3/4 inch diameter bore of the cell. Assuming that the molecules are uniformly distributed through the radial extent of the cell, we can estimate a total molecule number of $2\times 10^{11}$ YbOH molecules. For YbOH, we can validate our fluorescence molecule number estimates by comparing to absorption values recorded under the same production conditions. Using eq. \ref{beerslaw} and the earlier procedure for atom density calculation, we take $\Gamma_\text{rad} \sim 2\pi \times 7.95$ MHz and the initial Doppler broadening at the peak to be $\sim 400$ MHz. The peak density is therefore $1.8\times 10^9$ molecules cm$^{-3}$, implying a total number of $8\times 10^{10}$ molecules inside the cell, which is within a factor of unity of the initial fluorescence estimate. 

We now apply the same calculations to the RaOH data, recorded on the first spot of a new raster position. Given a pulse integrated count of 3.6$\times 10^6$ photons and fast collisional remixing, the optical collection efficiencies imply a total of $3.6\times10^8$ molecules addressed by the laser probe. Using the same beam waist, this yields an estimated total molecule number of $\sim 10^{10}$ distributed through the full cell volume. Incidentally, this is consistent with $\sim 1\%$ chemical conversion efficiency from $10^{12}$ atoms off of a new raster spot computed in the earlier section.

\end{document}